\documentclass{article}
\usepackage{iclr2027_conference,times}

\usepackage{amsmath,amsfonts,bm}

\def\eqref#1{equation~\ref{#1}}

\def\1{\bm{1}}

\DeclareMathAlphabet{\mathsfit}{\encodingdefault}{\sfdefault}{m}{sl}
\SetMathAlphabet{\mathsfit}{bold}{\encodingdefault}{\sfdefault}{bx}{n}

\iclrfinalcopy
\usepackage{hyperref}
\usepackage{url}
\usepackage{xcolor}
\usepackage{kotex}
\usepackage{colortbl}
\usepackage{booktabs}   
\usepackage{multirow}  
\usepackage{multicol}    
\usepackage{array}      
\usepackage{tabularx}    
\usepackage{makecell}    
\usepackage{siunitx}    
\usepackage{graphicx}  
\usepackage{bbm}
\usepackage{subcaption}

\title{DiffuPlex: Accelerating Full-Duplex Spoken Dialog Models via Rolling Masked Diffusion}

\author{Heeseung Kim \\
Department of Artificial Intelligence \\
University of Seoul \\
Seoul, Republic of Korea \\
\texttt{gmltmd789@uos.ac.kr}
}

\begin{document}

\maketitle
\lhead{Preprint. Under review.}

\begin{abstract}
Recent full-duplex spoken dialog models enable simultaneous listening and speaking, but fine-grained models still advance their backbone autoregressively at every interaction frame.
We introduce DiffuPlex, a rolling masked diffusion framework that reduces this sequential computation by predicting multiple future user and assistant frames in a single backbone wake.
DiffuPlex consumes only a confident prefix of each predicted future while interaction continues at the original frame rate.
As user speech arrives, it checks the corresponding user predictions and, when the interaction diverges, preserves already played assistant content while revising only the unplayed future.
We consider two inference policies over the same predictor: DiffuPlex-\textsc{Listen} consumes multiple future frames when they predict assistant silence, whereas DiffuPlex-\textsc{Speak} can also consume predicted assistant speech.
Across full-duplex interaction and spoken-language evaluations, DiffuPlex substantially reduces sequential backbone computation while largely preserving interaction behavior and general capability.
DiffuPlex-\textsc{Listen} and DiffuPlex-\textsc{Speak} achieve $1.46\times$ and $1.59\times$ deployment-path wall-clock speedups and $1.61\times$ and $1.80\times$ Core LM speedups, with all measured backbone invocations completing within the 80ms interaction interval.
Human evaluation shows that \textsc{Listen} preserves speech naturalness and conversational quality, while \textsc{Speak} retains conversational quality with some degradation in speech naturalness.
\end{abstract}
\section{Introduction}
\label{sec:intro}

With the advent of large language models (LLMs)~\citep{singh2026openaigpt5card,qiu2026designqwen38nextarchitectureevaluation}, voice assistants have shifted from cascaded pipelines of automatic speech recognition (ASR), text processing, and text-to-speech (TTS)~\citep{lin-etal-2024-advancing} toward end-to-end speech LLMs that directly take speech as input and generate speech as output~\citep{zhang-etal-2023-speechgpt,NEURIPS2024_ecfa5340,ICLR2025_90d1fc07,zeng2024glm4voiceintelligenthumanlikeendtoend}.
Most early systems remain half-duplex: the user and model take turns, and response generation begins only after the user is judged to have finished speaking~\citep{SileroVAD,liao2025flexduopluggableenablingfullduplex,11463929}.
Human conversation, however, includes interruptions, overlap, and backchannels.
This has motivated full-duplex spoken dialog models that continuously listen while generating speech, allowing both sides to speak and respond even when their speech overlaps~\citep{defossez2024moshispeechtextfoundationmodel,11463413,kim2026raonspeechtechnicalreport,cui2026minicpmo45realtimefullduplex}. 

\begin{figure}[t]
    \centering
    \includegraphics[width=0.9\textwidth]{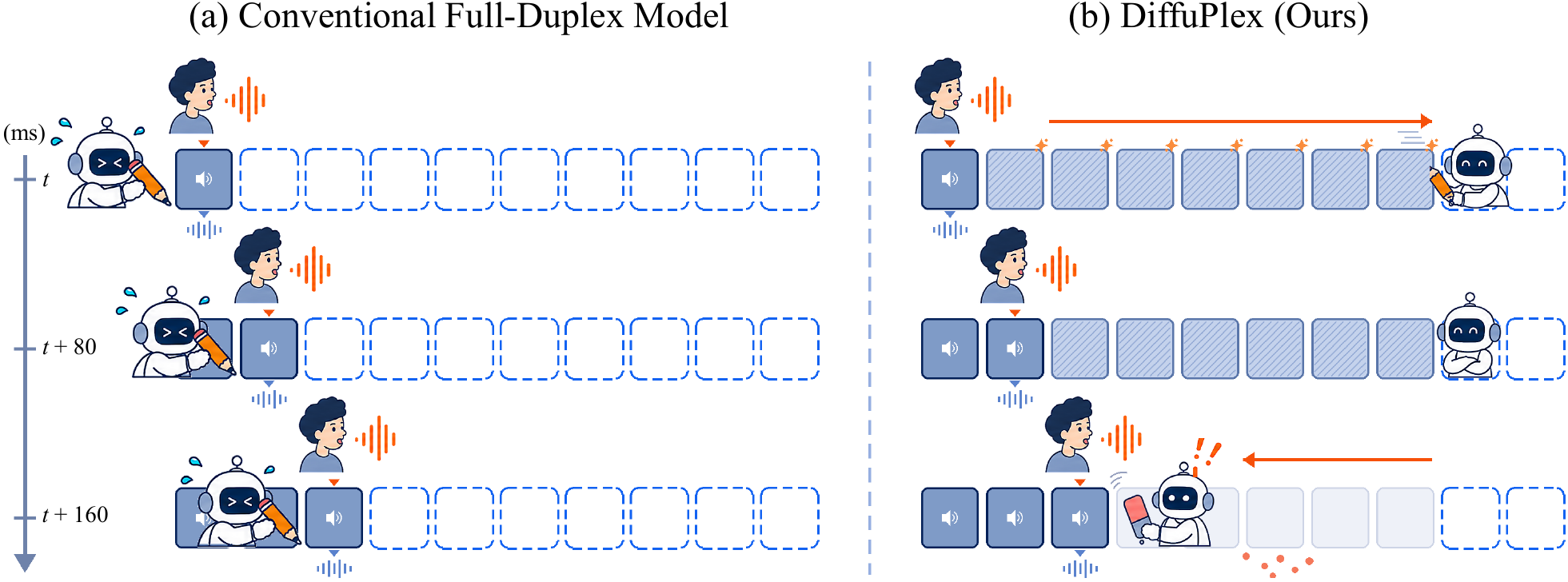}
    \caption{
    Overview of conventional full-duplex generation and DiffuPlex.
    Conventional models advance one interaction frame at a time, requiring a new backbone computation at every step (left).
    DiffuPlex predicts multiple future frames in a single backbone wake, consumes a retained prefix, and revises only the unplayed future when new user input diverges from the prediction (right).
    }
    \label{fig:1_overview}
    \vskip -15pt
\end{figure}

Despite enabling continuous interaction, full-duplex models remain fundamentally autoregressive.
Some serialize user and assistant speech into a single sequence and update relatively coarse chunks~\citep{veluri-etal-2024-beyond,zhang-etal-2025-omniflatten,cui2026minicpmo45realtimefullduplex,kim2026raonspeechtechnicalreport}, while multi-stream models keep parallel user and assistant channels and update both at every interaction step~\citep{defossez2024moshispeechtextfoundationmodel,11463413,ohashi2026multifacetedinteractivityalignmentfullduplex}.
This finer granularity enables prompt reactions to interruptions, overlaps, and backchannels, but increases backbone computation.
Moshi~\citep{defossez2024moshispeechtextfoundationmodel} and PersonaPlex~\citep{11463413}, for example, process one frame every 80ms, requiring a new temporal-backbone forward pass every 80ms throughout the conversation.

Fine-grained interaction, however, does not necessarily require a fresh backbone computation at every frame.
Across 38h of Switchboard dyadic telephone conversations~\citep{225858}, 95.3\% of active conversational time contains only one speaking participant~\citep{10.3389/fpsyg.2015.00731}, leaving long stretches in which the other side remains silent.
Even during speech, upcoming content is planned beyond individual frames: response preparation can begin before the interlocutor finishes speaking~\citep{10.3389/fpsyg.2017.00393}, and speech planning can extend beyond individual words~\citep{MARTIN2010177}.
These observations suggest that several future interaction frames may be predicted together.
Text generation has exploited similar redundancy through speculative decoding~\citep{pmlr-v202-leviathan23a,chen2023acceleratinglargelanguagemodel} and diffusion language models~\citep{NEURIPS2025_48b383b2,ICLR2025_7ede97c3,fu2026nemotronlabsdiffusiontrimodelanguagemodel}.
However, extending such parallel prediction to full-duplex speech is challenging because the future user speech needed to verify the assistant draft has not yet arrived.

We introduce \textbf{DiffuPlex}, which replaces frame-by-frame autoregressive rollout with masked prediction of a short future block.
At each temporal-backbone forward pass, which we call a \textit{backbone wake}, DiffuPlex predicts a block of future user and assistant frames by replacing unavailable future inputs with learned MASK tokens.
It retains a confident contiguous prefix of the assistant draft and plays it at the original interaction rate while user speech continues to arrive.
The user predictions are stored for consistency checking.
If the observed user input diverges from the prediction, DiffuPlex keeps all assistant frames that have already been played and revises only the remaining future.
Figure~\ref{fig:1_overview} illustrates this process of predicting a future block and revising its unplayed portion as the interaction unfolds.
We refer to this repeated process of predicting, consuming, and revising a masked future as \textit{rolling masked diffusion}.

We evaluate DiffuPlex on full-duplex interaction, spoken-language capability, human evaluation, and inference efficiency.
We consider two inference policies over the same predictor: DiffuPlex-\textsc{Listen} consumes multiple future frames only when they predict assistant silence, whereas DiffuPlex-\textsc{Speak} can also consume predicted assistant speech.
On Full-Duplex-Bench v1.0, the two policies advance the interaction by 1.91 and 2.23 frames per backbone wake on average while remaining close to the autoregressive backbone in interaction behavior.
On the fixed 60-clip runtime subset, they achieve 1.46$\times$ and 1.59$\times$ deployment-path wall-clock speedups and 1.61$\times$ and 1.80$\times$ Core LM speedups, with all measured invocations completing within 80ms.
The more conservative \textsc{Listen} policy also preserves human-rated speech and conversational quality, while extending prediction into assistant speech exposes a speech-quality trade-off that we analyze further.
To test whether speedups can be achieved without future prediction, we compare against training-free baselines that defer autoregressive execution and emit silence on skipped frames, at the cost of degraded full-duplex conversational dynamics.
DiffuPlex instead predicts the intervening interaction, reducing backbone calls while better preserving these dynamics.
\section{Related Work}
\label{sec:related_work}

\textbf{Full-Duplex Spoken Dialog Models.}
Early work such as dGSLM~\citep{nguyen-etal-2023-generative} jointly modeled two conversational speech streams to capture turn-taking and overlap.
More recent end-to-end full-duplex models use either single-stream generation, as in SyncLLM~\citep{veluri-etal-2024-beyond}, OmniFlatten~\citep{zhang-etal-2025-omniflatten}, Raon-SpeechChat~\citep{kim2026raonspeechtechnicalreport}, and MiniCPM-o 4.5~\citep{cui2026minicpmo45realtimefullduplex}, or parallel user and assistant streams, as in Moshi~\citep{defossez2024moshispeechtextfoundationmodel} and PersonaPlex~\citep{11463413}.
Recent work further improves turn-taking, interruption handling, and response timing through interactivity training~\citep{pmlr-v267-wu25t,hsiao2026aspirinactionspaceprojection,ohashi2026multifacetedinteractivityalignmentfullduplex,li2026decouplingconversationaldynamicsfullduplex}.
Despite these advances, these models still advance the interaction autoregressively, requiring sequential backbone computation as new frames or chunks are generated.

\textbf{Accelerating Autoregressive Generation.}
Speculative and parallel decoding reduce sequential generation by predicting multiple future tokens and processing them with fewer target-model calls~\citep{NEURIPS2018_c4127b91,pmlr-v202-leviathan23a,chen2023acceleratinglargelanguagemodel}.
Methods such as Medusa~\citep{pmlr-v235-cai24b} and EAGLE~\citep{pmlr-v235-li24bt} improve drafting through additional prediction heads or feature-level prediction, and related approaches have been extended to autoregressive speech generation~\citep{10887855,10888194,11464026}.
PredGen~\citep{li2025predgen} and RelayS2S~\citep{mai2026relays2sdualpathspeculativegeneration} instead begin response generation before the complete spoken input is available, using partial user speech to prepare future output while subsequent input continues to arrive.
However, their speculative responses remain autoregressively generated.
They therefore move response computation earlier rather than reducing the sequential backbone steps required to predict multiple future interaction frames within an end-to-end full-duplex model.

\textbf{Diffusion-Based Parallel Generation.}
Masked and block diffusion replace token-by-token generation with parallel prediction over multiple positions~\citep{NEURIPS2021_958c5305,pmlr-v235-lou24a,NEURIPS2024_eb0b13cc,ICLR2025_7ede97c3}.
Recent work adapts pretrained autoregressive models to masked or block prediction~\citep{ICLR2025_0fa81c3f,cheng-etal-2026-sdar,ICLR2026_d0865cbe,gat2025setblockdecodinglanguage,samragh2025llmknowsfutureuncovering},
while DFlash (Chen et al., 2026a) and DiffuSpec (Li et al., 2026a) use diffusion models as parallel
drafters for autoregressive verification.
Rolling Diffusion Models~\citep{pmlr-v235-ruhe24a} and Diffusion Forcing~\citep{NEURIPS2024_2aee1c41} further consider causal or rolling prediction over temporal sequences.
Related ideas have recently been applied to speech: VocalNet-MDM~\citep{cheng2026vocalnetmdmacceleratingstreamingspeech} accelerates streaming half-duplex speech LLMs, while Chatterbox-Flash~\citep{seo2026chatterboxflashpriorcalibratedblockdiffusion} and DELTA-TTS~\citep{moon2026deltattsadaptingautoregressivemodel} adapt autoregressive TTS models for block or masked diffusion.
In these settings, farther output positions can be predicted from already available input.
Full-duplex dialog introduces a different constraint: future assistant frames depend on user speech that has not yet arrived.
DiffuPlex therefore predicts the missing future interaction jointly and revises only the unplayed assistant future as the corresponding user speech is observed.

\section{DiffuPlex}
\label{sec:method}

\begin{figure}[t]
    \centering
    \includegraphics[width=\textwidth]{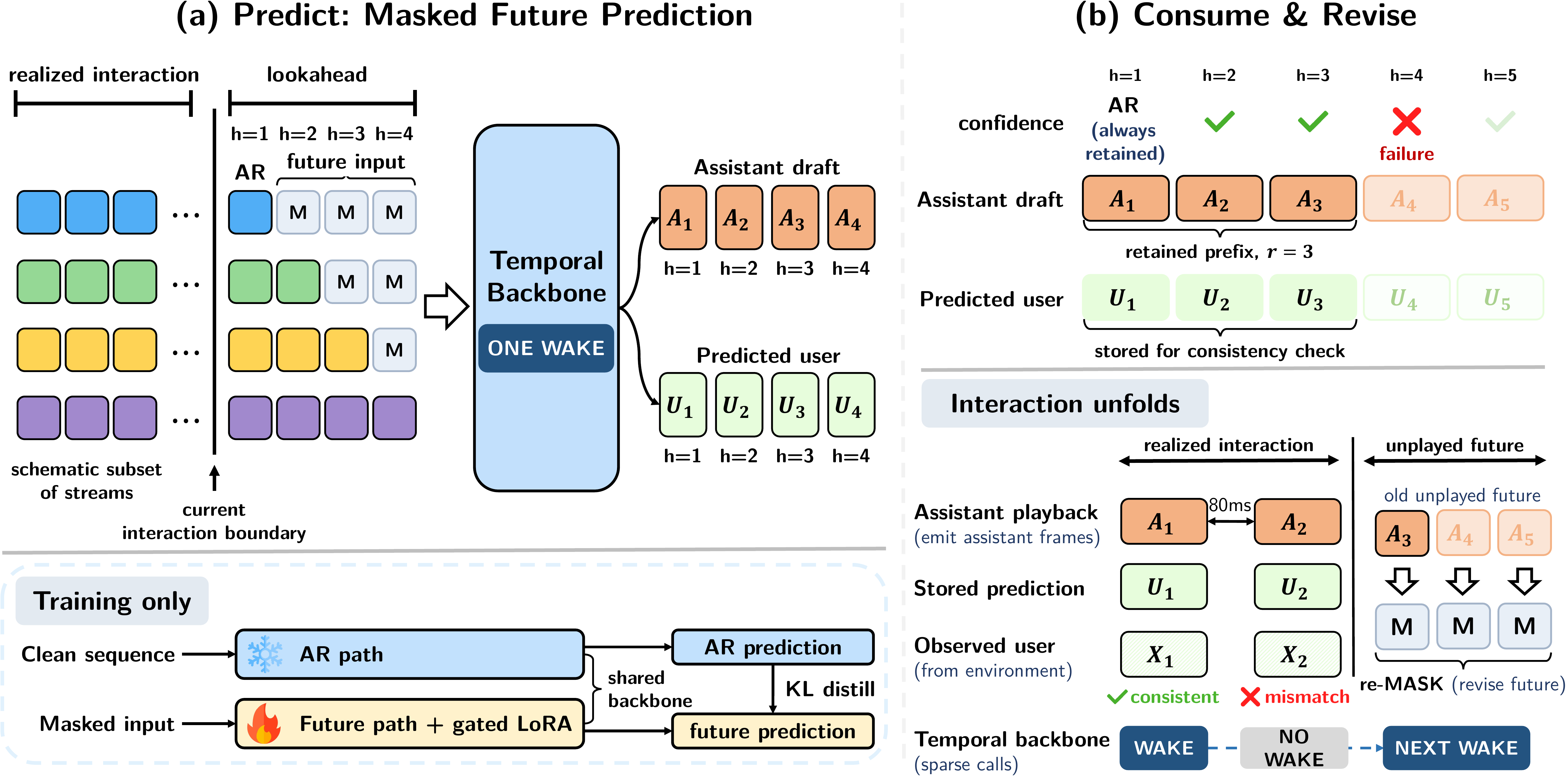}
    \caption{
    Overview of DiffuPlex.
    (a) One backbone wake predicts a masked future block with an assistant draft and user predictions.
    The staircase indicates stream-wise masking of inputs unavailable at that wake; the exact 17-stream layout is shown in Figure~\ref{afig:training_layout} in Appendix~\ref{asubsec:masked_training_layout}.
    (b) A confident prefix is played at the original 80ms rate without additional backbone wakes, while incoming user frames can trigger revision of only the unplayed future.
    }
    \label{fig:diffuplex_overview}
\end{figure}

We introduce \textbf{DiffuPlex}, an approach for reducing sequential temporal-backbone calls in full-duplex spoken dialog models.
We build on \texttt{personaplex-rl-seamless}~\citep{ohashi2026multifacetedinteractivityalignmentfullduplex}, which operates at 12.5Hz, or one interaction frame every 80ms.
At each frame, its temporal backbone jointly predicts the user and assistant streams, requiring one backbone call per frame.
Backbone details are in Appendix~\ref{asubsec:backbone_architecture}.
DiffuPlex instead predicts a short masked future at each backbone call, which we refer to as a \textit{backbone wake}.
As in Figure~\ref{fig:diffuplex_overview}, each wake produces an assistant draft together with user predictions over the same future interval.
DiffuPlex consumes only a confident prefix of the draft and revises only the unplayed future when newly arriving user input is inconsistent with the stored user predictions.
We call this repeated predict-consume-revise process \textit{rolling masked diffusion}, as successive masked-future predictions roll forward with the evolving interaction boundary.

\subsection{Masked Future Prediction}
\label{subsec:masked_future}

Predicting several future interaction frames in one backbone wake is difficult because farther positions depend on assistant outputs that have not yet been generated and user input that has not yet arrived.
DiffuPlex replaces these inputs with learned stream-specific MASK tokens.
At each wake, it predicts $W$ future positions indexed by lookahead $h\in\{1,\ldots,W\}$, where $h=1$ is the ordinary next-frame autoregressive prediction and larger $h$ denotes farther future positions.
As $h$ increases, additional streams become unavailable and are masked according to their roles and native delays.
The exact 17-stream layout is shown in Figure~\ref{afig:training_layout} in Appendix~\ref{asubsec:masked_training_layout}.

To reflect the variable number of frames advanced between backbone wakes at inference time, we divide the training dialog into consecutive blocks with lengths $L_{\mathrm{blk}}$ sampled uniformly from $\{1,\ldots,W\}$.
Lookahead resets to $h=1$ at the start of each block.
Following the clean-context construction of block diffusion~\citep{ICLR2025_7ede97c3}, we concatenate a clean copy of the sequence with a masked copy tiled by these variable-length blocks.
Each masked block conditions on the realized clean context preceding it and predicts a fresh future from that point in the interaction.

To preserve the pretrained backbone while adapting to future masking, we freeze the backbone and train only stream-specific MASK embeddings and future-gated low-rank adapters (LoRA)~\citep{hu2022lora}.
The adapters are activated only for future positions using $g(h)=\mathbf{1}[h\geq2]$, and the same gate is applied to the LoRA modules in the depth transformer, which decodes temporal states into audio codebooks.
Thus, $h=1$ follows the original autoregressive path, while $h\geq2$ uses the learned future path.
All lookaheads are processed in the same temporal-backbone pass, while the depth transformer remains autoregressive over the audio codebooks within each frame.

We train the masked future to match the frozen autoregressive model on the clean sequence.
Let $p^{\mathrm{AR}}$ and $p^{\mathrm{mask}}$ denote the clean-teacher and masked-student distributions.
We minimize
\begin{align}
\mathcal{L}_{\mathrm{distill}}
=&\;
\frac{1}{|\mathcal{V}_{\mathrm{text}}|}
\sum_{i\in\mathcal{V}_{\mathrm{text}}}
\left[
D_{\mathrm{KL}}
\left(
p^{\mathrm{AR}}_{\mathrm{text},i}
\Vert
p^{\mathrm{mask}}_{\mathrm{text},i}
\right)
+
D_{\mathrm{KL}}
\left(
\bar{p}^{\mathrm{AR}}_{\mathrm{text},i}
\Vert
\bar{p}^{\mathrm{mask}}_{\mathrm{text},i}
\right)
\right]
\nonumber\\
&+
\frac{1}{Z_{\mathrm{audio}}}
\sum_{q=1}^{Q} w_q
\sum_{i\in\mathcal{V}_q}
D_{\mathrm{KL}}
\left(
p^{\mathrm{AR}}_{q,i}
\Vert
p^{\mathrm{mask}}_{q,i}
\right),
\label{eq:distill}
\end{align}
where $\mathcal{V}_{\mathrm{text}}$ and $\mathcal{V}_q$ denote positions with valid text and audio targets,
$q\in\{1,\ldots,Q\}$ indexes audio streams with weights $w_q$,
$Z_{\mathrm{audio}}=\sum_{q=1}^{Q}w_q|\mathcal{V}_q|$,
and $\bar{p}_{\mathrm{text}}$ denotes the assistant text distribution grouped into the backbone-defined PAD, EPAD, and lexical states, described in Section~\ref{subsec:rolling_draft_generation}.
The remaining loss and stream details are provided in Appendix~\ref{asubsec:training}.

\subsection{Rolling Draft Generation}
\label{subsec:rolling_draft_generation}

Given a predicted future block, DiffuPlex determines how many positions can be retained before the backbone wakes again.
For a selected assistant text token $y_h$ at lookahead $h$, we define its confidence as
$C_h=\mathrm{softmax}(z_h)_{y_h}$,
where $z_h$ is the logits before temperature scaling or top-$k$ filtering.
Assistant text tokens fall into PAD (no lexical token), EPAD (the final PAD before lexical generation), or lexical states.
The next-frame prediction at $h=1$ is always retained, regardless of confidence.

To control how aggressively DiffuPlex commits to the predicted future, we use two retention policies.
DiffuPlex-\textsc{Listen} consumes future positions only during sustained listening.
We approximate this condition by requiring predicted assistant text states to remain PAD, which indicates the absence of a lexical token.
DiffuPlex-\textsc{Speak} removes this restriction, permitting retention during both listening and speaking.
Let $s(y_h)$ denote the state of $y_h$, and let $\mathcal{S}$ denote the states permitted by each policy: $\{\mathrm{PAD}\}$ for \textsc{Listen} and all three states for \textsc{Speak}.
We retain a contiguous prefix of the $W$ predictions until the first position whose state falls outside $\mathcal{S}$ or whose confidence falls below $\tau_{\mathrm{conf}}$:
\begin{equation}
\tilde r =
\min\left(
\{h\in\{2,\ldots,W\}: s(y_h)\notin\mathcal{S}\ \lor\ C_h<\tau_{\mathrm{conf}}\}
\cup \{W+1\}
\right)-1.
\label{eq:retained_prefix}
\end{equation}
The candidate prefix $\tilde r$ gives the longest run that satisfies both the state and confidence criteria.
For DiffuPlex-\textsc{Listen}, short PAD runs may reflect non-lexical gaps, such as pauses or transitions within speech, rather than a listening period, since PAD only indicates the absence of a lexical token.
We therefore require $\tilde r$ to reach a policy-specific minimum length $m$ before consuming future positions:
\begin{equation}
r=
\begin{cases}
\tilde r, & \tilde r\ge m,\\
1, & \text{otherwise},
\end{cases}
\label{eq:min_retained}
\end{equation}
where the values of $m$ and $\tau_{\mathrm{conf}}$ are given in Section~\ref{subsec:experimental_setup}.

The retained assistant frames are played at the original 80ms rate without additional backbone wakes, while the corresponding user predictions are stored for the revision check in Section~\ref{subsec:user_aware_revision}.
At the next wake, previously predicted positions are replaced by the received user and played assistant frames, which are processed together with a newly masked future in a single forward pass.

\subsection{User-Aware Draft Revision}
\label{subsec:user_aware_revision}

The retained draft may become inconsistent with the evolving user interaction.
DiffuPlex therefore compares stored user predictions with their corresponding observed frames,
without waking the backbone again.
At lookahead $h$, let $x_h$ denote the received user semantic token and $q_h$ its stored predicted distribution, and let $y_h$ and $P_h$ denote the selected assistant text token and its stored distribution.

We revise the draft using two signals: how unlikely the received user token $x_h$ was under the stored user prediction and how weakly the assistant preferred its selected token over valid alternatives.
Let $\mathcal{A}_h(y_h)$ denote the valid alternative assistant text tokens to $y_h$.
Revision is triggered when
\begin{equation}
-\log q_h(x_h)
+\log P_h(\mathcal{A}_h(y_h))
-\log P_h(y_h)
\geq \tau_{\mathrm{rev}}.
\label{eq:revision}
\end{equation}
Here, $\tau_{\mathrm{rev}}$ is the revision threshold, with its value given in Section~\ref{subsec:experimental_setup}.
The first term increases when the received user token is unlikely under the stored prediction, while the remaining terms make revision easier when the assistant is less committed to its selected token.
For $h\geq2$, this criterion can be motivated as an approximate posterior-odds test between
keeping and revising the current draft; its derivation and the construction of $\mathcal{A}_h(y_h)$ are provided in Appendix~\ref{asubsec:inference}.

If the criterion is not met, playback continues from the existing draft without another backbone wake.
Otherwise, DiffuPlex discards only the unplayed future and wakes the backbone from the updated interaction, while all received user frames and played assistant frames remain as context.

\section{Experiments}
\label{sec:experiments}

\subsection{Experimental Setup}
\label{subsec:experimental_setup}

We use \texttt{personaplex-rl-seamless}~\citep{ohashi2026multifacetedinteractivityalignmentfullduplex}, a reinforcement-learning post-trained variant of PersonaPlex~\citep{11463413}, as the backbone for DiffuPlex.
We freeze the original backbone and train only the stream-specific MASK embeddings and future-gated LoRA parameters on Seamless Interaction~\citep{agrawal2025seamlessinteractiondyadicaudiovisual}, a large-scale dataset of face-to-face dyadic interactions.
Unless otherwise noted, we use a maximum lookahead of $W=16$, a confidence threshold of $\tau_{\mathrm{conf}}=0.80$, and a revision threshold of $\tau_{\mathrm{rev}}=-0.5$.
We evaluate two inference policies over the same masked-future predictor.
DiffuPlex-\textsc{Listen} consumes future positions only during sustained listening, with a minimum retained length of $m=6$, whereas DiffuPlex-\textsc{Speak} removes this state restriction and uses $m=2$.
The prediction at $h=1$ is always retained, ensuring that each backbone wake advances the interaction by at least one 80ms frame.
We use the original backbone sampling configuration and fix the generation seed to 0.
Further data processing, training, and inference details are provided in Appendices~\ref{asubsec:data_processing}--\ref{asubsec:inference}.

\subsection{Evaluation}
\label{subsec:evaluation}

We evaluate DiffuPlex along four axes: full-duplex interaction, general spoken-language capability, human evaluation, and inference efficiency.
For full-duplex interaction, we use Full-Duplex-Bench v1.0~\citep{11433838} to evaluate pause handling, backchanneling, turn-taking, and user interruption, and additionally evaluate on Full-Duplex-Bench v1.5~\citep{11463576} under overlapping and interfering speech.
We evaluate general spoken-language capability with VoiceBench~\citep{chen-etal-2026-voicebench} and OpenAudioBench~\citep{li2025baichuanaudiounifiedframeworkendtoend}, reporting both the assistant text stream ($s{\rightarrow}t$) and transcripts of generated speech ($s{\rightarrow}s$).
For human evaluation, we measure speech quality (NMOS) on the assistant response following an interruption and conversational quality (DMOS) over the full interaction, using user-interruption examples from Full-Duplex-Bench v1.0.
Finally, we measure inference efficiency on a fixed 60-clip subset of Full-Duplex-Bench v1.0, with 15 clips per task on a single NVIDIA RTX A6000 GPU.
We report wall-clock streaming, end-to-end GPU, and Core LM time, together with frames advanced per backbone wake and wake-frame latency.
Detailed evaluation, runtime, and human-evaluation protocols are provided in Appendices~\ref{asec:evaluation_details} and~\ref{asec:human_evaluation}.

\subsection{Baselines}
\label{subsec:baselines}

We compare against Moshi~\citep{defossez2024moshispeechtextfoundationmodel}, PersonaPlex~\citep{11463413}, and \texttt{personaplex-rl-seamless}, the autoregressive backbone used by DiffuPlex.
To test whether similar computational savings can be obtained simply by skipping some autoregressive backbone calls, we additionally construct three training-free deferred-execution baselines using the same frozen backbone: \textsc{PAD}, \textsc{Silero}, and \textsc{Smart Turn}.
On a deferred 80ms interaction step, these baselines do not run the backbone or predict the skipped frame; instead, they emit silence for the assistant and resume autoregressive generation at a later step.
\textsc{PAD} skips backbone calls while the assistant predicts silence, \textsc{Silero} while the user is speaking, and \textsc{Smart Turn} until a turn-ending detector predicts that the user has finished.
We cap the interval between autoregressive backbone calls at $R\in\{2,4,6,8\}$ interaction steps.
Further implementation details are in Appendix~\ref{asubsec:deferred_baselines}.

\section{Results}
\label{sec:results}

\begin{table*}[t]
\caption{
Full-duplex interaction and human evaluation on Full-Duplex-Bench v1.0.
RL-Seamless denotes \texttt{personaplex-rl-seamless}, and $c$ is the average number of interaction frames advanced per backbone wake.
Syn. and Can. denote the Synthetic and Candor pause sets.
TOR denotes turn-over rate; Freq., backchannel frequency; JSD, Jensen--Shannon divergence of backchannel timing; Lat., response latency; and Judge, the \textit{gpt-5.6-luna} response-appropriateness score for interruption responses (reasoning effort: none).
NMOS and DMOS denote naturalness and dialog mean opinion scores, reported as mean $\pm$ 95\% confidence interval.
}
\label{tab:full_duplex_results}
\centering
\setlength{\tabcolsep}{3.2pt}
\renewcommand{\arraystretch}{1.05}
\resizebox{\textwidth}{!}{%
\begin{tabular}{l
                c
                cc
                ccc
                cc
                ccc
                cc}
\toprule
& \multicolumn{1}{c}{Efficiency}
& \multicolumn{2}{c}{Pause}
& \multicolumn{3}{c}{Backchannel}
& \multicolumn{2}{c}{Turn-taking}
& \multicolumn{3}{c}{Interruption}
& \multicolumn{2}{c}{Human} \\
\cmidrule(lr){2-2}
\cmidrule(lr){3-4}
\cmidrule(lr){5-7}
\cmidrule(lr){8-9}
\cmidrule(lr){10-12}
\cmidrule(lr){13-14}
Model
& $c \uparrow$
& Syn. $\downarrow$
& Can. $\downarrow$
& TOR $\downarrow$
& Freq.
& JSD $\downarrow$
& TOR $\uparrow$
& Lat. $\downarrow$
& TOR $\uparrow$
& Judge $\uparrow$
& Lat. $\downarrow$
& NMOS $\uparrow$
& DMOS $\uparrow$ \\
\midrule
Moshi
& 1.00
& 0.993 & 1.000
& 0.582 & 0.184 & 0.679
& 1.000 & 0.000
& 0.765 & 3.359 & 1.229
& $3.52 \pm 0.15$ & $2.86 \pm 0.20$ \\

PersonaPlex
& 1.00
& 0.307 & 0.343
& 0.200 & 0.058 & 0.824
& 0.941 & 0.285
& 0.930 & 4.328 & 0.218
& $3.90 \pm 0.12$ & $4.10 \pm 0.13$ \\

RL-Seamless
& 1.00
& 0.299 & 0.301
& 0.218 & 0.163 & 0.740
& 1.000 & 0.124
& 0.995 & 4.286 & 0.281
& $3.93 \pm 0.12$ & $4.09 \pm 0.14$ \\
\midrule
DiffuPlex-\textsc{Listen}
& 1.91
& 0.277 & 0.343
& 0.109 & 0.140 & 0.753
& 1.000 & 0.173
& 0.990 & 4.328 & 0.329
& $3.92 \pm 0.11$ & $4.08 \pm 0.13$ \\

DiffuPlex-\textsc{Speak}
& 2.23
& 0.263 & 0.347
& 0.145 & 0.154 & 0.738
& 1.000 & 0.156
& 0.995 & 4.357 & 0.347
& $3.54 \pm 0.12$ & $4.07 \pm 0.13$ \\
\bottomrule
\end{tabular}%
}
\end{table*}
\subsection{Full-Duplex Interaction}
\label{subsec:full_duplex_results}

Table~\ref{tab:full_duplex_results} shows that DiffuPlex reduces backbone wake frequency while remaining close to the autoregressive backbone in full-duplex interaction behavior.
We use $c$ to denote the average number of interaction frames advanced per backbone wake; standard autoregressive generation has $c=1$.
DiffuPlex-\textsc{Listen} and DiffuPlex-\textsc{Speak} increase this to $c=1.91$ and $c=2.23$, respectively, while remaining close to \texttt{personaplex-rl-seamless} across pause handling, backchanneling, turn-taking, and user interruption.
Results on Full-Duplex-Bench v1.5 are provided in Appendix~\ref{asubsec:additional_full_duplex}.

The two settings expose a trade-off in how aggressively predicted futures are consumed.
DiffuPlex-\textsc{Listen} closely preserves both speech quality and conversational quality, with an NMOS of $3.92\pm0.11$ and a DMOS of $4.08\pm0.13$, compared with $3.93\pm0.12$ and $4.09\pm0.14$ for \texttt{personaplex-rl-seamless}.
DiffuPlex-\textsc{Speak} increases $c$ to 2.23 while maintaining a DMOS of
$4.07\pm0.13$, but its NMOS decreases to $3.54\pm0.12$, remaining close to Moshi at
$3.52\pm0.15$.
We analyze the source of this speech-quality degradation in Section~\ref{subsec:draft_analysis}.

Fewer backbone wakes do not necessarily imply lower runtime, because each DiffuPlex wake predicts over a wider future span and is therefore more expensive than a single autoregressive wake.
We next examine whether the reduction in wake frequency outweighs this increased per-wake cost.

\begin{table*}[t]
\caption{
Measured inference efficiency on the fixed 60-clip runtime subset.
Streaming wall is the elapsed wall-clock time of the deployment-path streaming loop, including CPU/Python control, synchronization, and GPU execution.
E2E GPU and Core LM denote accumulated GPU time for the full inference pipeline and the language-model backbone, respectively.
Spd. denotes speedup relative to RL-Seamless, and Rel. time/wake is the Core LM time per backbone wake normalized by RL-Seamless.
Wake-frame p99 is the 99th-percentile wall-clock latency of frames that invoke the backbone, and $>80$ms is the fraction of such frames exceeding the 80ms interaction interval.
}
\label{tab:runtime_results}
\centering
\setlength{\tabcolsep}{4pt}
\renewcommand{\arraystretch}{1.05}
\resizebox{\textwidth}{!}{%
\begin{tabular}{l
                c
                cc
                cc
                cc
                ccc}
\toprule
& \multicolumn{1}{c}{Wake}
& \multicolumn{2}{c}{Streaming wall}
& \multicolumn{2}{c}{E2E GPU}
& \multicolumn{2}{c}{Core LM}
& \multicolumn{3}{c}{Wake behavior} \\
\cmidrule(lr){2-2}
\cmidrule(lr){3-4}
\cmidrule(lr){5-6}
\cmidrule(lr){7-8}
\cmidrule(lr){9-11}
Model
& $c \uparrow$
& Time (s) $\downarrow$
& Spd. $\uparrow$
& Time (s) $\downarrow$
& Spd. $\uparrow$
& Time (s) $\downarrow$
& Spd. $\uparrow$
& Rel. time/wake
& p99 (ms) $\downarrow$
& $>80$ms $\downarrow$ \\
\midrule
Moshi
& 1.00
& 638.7 & 1.168$\times$
& 638.3 & 1.168$\times$
& 550.1 & 1.194$\times$
& 0.84$\times$
& -- & -- \\

PersonaPlex
& 1.00
& 744.1 & 1.002$\times$
& 743.7 & 1.002$\times$
& 656.8 & 1.000$\times$
& 1.00$\times$
& -- & -- \\

RL-Seamless
& 1.00
& 745.7 & 1.000$\times$
& 745.3 & 1.000$\times$
& 656.8 & 1.000$\times$
& 1.00$\times$
& 46.85 & 0\% \\
\midrule
DiffuPlex-\textsc{Listen}
& 1.98
& 512.4 & 1.455$\times$
& 511.1 & 1.458$\times$
& 408.4 & 1.608$\times$
& 1.23$\times$
& 61.34 & 0\% \\

DiffuPlex-\textsc{Speak}
& 2.26
& 469.1 & 1.590$\times$
& 467.8 & 1.593$\times$
& 365.5 & 1.797$\times$
& 1.26$\times$
& 61.41 & 0\% \\
\bottomrule
\end{tabular}
}
\end{table*}

\subsection{Runtime Efficiency}
\label{subsec:runtime_results}

Table~\ref{tab:runtime_results} shows that the reduction in backbone wakes translates into substantial measured speedups.
On the fixed 60-clip runtime subset, DiffuPlex-\textsc{Listen} and DiffuPlex-\textsc{Speak} advance $c=1.98$ and $c=2.26$ interaction frames per backbone wake, respectively.
This reduces accumulated Core-LM time from 656.8s for \texttt{personaplex-rl-seamless} to 408.4s and 365.5s, corresponding to $1.608\times$ and $1.797\times$ speedups.
The reduction also translates into deployment-path wall-clock speedups of $1.455\times$ and $1.590\times$ for DiffuPlex-\textsc{Listen} and DiffuPlex-\textsc{Speak}, respectively.
Considering device-side computation alone, this yields end-to-end GPU speedups of $1.458\times$ and $1.593\times$.
Moshi predicts only assistant-side tokens, whereas the PersonaPlex family jointly models both user and assistant streams, giving Moshi a lighter per-frame prediction workload.

These gains do not come from making each DiffuPlex backbone invocation cheaper.
The pooled measurements imply 1.23$\times$ and 1.26$\times$ as much Core LM time per
backbone wake for DiffuPlex-\textsc{Listen} and DiffuPlex-\textsc{Speak}, respectively.
Instead, masked-future prediction amortizes this larger per-wake cost over multiple interaction
frames, reducing accumulated Core LM time while sustaining the same 12.5Hz interaction stream.
Meeting the 80ms deadline establishes real-time feasibility, but systems that satisfy the same
deadline can still differ substantially in accumulated GPU time.
DiffuPlex therefore reduces accumulated GPU and Core LM time over the course of an interaction.

The larger individual wakes must still satisfy this real-time constraint.
The 99th-percentile wall-clock latency of frames that invoke the backbone is 61.34ms for DiffuPlex-\textsc{Listen} and 61.41ms for DiffuPlex-\textsc{Speak}, and no measured wake frame exceeds 80ms.
Thus, DiffuPlex reduces accumulated Core LM time while preserving the original real-time interaction cadence.

\subsection{Masked Future Prediction vs.\ Deferred Execution}
\label{subsec:deferred_results}

\begin{figure}[t]
    \centering
    \begin{subfigure}[t]{0.32\columnwidth}
        \centering
        \includegraphics[width=0.90\linewidth]{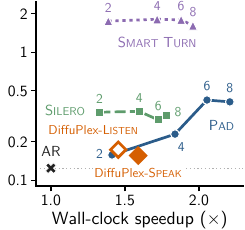}
        \caption{Turn-taking latency ($\downarrow$)}
        \label{fig:deferred_turn}
    \end{subfigure}
    \hfill
    \begin{subfigure}[t]{0.32\columnwidth}
        \centering
        \includegraphics[width=0.90\linewidth]{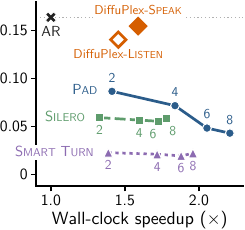}
        \caption{Backchannel frequency}
        \label{fig:deferred_bc}
    \end{subfigure}
    \hfill
    \begin{subfigure}[t]{0.32\columnwidth}
        \centering
        \includegraphics[width=0.90\linewidth]{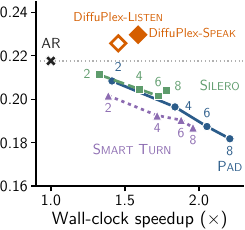}
        \caption{Assistant speaking fraction}
        \label{fig:deferred_speaking}
    \end{subfigure}
\caption{
Efficiency-interaction trade-offs under deferred execution.
Deferred baselines vary $R\in\{2,4,6,8\}$, and deferred frames are forced to silence.
Silero and Smart Turn replay precomputed controller decisions; controller inference cost is measured separately.
}
    \label{fig:deferred_tradeoff}
\end{figure}

We next ask whether similar efficiency gains can be obtained without predicting future interaction, simply by skipping autoregressive backbone calls and emitting silence during skipped frames.
Figure~\ref{fig:deferred_tradeoff} compares DiffuPlex with the \textsc{PAD}, \textsc{Silero}, and \textsc{Smart Turn} deferred-execution baselines over $R\in\{2,4,6,8\}$.
Increasing $R$ allows up to $R-1$ frames to be skipped between backbone calls, improving wall-clock speedup but increasingly degrading interaction behavior.
In particular, deferred execution suppresses backchanneling and reduces assistant speaking activity, while turn-taking latency increases substantially for several controllers.
For example, at $R=6$, \textsc{PAD} reaches a $2.05\times$ wall-clock speedup, but backchannel frequency drops from 0.163 to 0.048 and turn-taking latency increases from 0.124s to 0.423s relative to the backbone.

DiffuPlex largely mitigates this trade-off by predicting interaction between backbone invocations rather than replacing skipped frames with silence.
DiffuPlex-\textsc{Listen} and DiffuPlex-\textsc{Speak} achieve $1.455\times$ and $1.590\times$ wall-clock speedups while remaining much closer to the original backchannel, turn-taking, and speaking behavior.
These results show that deferring backbone calls while emitting silence is insufficient in these baselines; predicting what occurs between backbone invocations better preserves full-duplex behavior.
DiffuPlex can further adjust this efficiency-interaction trade-off through its confidence and revision thresholds, which we analyze in Appendices~\ref{asubsec:confidence_retention} and~\ref{asubsec:revision}.

\begin{table}[t]
\caption{
General spoken-language capability. Higher is better.
}
\label{tab:general_results}
\centering
\small
\setlength{\tabcolsep}{4.5pt}
\renewcommand{\arraystretch}{1.05}
\begin{tabular}{lcccc}
\toprule
& \multicolumn{2}{c}{VoiceBench}
& \multicolumn{2}{c}{OpenAudioBench} \\
\cmidrule(lr){2-3}
\cmidrule(lr){4-5}
Model
& $s{\rightarrow}t$
& $s{\rightarrow}s$
& $s{\rightarrow}t$
& $s{\rightarrow}s$ \\
\midrule
\texttt{personaplex-rl-seamless}
& 31.95 & 31.61
& 35.66 & 34.03 \\

DiffuPlex-\textsc{Listen}
& 31.93 & 31.52
& 36.71 & 33.53 \\

DiffuPlex-\textsc{Speak}
& 31.20 & 31.04
& 35.82 & 32.12 \\
\bottomrule
\end{tabular}
\end{table}
\begin{table}[t]
\caption{
Ablation of confidence-based retention and draft revision on Full-Duplex-Bench v1.0.
}
\label{tab:inference_ablation}
\centering
\small
\setlength{\tabcolsep}{3.5pt}
\renewcommand{\arraystretch}{1.03}
\begin{tabular}{llccccc}
\toprule
Policy & Variant
& $c\uparrow$
& BC Freq.
& Turn Lat.$\downarrow$
& Int. Judge$\uparrow$
& Int. Lat.$\downarrow$ \\
\midrule
\multirow{3}{*}{\textsc{Listen}}
& DiffuPlex
& 1.91 & .140 & .173 & 4.328 & .329 \\
& w/o confidence
& 2.11 & .191 & .170 & 4.398 & .314 \\
& w/o revision
& 2.49 & .078 & .440 & 4.372 & .744 \\
\midrule
\multirow{3}{*}{\textsc{Speak}}
& DiffuPlex
& 2.23 & .154 & .156 & 4.357 & .347 \\
& w/o confidence
& 2.80 & .191 & .209 & 3.550 & .281 \\
& w/o revision
& 3.08 & .080 & .480 & 4.251 & .515 \\
\bottomrule
\end{tabular}
\end{table}

\subsection{General Spoken-Language Capability}
\label{subsec:general_results}

Table~\ref{tab:general_results} shows that DiffuPlex largely preserves the general spoken-language capability of \texttt{personaplex-rl-seamless}.
DiffuPlex-\textsc{Listen} remains nearly unchanged on
VoiceBench and OpenAudioBench, while DiffuPlex-\textsc{Speak} shows moderate degradation,
most notably on OpenAudioBench $s\!\to\!s$.
Full per-task results are provided in Appendix~\ref{asubsec:additional_general}.

\subsection{Analysis of Future Drafts}
\label{subsec:draft_analysis}

\textbf{Rolling draft behavior.}
We first examine how confidence filtering and revision affect the use of predicted futures.
As shown in Table~\ref{tab:inference_ablation}, the policy-specific no-confidence variants allow more aggressive
retention, with different effects across the two policies.
For \textsc{Speak}, $c$ increases from 2.23 to 2.80, but the interruption Judge score decreases from 4.357 to 3.550.
For \textsc{Listen}, where retention remains restricted to PAD states, removing confidence increases $c$ more modestly from 1.91 to 2.11 while leaving interaction behavior similar.
Removing revision increases $c$ from 1.91 to 2.49 for \textsc{Listen} and from 2.23 to 3.08 for \textsc{Speak}, but substantially reduces responsiveness: backchannel frequency falls to 0.078 and 0.080, while turn-taking latency increases to 0.440s and 0.480s, respectively.
Thus, the retention constraints limit aggressive future consumption, particularly for \textsc{Speak},
while revision is important for preserving responsiveness to newly observed user speech.
Sensitivity to the confidence threshold $\tau_{\rm conf}$ and revision threshold $\tau_{\rm rev}$ is reported in Appendices~\ref{asubsec:confidence_retention} and~\ref{asubsec:revision}.

We next examine why DiffuPlex wakes the backbone again after producing a draft.
A wake occurs either when the retained draft is exhausted or when new user speech triggers revision.
When the draft is exhausted, DiffuPlex advances only 1.17 frames on average for \textsc{Listen} and 1.48 for \textsc{Speak}; 98.2\% and 81.5\% of these wakes, respectively, occur after a single frame.
Revision-triggered wakes instead occur after several draft frames have already been consumed, with mean advancements of 4.24 and 3.79 frames and medians of 4 and 3.
In other words, a short wake interval usually means that only a short prefix was retained in the first place, rather than that a longer retained draft was quickly revised.
When revision does occur, several retained frames have typically been consumed.
Additional wake-level diagnostics and revision analyses are provided in
Appendix~\ref{asubsec:wake_dynamics}.

\textbf{Why does future speech degrade?}
We next investigate the speech-quality degradation of DiffuPlex-\textsc{Speak}.
\texttt{personaplex-rl-seamless} models assistant audio using a semantic codebook, cb0, followed by seven acoustic codebooks, cb1--7.
Because cb1--7 are natively delayed by one temporal step, the cb0 corresponding to a given physical audio frame appears one temporal position earlier in the model sequence and is therefore already available when the autoregressive model predicts cb1--7 for that frame.
DiffuPlex preserves this native delay pattern, but at future depths $h\geq2$ all assistant streams are masked, so future cb1--7 must be predicted without the same-frame cb0 that would normally condition them.
To test whether this missing conditioning contributes to the degradation, we use the 199 valid user-interruption examples from Full-Duplex-Bench v1.0 and replace only the masked cb0 with the ground-truth cb0 for the physical frame whose cb1--7 are being predicted.
During assistant speech, this reduces the cb1--7 negative log-likelihood from 5.702 to 5.611
($\Delta=0.091$, 95\% CI [0.081, 0.101]), compared with a much smaller reduction during silence ($\Delta=0.029$).
These results show that same-frame semantic conditioning has a substantially larger effect on acoustic-token prediction during speech than during silence.

We then test whether restoring autoregressive audio-token conditioning improves generated audio.
Starting from each DiffuPlex-\textsc{Speak} output, we teacher-force its generated text and cb0 while recomputing the sequence frame by frame through the original autoregressive backbone.
We then use the resulting temporal representations to regenerate only cb1--7 with the unchanged depth transformer.
UTMOSv2, a neural predictor of perceptual speech quality on a 5-scale~\citep{10832315}, increases from 2.646 to 2.894.
Regenerating cb0--7 while keeping only the generated text fixed yields 2.903, compared with 2.919 for fully autoregressive generation.
Thus, most of the measured quality loss is recovered by restoring autoregressive audio-token conditioning even when the text and cb0 generated by DiffuPlex are unchanged.
Additional intervention controls are in
Appendix~\ref{asubsec:future_speech}.

\section{Conclusion}

We introduced DiffuPlex, a masked-future prediction framework for reducing sequential forward passes in autoregressive full-duplex speech models.
Rather than invoking the backbone at every interaction frame, DiffuPlex predicts multiple future positions in a single wake, consumes a confident prefix, and revises only the unplayed future when observed user speech deviates from the prediction.
Experiments show that this rolling generation reduces backbone wakes and measured Core LM time
while largely preserving full-duplex interaction behavior and general spoken-language capability.
Our analyses show that confidence-based retention and revision play complementary roles in controlling future consumption, and identify missing autoregressive audio-token conditioning as a source of degradation in deeper future speech.
These results suggest that predicting interaction ahead can provide a path toward more compute-efficient real-time full-duplex generation.

\subsection*{AI Use Statement}

The research problem, core ideas, and methodological development in this work were conceived and developed by the authors.
We used Claude Code as a coding assistant to support implementation of the proposed method.
The first author manually reviewed and tested the resulting implementation and verified the experimental outputs and reported results.
Claude Code was also used to assist with visualization, including developing figure layouts and producing or refining scientific illustrations, as well as with implementation of the project page.

We used ChatGPT to assist with writing after the authors had developed the paper outline, organization, and initial drafts.
Its use was primarily limited to improving clarity and readability, paraphrasing, and revising author-written text.
For related work, the authors first conducted the literature review independently, primarily by following citations from relevant prior work.
Generative AI tools were subsequently used as a supplementary check to identify potentially relevant work that might have been overlooked and to help organize the related-work discussion.
The authors independently inspected the original papers before including or citing them.

Generative AI tools were not used to originate the research problem or the core methodological ideas presented in this work.
All AI-assisted code, text, figures, and other artifacts were reviewed by the authors.
The authors take responsibility for the final content of the paper, including all claims, results, and artifacts produced with the assistance of generative AI.
\subsection*{Ethics Statement}

This work studies computational acceleration for full-duplex spoken dialog models.
Because the backbone supports voice-conditioned generation, our evaluation requires an acoustic voice prompt;
we use the official voice prompt provided with the backbone, without constructing new speaker
identities or attempting to imitate specific individuals.
For human evaluation, participants recruited through Amazon Mechanical Turk listened to generated audio samples and provided perceptual ratings.
Our work does not aim to develop or evaluate capabilities for impersonation or other deceptive uses.
Nevertheless, as with other spoken-output generative models, the underlying systems could potentially be misused to produce misleading synthetic speech.
Their use should therefore follow appropriate safeguards as well as the licenses and usage conditions of the underlying models and datasets.
The external models, datasets, and evaluation resources used in this work, together with their licenses and sources, are documented in Table~\ref{atab:artifacts}.
\subsection*{Reproducibility statement}
We provide reproducibility information throughout the main paper and Appendices.
Section~\ref{sec:method} specifies the DiffuPlex training and inference procedures, including masked-future prediction, confidence-based retention, and user-aware revision, while Appendices provide implementation, data-processing, evaluation, ablation, and runtime-measurement details.
Table~\ref{atab:artifacts} lists the external datasets, pretrained models, evaluation tools, licenses, and public sources used in this work.
Selected generated audio and qualitative examples are available on the anonymous project page at
\url{https://diffuplex.github.io/}.
We plan to release the implementation and evaluation scripts after the review process.

\bibliography{iclr2027_conference}

@misc{singh2026openaigpt5card,
      title={OpenAI GPT-5 System Card}, 
      author={OpenAI},
      year={2026},
      eprint={2601.03267},
      archivePrefix={arXiv},
      primaryClass={cs.CL},
      url={https://arxiv.org/abs/2601.03267}, 
}

@misc{qiu2026designqwen38nextarchitectureevaluation,
      title={On the Design of Qwen3.8-Next Architecture: Evaluation, Efficiency, and Training Stability}, 
      author={Zihan Qiu and Zekun Wang and Xiao Li and Yanpeng Li and Yang Xu and Yixuan Wang and Huaqing Zhang and Rui Men and Bochao Mao and Chengruidong Zhang and Fan Zhou and Hao Luo and Haofeng Huang and Haoran Lian and Haoyan Huang and Hongqing Chen and Jianwei Zhang and Jing Xu and Junjie Wang and Langshi Chen and Liangyu Wang and Linlang Jiang and Man Yuan and Minmin Sun and Peng Jin and Siqi Zhang and Siyu Wang and Xingzhang Ren and Yakai Wang and Yi Zhang and Yiming Dong and Yizhong Cao and Yubo Ma and Yunfei Mao and Bo Zheng and Dayiheng Liu},
      year={2026},
      eprint={2608.30320},
      archivePrefix={arXiv},
      primaryClass={cs.CL},
      url={https://arxiv.org/abs/2608.30320}, 
}

@inproceedings{lin-etal-2024-advancing,
    title = "Advancing Large Language Models to Capture Varied Speaking Styles and Respond Properly in Spoken Conversations",
    author = "Lin, Guan-Ting  and
      Chiang, Cheng-Han  and
      Lee, Hung-yi",
    editor = "Ku, Lun-Wei  and
      Martins, Andre  and
      Srikumar, Vivek",
    booktitle = "Proceedings of the 62nd Annual Meeting of the Association for Computational Linguistics (Volume 1: Long Papers)",
    month = aug,
    year = "2024",
    address = "Bangkok, Thailand",
    publisher = "Association for Computational Linguistics",
    url = "https://aclanthology.org/2024.acl-long.358/",
    doi = "10.18653/v1/2024.acl-long.358",
    pages = "6626--6642"
}

@inproceedings{zhang-etal-2023-speechgpt,
    title = "{S}peech{GPT}: Empowering Large Language Models with Intrinsic Cross-Modal Conversational Abilities",
    author = "Zhang, Dong  and
      Li, Shimin  and
      Zhang, Xin  and
      Zhan, Jun  and
      Wang, Pengyu  and
      Zhou, Yaqian  and
      Qiu, Xipeng",
    editor = "Bouamor, Houda  and
      Pino, Juan  and
      Bali, Kalika",
    booktitle = "Findings of the Association for Computational Linguistics: EMNLP 2023",
    month = dec,
    year = "2023",
    address = "Singapore",
    publisher = "Association for Computational Linguistics",
    url = "https://aclanthology.org/2023.findings-emnlp.1055/",
    doi = "10.18653/v1/2023.findings-emnlp.1055",
    pages = "15757--15773"
}

@inproceedings{NEURIPS2024_ecfa5340,
 author = {Kim, Heeseung and Seo, Soonshin and Jeong, Kyeongseok and Kwon, Ohsung and Kim, Soyoon and Kim, Jungwhan and Lee, Jaehong and Song, Eunwoo and Oh, Myungwoo and Ha, Jung-Woo and Yoon, Sungroh and Yoo, Kang Min},
 booktitle = {Advances in Neural Information Processing Systems},
 doi = {10.52202/079017-4166},
 editor = {A. Globerson and L. Mackey and D. Belgrave and A. Fan and U. Paquet and J. Tomczak and C. Zhang},
 pages = {131072--131103},
 publisher = {Curran Associates, Inc.},
 title = {Paralinguistics-Aware Speech-Empowered Large Language Models for Natural Conversation},
 url = {https://proceedings.neurips.cc/paper_files/paper/2024/file/ecfa5340fd896e5314bc5e132b5dd5ca-Paper-Conference.pdf},
 volume = {37},
 year = {2024}
}

@inproceedings{ICLR2025_90d1fc07,
 author = {Fang, Qingkai and Guo, Shoutao and Zhou, Yan and Ma, Zhengrui and Zhang, Shaolei and Feng, Yang},
 booktitle = {International Conference on Learning Representations},
 editor = {Y. Yue and A. Garg and N. Peng and F. Sha and R. Yu},
 pages = {57607--57624},
 title = {LLaMA-Omni: Seamless Speech Interaction with Large Language Models},
 url = {https://proceedings.iclr.cc/paper_files/paper/2025/file/90d1fc07f46e31387978b88e7e057a31-Paper-Conference.pdf},
 volume = {2025},
 year = {2025}
}

@misc{zeng2024glm4voiceintelligenthumanlikeendtoend,
      title={GLM-4-Voice: Towards Intelligent and Human-Like End-to-End Spoken Chatbot}, 
      author={Aohan Zeng and Zhengxiao Du and Mingdao Liu and Kedong Wang and Shengmin Jiang and Lei Zhao and Yuxiao Dong and Jie Tang},
      year={2024},
      eprint={2412.02612},
      archivePrefix={arXiv},
      primaryClass={cs.CL},
      url={https://arxiv.org/abs/2412.02612}, 
}

@misc{defossez2024moshispeechtextfoundationmodel,
      title={Moshi: a speech-text foundation model for real-time dialogue}, 
      author={Alexandre Défossez and Laurent Mazaré and Manu Orsini and Amélie Royer and Patrick Pérez and Hervé Jégou and Edouard Grave and Neil Zeghidour},
      year={2024},
      eprint={2410.00037},
      archivePrefix={arXiv},
      primaryClass={eess.AS},
      url={https://arxiv.org/abs/2410.00037}, 
}

@INPROCEEDINGS{11463413,
  author={Roy, Rajarshi and Raiman, Jonathan and Lee, Sang-Gil and Ene, Teodor-Dumitru and Kirby, Robert and Kim, Sungwon and Kim, Jaehyeon and Catanzaro, Bryan},
  booktitle={ICASSP 2026 - 2026 IEEE International Conference on Acoustics, Speech and Signal Processing (ICASSP)}, 
  title={PersonaPlex: Voice and Role Control for Full Duplex Conversational Speech Models}, 
  year={2026},
  volume={},
  number={},
  pages={16137-16141},
  doi={10.1109/ICASSP55912.2026.11463413}}

@misc{kim2026raonspeechtechnicalreport,
      title={Raon-Speech Technical Report}, 
      author={Beomsoo Kim and Changho Choi and Dohyun Kim and Dongki Lee and Ethan Ewer and Eunchong Kim and Gyeongman Kim and Haechan Kim and Hyeonghwan Kim and Inkyu Park and Jihun Yun and Jihwan Moon and Jiyun Kim and Joonghyun Bae and Junhyuck Kim and Minkyu Kim and Sehun Lee and Seungjun Chung and Sungwoo Cho and Dongmin Park and Dongwon Kim and Hara Kang and Jonghyun Lee and Keon Lee and Kangwook Lee and Jaewoong Cho},
      year={2026},
      eprint={2605.23912},
      archivePrefix={arXiv},
      primaryClass={cs.CL},
      url={https://arxiv.org/abs/2605.23912}, 
}

@misc{cui2026minicpmo45realtimefullduplex,
      title={MiniCPM-o 4.5: Towards Real-Time Full-Duplex Omni-Modal Interaction}, 
      author={Junbo Cui and Bokai Xu and Chongyi Wang and Tianyu Yu and Weiyue Sun and Yingjing Xu and Tianran Wang and Zhihui He and Wenshuo Ma and Tianchi Cai and Jiancheng Gui and Luoyuan Zhang and Xian Sun and Fuwei Huang and Moye Chen and Zhuo Lin and Hanyu Liu and Qingxin Gui and Qingzhe Han and Yuyang Wen and Huiping Liu and Rongkang Wang and Yaqi Zhang and Hongliang Wei and Chi Chen and You Li and Kechen Fang and Jie Zhou and Yuxuan Li and Guoyang Zeng and Chaojun Xiao and Yankai Lin and Xu Han and Maosong Sun and Zhiyuan Liu and Yuan Yao},
      year={2026},
      eprint={2604.27393},
      archivePrefix={arXiv},
      primaryClass={cs.CL},
      url={https://arxiv.org/abs/2604.27393}, 
}

@misc{ohashi2026multifacetedinteractivityalignmentfullduplex,
      title={Multi-Faceted Interactivity Alignment in Full-Duplex Speech Models}, 
      author={Atsumoto Ohashi and Neil Zeghidour and Alexandre Défossez and Eugene Kharitonov},
      year={2026},
      eprint={2606.11167},
      archivePrefix={arXiv},
      primaryClass={cs.CL},
      url={https://arxiv.org/abs/2606.11167}, 
}

@misc{SileroVAD,
  author = {{Silero Team}},
  title = {Silero VAD: pre-trained enterprise-grade Voice Activity Detector (VAD), Number Detector and Language Classifier},
  year = {2024},
  publisher = {GitHub},
  journal = {GitHub repository},
  howpublished = {\url{https://github.com/snakers4/silero-vad}},
  commit = {insert_some_commit_here},
  email = {hello@silero.ai}
}

@misc{li2026decouplingconversationaldynamicsfullduplex,
      title={Decoupling Conversational Dynamics in Full-Duplex Spoken Models through Reinforcement Learning}, 
      author={Yuxin Li and Donghang Wu and Guan-Ting Lin and Hung-yi Lee and Chengwei Qin and Zhehuai Chen and Chen Chen},
      year={2026},
      eprint={2607.07148},
      archivePrefix={arXiv},
      primaryClass={eess.AS},
      url={https://arxiv.org/abs/2607.07148}, 
}

@misc{liao2025flexduopluggableenablingfullduplex,
      title={FlexDuo: A Pluggable System for Enabling Full-Duplex Capabilities in Speech Dialogue Systems}, 
      author={Borui Liao and Yulong Xu and Jiao Ou and Kaiyuan Yang and Weihua Jian and Pengfei Wan and Di Zhang},
      year={2025},
      eprint={2502.13472},
      archivePrefix={arXiv},
      primaryClass={cs.CL},
      url={https://arxiv.org/abs/2502.13472}, 
}

@INPROCEEDINGS{11463929,
  author={Li, Guojian and Wang, Chengyou and Xue, Hongfei and Wang, Shuiyuan and Gao, Dehui and Zhang, Zihan and Lin, Yuke and Li, Wenjie and Xiao, Longshuai and Fu, Zhonghua and Xie, Lei},
  booktitle={ICASSP 2026 - 2026 IEEE International Conference on Acoustics, Speech and Signal Processing (ICASSP)}, 
  title={Easy Turn: Integrating Acoustic and Linguistic Modalities for Robust Turn-Taking in Full-Duplex Spoken Dialogue Systems}, 
  year={2026},
  volume={},
  number={},
  pages={16957-16961},
  doi={10.1109/ICASSP55912.2026.11463929}}

@inproceedings{veluri-etal-2024-beyond,
    title = "Beyond Turn-Based Interfaces: Synchronous {LLM}s as Full-Duplex Dialogue Agents",
    author = "Veluri, Bandhav  and
      Peloquin, Benjamin N  and
      Yu, Bokai  and
      Gong, Hongyu  and
      Gollakota, Shyamnath",
    editor = "Al-Onaizan, Yaser  and
      Bansal, Mohit  and
      Chen, Yun-Nung",
    booktitle = "Proceedings of the 2024 Conference on Empirical Methods in Natural Language Processing",
    month = nov,
    year = "2024",
    address = "Miami, Florida, USA",
    publisher = "Association for Computational Linguistics",
    url = "https://aclanthology.org/2024.emnlp-main.1192/",
    doi = "10.18653/v1/2024.emnlp-main.1192",
    pages = "21390--21402"
}

@inproceedings{zhang-etal-2025-omniflatten,
    title = "{O}mni{F}latten: An End-to-end {GPT} Model for Seamless Voice Conversation",
    author = "Zhang, Qinglin  and
      Cheng, Luyao  and
      Deng, Chong  and
      Chen, Qian  and
      Wang, Wen  and
      Zheng, Siqi  and
      Liu, Jiaqing  and
      Yu, Hai  and
      Tan, Chao-Hong  and
      Du, Zhihao  and
      Zhang, ShiLiang",
    editor = "Che, Wanxiang  and
      Nabende, Joyce  and
      Shutova, Ekaterina  and
      Pilehvar, Mohammad Taher",
    booktitle = "Proceedings of the 63rd Annual Meeting of the Association for Computational Linguistics (Volume 1: Long Papers)",
    month = jul,
    year = "2025",
    address = "Vienna, Austria",
    publisher = "Association for Computational Linguistics",
    url = "https://aclanthology.org/2025.acl-long.709/",
    doi = "10.18653/v1/2025.acl-long.709",
    pages = "14570--14580",
    ISBN = "979-8-89176-251-0"
}

@ARTICLE{10.3389/fpsyg.2015.00731,
    
AUTHOR={Levinson, Stephen C.  and Torreira, Francisco },
           
TITLE={Timing in turn-taking and its implications for processing models of language},
          
JOURNAL={Frontiers in Psychology},
          
VOLUME={Volume 6 - 2015},
  
YEAR={2015},
  
URL={https://www.frontiersin.org/journals/psychology/articles/10.3389/fpsyg.2015.00731},
  
DOI={10.3389/fpsyg.2015.00731},
  
ISSN={1664-1078}}

@article{MARTIN2010177,
title = {Planning in sentence production: Evidence for the phrase as a default planning scope},
journal = {Cognition},
volume = {116},
number = {2},
pages = {177-192},
year = {2010},
issn = {0010-0277},
doi = {https://doi.org/10.1016/j.cognition.2010.04.010},
url = {https://www.sciencedirect.com/science/article/pii/S0010027710001022},
author = {Randi C. Martin and Jason E. Crowther and Meredith Knight and Franklin P. Tamborello and Chin-Lung Yang}
}

@ARTICLE{10.3389/fpsyg.2017.00393,
    
AUTHOR={Barthel, Mathias  and Meyer, Antje S.  and Levinson, Stephen C. },
           
TITLE={Next Speakers Plan Their Turn Early and Speak after Turn-Final “Go-Signals”},
          
JOURNAL={Frontiers in Psychology},
          
VOLUME={Volume 8 - 2017},
  
YEAR={2017},
  
URL={https://www.frontiersin.org/journals/psychology/articles/10.3389/fpsyg.2017.00393},
  
DOI={10.3389/fpsyg.2017.00393},
  
ISSN={1664-1078}}

@InProceedings{pmlr-v202-leviathan23a,
  title = 	 {Fast Inference from Transformers via Speculative Decoding},
  author =       {Leviathan, Yaniv and Kalman, Matan and Matias, Yossi},
  booktitle = 	 {Proceedings of the 40th International Conference on Machine Learning},
  pages = 	 {19274--19286},
  year = 	 {2023},
  editor = 	 {Krause, Andreas and Brunskill, Emma and Cho, Kyunghyun and Engelhardt, Barbara and Sabato, Sivan and Scarlett, Jonathan},
  volume = 	 {202},
  series = 	 {Proceedings of Machine Learning Research},
  month = 	 {23--29 Jul},
  publisher =    {PMLR},
  url = 	 {https://proceedings.mlr.press/v202/leviathan23a.html}
}

@misc{chen2023acceleratinglargelanguagemodel,
      title={Accelerating Large Language Model Decoding with Speculative Sampling}, 
      author={Charlie Chen and Sebastian Borgeaud and Geoffrey Irving and Jean-Baptiste Lespiau and Laurent Sifre and John Jumper},
      year={2023},
      eprint={2302.01318},
      archivePrefix={arXiv},
      primaryClass={cs.CL},
      url={https://arxiv.org/abs/2302.01318}, 
}

@inproceedings{NEURIPS2025_48b383b2,
 author = {Nie, Shen and Zhu, Fengqi and You, Zebin and Zhang, Xiaolu and Ou, Jingyang and Hu, Jun and Zhou, Jun and Lin, Yankai and Wen, Ji-Rong and LI, Chongxuan},
 booktitle = {Advances in Neural Information Processing Systems},
 doi = {10.52202/085713-1689},
 editor = {D. Belgrave and C. Zhang and H. Lin and R. Pascanu and P. Koniusz and M. Ghassemi and N. Chen},
 pages = {50608--50646},
 publisher = {Curran Associates, Inc.},
 title = {Large Language Diffusion Models},
 url = {https://proceedings.neurips.cc/paper_files/paper/2025/file/48b383b24230e0e6e649d9c98dae4d8c-Paper-Conference.pdf},
 volume = {38, Main Conference},
 year = {2025}
}

@inproceedings{ICLR2025_7ede97c3,
 author = {Arriola, Marianne and Gokaslan, Aaron and Chiu, Justin and Yang, Zhihan and Qi, Zhixuan and Han, Jiaqi and Sahoo, Subham and Kuleshov, Volodymyr},
 booktitle = {International Conference on Learning Representations},
 editor = {Y. Yue and A. Garg and N. Peng and F. Sha and R. Yu},
 pages = {50726--50753},
 title = {Block Diffusion: Interpolating Between Autoregressive and Diffusion Language Models},
 url = {https://proceedings.iclr.cc/paper_files/paper/2025/file/7ede97c3e082c6df10a8d6103a2eebd2-Paper-Conference.pdf},
 volume = {2025},
 year = {2025}
}

@misc{fu2026nemotronlabsdiffusiontrimodelanguagemodel,
      title={Nemotron-Labs-Diffusion: A Tri-Mode Language Model Unifying Autoregressive, Diffusion, and Self-Speculation Decoding}, 
      author={Yonggan Fu and Lexington Whalen and Abhinav Garg and Chengyue Wu and Maksim Khadkevich and Nicolai Oswald and Enze Xie and Daniel Egert and Sharath Turuvekere Sreenivas and Shizhe Diao and Chenhan Yu and Ye Yu and Weijia Chen and Sajad Norouzi and Jingyu Liu and Shiyi Lan and Ligeng Zhu and Jin Wang and Jindong Jiang and Morteza Mardani and Mehran Maghoumi and Song Han and Ante Jukić and Nima Tajbakhsh and Jan Kautz and Pavlo Molchanov},
      year={2026},
      eprint={2607.05722},
      archivePrefix={arXiv},
      primaryClass={cs.CL},
      url={https://arxiv.org/abs/2607.05722}, 
}

@INPROCEEDINGS{225858,
  author={Godfrey, J.J. and Holliman, E.C. and McDaniel, J.},
  booktitle={[Proceedings] ICASSP-92: 1992 IEEE International Conference on Acoustics, Speech, and Signal Processing}, 
  title={SWITCHBOARD: telephone speech corpus for research and development}, 
  year={1992},
  volume={1},
  number={},
  pages={517-520 vol.1},
  doi={10.1109/ICASSP.1992.225858}}

@article{nguyen-etal-2023-generative,
    title = "Generative Spoken Dialogue Language Modeling",
    author = "Nguyen, Tu Anh  and
      Kharitonov, Eugene  and
      Copet, Jade  and
      Adi, Yossi  and
      Hsu, Wei-Ning  and
      Elkahky, Ali  and
      Tomasello, Paden  and
      Algayres, Robin  and
      Sagot, Beno{\^i}t  and
      Mohamed, Abdelrahman  and
      Dupoux, Emmanuel",
    journal = "Transactions of the Association for Computational Linguistics",
    volume = "11",
    year = "2023",
    address = "Cambridge, MA",
    publisher = "MIT Press",
    url = "https://aclanthology.org/2023.tacl-1.15/",
    doi = "10.1162/tacl_a_00545",
    pages = "250--266"
}

@inproceedings{cheng-etal-2026-sdar,
    title = "{SDAR}: A Synergistic Diffusion-{A}uto{R}egression Paradigm for Scalable Sequence Generation",
    author = "Cheng, Shuang  and
      Bian, Yihan  and
      Liu, Dawei  and
      Jiang, Yuhua  and
      Liu, Yihao  and
      Zhang, Linfeng  and
      Yao, Qian  and
      Tian, Zhongbo  and
      Wang, Wenhai  and
      Guo, Qipeng  and
      Chen, Kai  and
      Qi, Biqing  and
      Zhou, Bowen",
    editor = "Liakata, Maria  and
      Moreira, Viviane P.  and
      Zhang, Jiajun  and
      Jurgens, David",
    booktitle = "Findings of the {A}ssociation for {C}omputational {L}inguistics: {ACL} 2026",
    month = jul,
    year = "2026",
    address = "San Diego, California, United States",
    publisher = "Association for Computational Linguistics",
    url = "https://aclanthology.org/2026.findings-acl.1110/",
    doi = "10.18653/v1/2026.findings-acl.1110",
    pages = "22058--22075",
    ISBN = "979-8-89176-395-1"
}

@misc{gat2025setblockdecodinglanguage,
      title={Set Block Decoding is a Language Model Inference Accelerator}, 
      author={Itai Gat and Heli Ben-Hamu and Marton Havasi and Daniel Haziza and Jeremy Reizenstein and Gabriel Synnaeve and David Lopez-Paz and Brian Karrer and Yaron Lipman},
      year={2025},
      eprint={2509.04185},
      archivePrefix={arXiv},
      primaryClass={cs.LG},
      url={https://arxiv.org/abs/2509.04185}, 
}

@misc{cheng2026vocalnetmdmacceleratingstreamingspeech,
      title={VocalNet-MDM: Accelerating Streaming Speech LLM via Self-Distilled Masked Diffusion Modeling}, 
      author={Ziyang Cheng and Yuhao Wang and Heyang Liu and Ronghua Wu and Qunshan Gu and Yanfeng Wang and Yu Wang},
      year={2026},
      eprint={2602.08607},
      archivePrefix={arXiv},
      primaryClass={cs.CL},
      url={https://arxiv.org/abs/2602.08607}, 
}

@misc{seo2026chatterboxflashpriorcalibratedblockdiffusion,
      title={Chatterbox-Flash: Prior-Calibrated Block Diffusion for Streaming Zero-Shot TTS}, 
      author={Deokjin Seo and Gangin Park and Kihyun Nam},
      year={2026},
      eprint={2605.30748},
      archivePrefix={arXiv},
      primaryClass={cs.SD},
      url={https://arxiv.org/abs/2605.30748}, 
}

@misc{moon2026deltattsadaptingautoregressivemodel,
      title={DELTA-TTS: Adapting Autoregressive Model into Diffusion Language Model for Text-to-Speech}, 
      author={Junwon Moon and Yejin Lee and Seungbeom Kim and Hoseong Ahn and Sewoong Park and Heeseung Kim and Kyuhong Shim},
      year={2026},
      eprint={2607.04140},
      archivePrefix={arXiv},
      primaryClass={eess.AS},
      url={https://arxiv.org/abs/2607.04140}, 
}

@inproceedings{NEURIPS2018_c4127b91,
 author = {Stern, Mitchell and Shazeer, Noam and Uszkoreit, Jakob},
 booktitle = {Advances in Neural Information Processing Systems},
 editor = {S. Bengio and H. Wallach and H. Larochelle and K. Grauman and N. Cesa-Bianchi and R. Garnett},
 pages = {},
 publisher = {Curran Associates, Inc.},
 title = {Blockwise Parallel Decoding for Deep Autoregressive Models},
 url = {https://proceedings.neurips.cc/paper_files/paper/2018/file/c4127b9194fe8562c64dc0f5bf2c93bc-Paper.pdf},
 volume = {31},
 year = {2018}
}

@inproceedings{
li2025predgen,
title={PredGen: Accelerated Inference of Large Language Models through Input-Time Speculation for Real-Time Speech Interaction},
author={Shufan Li and Aditya Grover},
booktitle={Second Conference on Language Modeling},
year={2025},
url={https://openreview.net/forum?id=2Kl8Ztw6wk}
}

@misc{mai2026relays2sdualpathspeculativegeneration,
      title={RelayS2S: A Dual-Path Speculative Generation for Real-Time Dialogue}, 
      author={Long Mai and Junli Liang},
      year={2026},
      eprint={2603.23346},
      archivePrefix={arXiv},
      primaryClass={cs.AI},
      url={https://arxiv.org/abs/2603.23346}, 
}

@misc{hsiao2026aspirinactionspaceprojection,
      title={ASPIRin: Action Space Projection for Interactivity-Optimized Reinforcement Learning in Full-Duplex Speech Language Models}, 
      author={Chi-Yuan Hsiao and Ke-Han Lu and Yu-Kuan Fu and Guan-Ting Lin and Hsiao-Tsung Hung and Hung-yi Lee},
      year={2026},
      eprint={2604.10065},
      archivePrefix={arXiv},
      primaryClass={cs.CL},
      url={https://arxiv.org/abs/2604.10065}, 
}

@inproceedings{NEURIPS2021_958c5305,
 author = {Austin, Jacob and Johnson, Daniel D. and Ho, Jonathan and Tarlow, Daniel and van den Berg, Rianne},
 booktitle = {Advances in Neural Information Processing Systems},
 editor = {M. Ranzato and A. Beygelzimer and Y. Dauphin and P.S. Liang and J. Wortman Vaughan},
 pages = {17981--17993},
 publisher = {Curran Associates, Inc.},
 title = {Structured Denoising Diffusion Models in Discrete State-Spaces},
 url = {https://proceedings.neurips.cc/paper_files/paper/2021/file/958c530554f78bcd8e97125b70e6973d-Paper.pdf},
 volume = {34},
 year = {2021}
}

@InProceedings{pmlr-v235-lou24a,
  title = 	 {Discrete Diffusion Modeling by Estimating the Ratios of the Data Distribution},
  author =       {Lou, Aaron and Meng, Chenlin and Ermon, Stefano},
  booktitle = 	 {Proceedings of the 41st International Conference on Machine Learning},
  pages = 	 {32819--32848},
  year = 	 {2024},
  editor = 	 {Salakhutdinov, Ruslan and Kolter, Zico and Heller, Katherine and Weller, Adrian and Oliver, Nuria and Scarlett, Jonathan and Berkenkamp, Felix},
  volume = 	 {235},
  series = 	 {Proceedings of Machine Learning Research},
  month = 	 {21--27 Jul},
  publisher =    {PMLR},
  url = 	 {https://proceedings.mlr.press/v235/lou24a.html}
}

@inproceedings{NEURIPS2024_eb0b13cc,
 author = {Sahoo, Subham Sekhar and Arriola, Marianne and Schiff, Yair and Gokaslan, Aaron and Marroquin, Edgar and Chiu, Justin T and Rush, Alexander and Kuleshov, Volodymyr},
 booktitle = {Advances in Neural Information Processing Systems},
 doi = {10.52202/079017-4135},
 editor = {A. Globerson and L. Mackey and D. Belgrave and A. Fan and U. Paquet and J. Tomczak and C. Zhang},
 pages = {130136--130184},
 publisher = {Curran Associates, Inc.},
 title = {Simple and Effective Masked Diffusion Language Models},
 url = {https://proceedings.neurips.cc/paper_files/paper/2024/file/eb0b13cc515724ab8015bc978fdde0ad-Paper-Conference.pdf},
 volume = {37},
 year = {2024}
}

@inproceedings{ICLR2025_0fa81c3f,
 author = {Gong, Shansan and Agarwal, Shivam and Zhang, Yizhe and Ye, Jiacheng and Zheng, Lin and Li, Mukai and An, Chenxin and Zhao, Peilin and BI, Wei and Han, Jiawei and Peng, Hao and Kong, Lingpeng},
 booktitle = {International Conference on Learning Representations},
 editor = {Y. Yue and A. Garg and N. Peng and F. Sha and R. Yu},
 pages = {5046--5073},
 title = {Scaling Diffusion Language Models via Adaptation from Autoregressive Models},
 url = {https://proceedings.iclr.cc/paper_files/paper/2025/file/0fa81c3f0d57f95b8776de3a248ef0ed-Paper-Conference.pdf},
 volume = {2025},
 year = {2025}
}

@inproceedings{ICLR2026_d0865cbe,
 author = {Wu, Chengyue and Zhang, Hao and Xue, Shuchen and Diao, Shizhe and Fu, Yonggan and Liu, Zhijian and Molchanov, Pavlo and Luo, Ping and Han, Song and Xie, Enze},
 booktitle = {International Conference on Learning Representations},
 editor = {C. Vondrick and B. Hariharan and C. Raffel and L. Pinto and D. Yang and A. Faust},
 pages = {128353--128370},
 title = {Fast-dLLM v2: Efficient Block-Diffusion LLM},
 url = {https://proceedings.iclr.cc/paper_files/paper/2026/file/d0865cbe51d35ec322f9af9db7806fc7-Paper-Conference.pdf},
 volume = {2026},
 year = {2026}
}

@InProceedings{pmlr-v235-cai24b,
  title = 	 {Medusa: Simple {LLM} Inference Acceleration Framework with Multiple Decoding Heads},
  author =       {Cai, Tianle and Li, Yuhong and Geng, Zhengyang and Peng, Hongwu and Lee, Jason D. and Chen, Deming and Dao, Tri},
  booktitle = 	 {Proceedings of the 41st International Conference on Machine Learning},
  pages = 	 {5209--5235},
  year = 	 {2024},
  editor = 	 {Salakhutdinov, Ruslan and Kolter, Zico and Heller, Katherine and Weller, Adrian and Oliver, Nuria and Scarlett, Jonathan and Berkenkamp, Felix},
  volume = 	 {235},
  series = 	 {Proceedings of Machine Learning Research},
  month = 	 {21--27 Jul},
  publisher =    {PMLR},
  url = 	 {https://proceedings.mlr.press/v235/cai24b.html}
}

@InProceedings{pmlr-v235-li24bt,
  title = 	 {{EAGLE}: Speculative Sampling Requires Rethinking Feature Uncertainty},
  author =       {Li, Yuhui and Wei, Fangyun and Zhang, Chao and Zhang, Hongyang},
  booktitle = 	 {Proceedings of the 41st International Conference on Machine Learning},
  pages = 	 {28935--28948},
  year = 	 {2024},
  editor = 	 {Salakhutdinov, Ruslan and Kolter, Zico and Heller, Katherine and Weller, Adrian and Oliver, Nuria and Scarlett, Jonathan and Berkenkamp, Felix},
  volume = 	 {235},
  series = 	 {Proceedings of Machine Learning Research},
  month = 	 {21--27 Jul},
  publisher =    {PMLR},
  url = 	 {https://proceedings.mlr.press/v235/li24bt.html}
}

@INPROCEEDINGS{10887855,
  author={Nguyen, Tan Dat and Kim, Ji-Hoon and Choi, Jeongsoo and Choi, Shukjae and Park, Jinseok and Lee, Younglo and Chung, Joon Son},
  booktitle={ICASSP 2025 - 2025 IEEE International Conference on Acoustics, Speech and Signal Processing (ICASSP)}, 
  title={Accelerating Codec-based Speech Synthesis with Multi-Token Prediction and Speculative Decoding}, 
  year={2025},
  volume={},
  number={},
  pages={1-5},
  doi={10.1109/ICASSP49660.2025.10887855}}

@INPROCEEDINGS{10888194,
  author={Li, Bohan and Wang, Hankun and Zhang, Situo and Guo, Yiwei and Yu, Kai},
  booktitle={ICASSP 2025 - 2025 IEEE International Conference on Acoustics, Speech and Signal Processing (ICASSP)}, 
  title={Fast and High-Quality Auto-Regressive Speech Synthesis via Speculative Decoding}, 
  year={2025},
  volume={},
  number={},
  pages={1-5},
  doi={10.1109/ICASSP49660.2025.10888194}}

@INPROCEEDINGS{11464026,
  author={Yanuka, Moran and Dixon, Paul and Finkelshtein, Eyal and Rotman, Daniel and Giryes, Raja},
  booktitle={ICASSP 2026 - 2026 IEEE International Conference on Acoustics, Speech and Signal Processing (ICASSP)}, 
  title={Principled Coarse-Grained Acceptance For Speculative Decoding In Speech}, 
  year={2026},
  volume={},
  number={},
  pages={16692-16696},
  doi={10.1109/ICASSP55912.2026.11464026}}

@InProceedings{pmlr-v267-wu25t,
  title = 	 {Aligning Spoken Dialogue Models from User Interactions},
  author =       {Wu, Anne and Mazar\'{e}, Laurent and Zeghidour, Neil and D\'{e}fossez, Alexandre},
  booktitle = 	 {Proceedings of the 42nd International Conference on Machine Learning},
  pages = 	 {67476--67498},
  year = 	 {2025},
  editor = 	 {Singh, Aarti and Fazel, Maryam and Hsu, Daniel and Lacoste-Julien, Simon and Berkenkamp, Felix and Maharaj, Tegan and Wagstaff, Kiri and Zhu, Jerry},
  volume = 	 {267},
  series = 	 {Proceedings of Machine Learning Research},
  month = 	 {13--19 Jul},
  publisher =    {PMLR},
  url = 	 {https://proceedings.mlr.press/v267/wu25t.html}
}

@inproceedings{
hu2022lora,
title={Lo{RA}: Low-Rank Adaptation of Large Language Models},
author={Edward J Hu and yelong shen and Phillip Wallis and Zeyuan Allen-Zhu and Yuanzhi Li and Shean Wang and Lu Wang and Weizhu Chen},
booktitle={International Conference on Learning Representations},
year={2022},
url={https://openreview.net/forum?id=nZeVKeeFYf9}
}

@InProceedings{pmlr-v235-ruhe24a,
  title = 	 {Rolling Diffusion Models},
  author =       {Ruhe, David and Heek, Jonathan and Salimans, Tim and Hoogeboom, Emiel},
  booktitle = 	 {Proceedings of the 41st International Conference on Machine Learning},
  pages = 	 {42818--42835},
  year = 	 {2024},
  editor = 	 {Salakhutdinov, Ruslan and Kolter, Zico and Heller, Katherine and Weller, Adrian and Oliver, Nuria and Scarlett, Jonathan and Berkenkamp, Felix},
  volume = 	 {235},
  series = 	 {Proceedings of Machine Learning Research},
  month = 	 {21--27 Jul},
  publisher =    {PMLR},
  url = 	 {https://proceedings.mlr.press/v235/ruhe24a.html}
}

@inproceedings{NEURIPS2024_2aee1c41,
 author = {Chen, Boyuan and Mart\'{\i} Mons\'{o}, Diego and Du, Yilun and Simchowitz, Max and Tedrake, Russ and Sitzmann, Vincent},
 booktitle = {Advances in Neural Information Processing Systems},
 doi = {10.52202/079017-0759},
 editor = {A. Globerson and L. Mackey and D. Belgrave and A. Fan and U. Paquet and J. Tomczak and C. Zhang},
 pages = {24081--24125},
 publisher = {Curran Associates, Inc.},
 title = {Diffusion Forcing: Next-token Prediction Meets Full-Sequence Diffusion},
 url = {https://proceedings.neurips.cc/paper_files/paper/2024/file/2aee1c4159e48407d68fe16ae8e6e49e-Paper-Conference.pdf},
 volume = {37},
 year = {2024}
}

@INPROCEEDINGS{11433838,
  author={Lin, Guan-Ting and Lian, Jiachen and Li, Tingle and Wang, Qirui and Anumanchipalli, Gopala and Liu, Alexander H. and Lee, Hung-Yi},
  booktitle={2025 IEEE Automatic Speech Recognition and Understanding Workshop (ASRU)}, 
  title={Full-Duplex-Bench: A Benchmark to Evaluate Full-Duplex Spoken Dialogue Models on Turn-taking Capabilities}, 
  year={2025},
  volume={},
  number={},
  pages={1-8},
  doi={10.1109/ASRU65441.2025.11433838}}

@INPROCEEDINGS{11463576,
  author={Lin, Guan-Ting and Kuan, Shih-Yun Shan and Wang, Qirui and Lian, Jiachen and Li, Tingle and Watanabe, Shinji and Lee, Hung-Yi},
  booktitle={ICASSP 2026 - 2026 IEEE International Conference on Acoustics, Speech and Signal Processing (ICASSP)}, 
  title={Full-Duplex-Bench v1.5: Evaluating Overlap Handling for Full-Duplex Speech Models}, 
  year={2026},
  volume={},
  number={},
  pages={19447-19451},
  doi={10.1109/ICASSP55912.2026.11463576}}

@article{chen-etal-2026-voicebench,
    title = "{V}oice{B}ench: Benchmarking {LLM}-Based Voice Assistants",
    author = "Chen, Yiming  and
      Yue, Xianghu  and
      Zhang, Chen  and
      Gao, Xiaoxue  and
      Tan, Robby T.  and
      Li, Haizhou",
    journal = "Transactions of the Association for Computational Linguistics",
    volume = "14",
    year = "2026",
    address = "Cambridge, MA",
    publisher = "MIT Press",
    url = "https://aclanthology.org/2026.tacl-1.18/",
    doi = "10.1162/tacl.a.628",
    pages = "378--398"
}

@misc{li2025baichuanaudiounifiedframeworkendtoend,
      title={Baichuan-Audio: A Unified Framework for End-to-End Speech Interaction}, 
      author={Tianpeng Li and Jun Liu and Tao Zhang and Yuanbo Fang and Da Pan and Mingrui Wang and Zheng Liang and Zehuan Li and Mingan Lin and Guosheng Dong and Jianhua Xu and Haoze Sun and Zenan Zhou and Weipeng Chen},
      year={2025},
      eprint={2502.17239},
      archivePrefix={arXiv},
      primaryClass={cs.CL},
      url={https://arxiv.org/abs/2502.17239}, 
}

@InProceedings{pmlr-v202-xu23g,
  title = 	 {Efficient Sequence Transduction by Jointly Predicting Tokens and Durations},
  author =       {Xu, Hainan and Jia, Fei and Majumdar, Somshubra and Huang, He and Watanabe, Shinji and Ginsburg, Boris},
  booktitle = 	 {Proceedings of the 40th International Conference on Machine Learning},
  pages = 	 {38462--38484},
  year = 	 {2023},
  editor = 	 {Krause, Andreas and Brunskill, Emma and Cho, Kyunghyun and Engelhardt, Barbara and Sabato, Sivan and Scarlett, Jonathan},
  volume = 	 {202},
  series = 	 {Proceedings of Machine Learning Research},
  month = 	 {23--29 Jul},
  publisher =    {PMLR},
  url = 	 {https://proceedings.mlr.press/v202/xu23g.html}
}

@misc{agrawal2025seamlessinteractiondyadicaudiovisual,
      title={Seamless Interaction: Dyadic Audiovisual Motion Modeling and Large-Scale Dataset}, 
      author={Vasu Agrawal and Akinniyi Akinyemi and Kathryn Alvero and Morteza Behrooz and Julia Buffalini and Fabio Maria Carlucci and Joy Chen and Junming Chen and Zhang Chen and Shiyang Cheng and Praveen Chowdary and Joe Chuang and Antony D'Avirro and Jon Daly and Ning Dong and Mark Duppenthaler and Cynthia Gao and Jeff Girard and Martin Gleize and Sahir Gomez and Hongyu Gong and Srivathsan Govindarajan and Brandon Han and Sen He and Denise Hernandez and Yordan Hristov and Rongjie Huang and Hirofumi Inaguma and Somya Jain and Raj Janardhan and Qingyao Jia and Christopher Klaiber and Dejan Kovachev and Moneish Kumar and Hang Li and Yilei Li and Pavel Litvin and Wei Liu and Guangyao Ma and Jing Ma and Martin Ma and Xutai Ma and Lucas Mantovani and Sagar Miglani and Sreyas Mohan and Louis-Philippe Morency and Evonne Ng and Kam-Woh Ng and Tu Anh Nguyen and Amia Oberai and Benjamin Peloquin and Juan Pino and Jovan Popovic and Omid Poursaeed and Fabian Prada and Alice Rakotoarison and Rakesh Ranjan and Alexander Richard and Christophe Ropers and Safiyyah Saleem and Vasu Sharma and Alex Shcherbyna and Jia Shen and Jie Shen and Anastasis Stathopoulos and Anna Sun and Paden Tomasello and Tuan Tran and Arina Turkatenko and Bo Wan and Chao Wang and Jeff Wang and Mary Williamson and Carleigh Wood and Tao Xiang and Yilin Yang and Julien Yao and Chen Zhang and Jiemin Zhang and Xinyue Zhang and Jason Zheng and Pavlo Zhyzheria and Jan Zikes and Michael Zollhoefer},
      year={2025},
      eprint={2506.22554},
      archivePrefix={arXiv},
      primaryClass={cs.CV},
      url={https://arxiv.org/abs/2506.22554}, 
}

@INPROCEEDINGS{10832315,
  author={Baba, Kaito and Nakata, Wataru and Saito, Yuki and Saruwatari, Hiroshi},
  booktitle={2024 IEEE Spoken Language Technology Workshop (SLT)}, 
  title={The T05 System for the voicemos challenge 2024: Transfer Learning from Deep Image Classifier to Naturalness MOS Prediction of High-Quality Synthetic Speech}, 
  year={2024},
  volume={},
  number={},
  pages={818-824},
  doi={10.1109/SLT61566.2024.10832315}}

@misc{shi2026qwen3asrtechnicalreport,
      title={Qwen3-ASR Technical Report}, 
      author={Xian Shi and Xiong Wang and Zhifang Guo and Yongqi Wang and Pei Zhang and Xinyu Zhang and Zishan Guo and Hongkun Hao and Yu Xi and Baosong Yang and Jin Xu and Jingren Zhou and Junyang Lin},
      year={2026},
      eprint={2601.21337},
      archivePrefix={arXiv},
      primaryClass={cs.CL},
      url={https://arxiv.org/abs/2601.21337}, 
}

@misc{loshchilov2019decoupledweightdecayregularization,
      title={Decoupled Weight Decay Regularization}, 
      author={Ilya Loshchilov and Frank Hutter},
      year={2019},
      eprint={1711.05101},
      archivePrefix={arXiv},
      primaryClass={cs.LG},
      url={https://arxiv.org/abs/1711.05101}, 
}

@article{
doi:10.1126/sciadv.adf3197,
author = {Andrew Reece  and Gus Cooney  and Peter Bull  and Christine Chung  and Bryn Dawson  and Casey Fitzpatrick  and Tamara Glazer  and Dean Knox  and Alex Liebscher  and Sebastian Marin },
title = {The CANDOR corpus: Insights from a large multimodal dataset of naturalistic conversation},
journal = {Science Advances},
volume = {9},
number = {13},
pages = {eadf3197},
year = {2023},
doi = {10.1126/sciadv.adf3197},
URL = {https://www.science.org/doi/abs/10.1126/sciadv.adf3197},
eprint = {https://www.science.org/doi/pdf/10.1126/sciadv.adf3197}}

@inproceedings{umair-etal-2024-large,
    title = "Large Language Models Know What To Say But Not When To Speak",
    author = "Umair, Muhammad  and
      Sarathy, Vasanth  and
      Ruiter, Jan",
    editor = "Al-Onaizan, Yaser  and
      Bansal, Mohit  and
      Chen, Yun-Nung",
    booktitle = "Findings of the Association for Computational Linguistics: EMNLP 2024",
    month = nov,
    year = "2024",
    address = "Miami, Florida, USA",
    publisher = "Association for Computational Linguistics",
    url = "https://aclanthology.org/2024.findings-emnlp.909/",
    doi = "10.18653/v1/2024.findings-emnlp.909",
    pages = "15503--15514"
}

@misc{samragh2025llmknowsfutureuncovering,
      title={Your LLM Knows the Future: Uncovering Its Multi-Token Prediction Potential}, 
      author={Mohammad Samragh and Arnav Kundu and David Harrison and Kumari Nishu and Devang Naik and Minsik Cho and Mehrdad Farajtabar},
      year={2025},
      eprint={2507.11851},
      archivePrefix={arXiv},
      primaryClass={cs.CL},
      url={https://arxiv.org/abs/2507.11851}, 
}
\bibliographystyle{iclr2027_conference}

\appendix

\appendix
\newpage

\section{Limitations}
\label{asec:limitations}

DiffuPlex has several limitations that suggest directions for future work.
First, DiffuPlex is adapted from a pretrained autoregressive backbone rather than trained natively for masked-future generation.
We freeze the original backbone and train future-gated LoRA adapters together with learned MASK embeddings using self-distillation from the autoregressive model.
This design makes it possible to add masked-future prediction without retraining the full model, but it also means that the backbone was not originally optimized to represent several uncertain future interaction frames at once.
A model trained specifically for masked-future generation may therefore make better use of the additional future context.

Second, DiffuPlex separates future prediction from the decisions about how much of that prediction to use.
The predictor is trained to model future user and assistant frames, whereas retention and revision are controlled by inference-time rules based on prediction confidence and a fixed revision threshold.
These rules work well in our experiments, but they are not learned jointly with the future predictor.
Jointly learning future prediction, retention, and revision could allow the model to better trade off computation, interaction responsiveness, and prediction reliability.

Third, aggressive future consumption remains more difficult during assistant speech than during listening.
DiffuPlex-\textsc{Speak} reduces backbone wake frequency more aggressively, but this comes with lower speech naturalness.
As analyzed in Section~\ref{subsec:draft_analysis} and Appendix~\ref{asubsec:future_speech}, one contributor is the loss of same-frame semantic conditioning for deeper future acoustic codebooks: when predicting farther into the future, the acoustic codebooks can no longer condition on the assistant semantic codebook that would normally already be available under autoregressive generation.
Our interventions show that restoring same-frame cb0 improves acoustic-token prediction, while autoregressive replay recovers much of the measured quality gap; neither provides a practical solution to future speech generation.
Improved future audio modeling is therefore still needed before long assistant-speech segments can be consumed as reliably as silent listening segments.

Finally, our main experiments use \texttt{personaplex-rl-seamless} as the backbone.
We additionally apply the same training and inference framework to PersonaPlex~\citep{11463413} in Appendix~\ref{asubsec:backbone_ablation}.
Although this variant still reduces the number of backbone wakes, its full-duplex interaction behavior degrades substantially, particularly on pause handling and interruption behavior.
This result suggests that transferring DiffuPlex across backbone checkpoints is not automatic and may depend on the interaction behavior already learned by the underlying model.
We have not yet evaluated the method on substantially different full-duplex architectures.
Moreover, the measured wall-clock and Core LM speedups depend on the evaluated hardware, software stack, and implementation.
Evaluating DiffuPlex across additional architectures and deployment environments remains important future work.

\section{Demo Page}
\label{asec:demo}
Selected audio examples and qualitative demonstrations of DiffuPlex are available at \url{https://diffuplex.github.io/}.

\section{Full-Duplex Backbone and Stream Alignment}
\label{asec:backbone}

\subsection{Backbone Architecture}
\label{asubsec:backbone_architecture}

We build DiffuPlex on \texttt{personaplex-rl-seamless}~\citep{ohashi2026multifacetedinteractivityalignmentfullduplex}, a post-trained PersonaPlex checkpoint that follows the Moshi-style full-duplex architecture~\citep{defossez2024moshispeechtextfoundationmodel}.
The model operates at 12.5Hz, corresponding to one interaction frame every 80ms, and represents the two sides of the conversation using parallel user and assistant streams.
Its temporal backbone models the interaction across successive positions, while a depth transformer autoregressively predicts the audio codebooks associated with each position.

The model contains one assistant text stream together with eight Mimi~\citep{defossez2024moshispeechtextfoundationmodel} audio codebooks for the assistant and eight for the user.
Following the terminology and training objective of the backbone, we refer to codebook 0 (cb0) as the semantic codebook and codebooks 1--7 (cb1--7) as the acoustic codebooks.
An important difference from the original Moshi architecture is that PersonaPlex~\citep{11463413} predicts audio codebooks for both sides of the interaction: whereas Moshi's depth transformer predicts eight assistant audio codebooks, PersonaPlex predicts eight assistant and eight user codebooks, for 16 audio streams in total.
This joint user-assistant prediction structure is retained by the post-trained \texttt{personaplex-rl-seamless} backbone and, consequently, by DiffuPlex.
The predicted user audio is not played back; during streaming inference, the actual user codebooks are supplied by the incoming audio stream, while the generated assistant codebooks are used for playback.
In our measurements using the official implementations, PersonaPlex and \texttt{personaplex-rl-seamless} require more Core LM time and wall-clock time than Moshi on the same runtime subset, as shown in Table~\ref{tab:runtime_results}.

The backbone uses a native one-position delay between cb0 and cb1--7.
As a result, when the temporal backbone computes the representation used to predict cb1--7 for a physical frame, cb0 from that same frame is already available as an input.
This allows same-frame semantic information to influence the temporal representation itself before the depth transformer predicts the acoustic codebooks.
Without this delay, cb0 and cb1--7 of the same frame would be generated from the same temporal state, leaving their within-frame dependency to be modeled only by the depth transformer.
We describe this alignment precisely in Appendix~\ref{asubsec:stream_alignment}; Moshi reports a theoretical algorithmic latency of 160ms for this streaming architecture~\citep{defossez2024moshispeechtextfoundationmodel}.
Despite jointly modeling both sides of the interaction, temporal generation remains autoregressive: under ordinary inference, the temporal backbone is evaluated once every 80ms as the interaction advances.
DiffuPlex targets this sequential temporal-backbone schedule while leaving the underlying stream organization, native delays, and the backbone's original next-frame prediction shift unchanged.

\subsection{Exact Stream Alignment}
\label{asubsec:stream_alignment}

The backbone uses 17 streams in the following order:
one assistant text stream, eight assistant Mimi codebooks, and eight user Mimi codebooks.
Their native delays are
\begin{equation}
\Delta =
[0,\,
0,\underbrace{1,\ldots,1}_{7},\,
0,\underbrace{1,\ldots,1}_{7}],
\label{aeq:stream_delays}
\end{equation}
where $\Delta_k=0$ for assistant text, assistant cb0, and user cb0, and $\Delta_k=1$ for cb1--7 of both speakers.

These delays mean that a model position should not be interpreted as containing all codebooks from the same physical interaction frame.
We use zero-based indices for both model positions $p=0,1,\ldots$ and physical interaction frames $f=0,1,\ldots$; references to negative frame indices correspond to the backbone's initial tokens.
For stream $k$ with native delay $\Delta_k$, the input and prediction target at model position $p$ are
\begin{equation}
\mathrm{input}[k,p]
=
\mathrm{frame}(p-1-\Delta_k),
\qquad
\mathrm{target}[k,p]
=
\mathrm{frame}(p-\Delta_k).
\label{aeq:stream_alignment}
\end{equation}
Thus, for streams with $\Delta_k=0$, position $p$ receives frame $p-1$ and predicts frame $p$.
For streams with $\Delta_k=1$, the same position receives frame $p-2$ and predicts frame $p-1$.
Equivalently, a physical user frame $f$ becomes available to the temporal backbone at position $f+1$ for cb0 and at position $f+2$ for cb1--7.
The one-step autoregressive input-to-target shift is preserved for every stream, while the physical frame represented at a given model position depends on its stream-specific delay.

This alignment is particularly important for the assistant acoustic codebooks.
To predict cb1--7 of physical frame $f$, ordinary autoregressive generation uses model position $f+1$.
At this position, assistant cb0 of the same physical frame $f$ is already part of the temporal-backbone input.
Same-frame semantic information can therefore influence the temporal representation before that representation is passed to the depth transformer.
The depth transformer remains autoregressive over the codebooks and can also model dependencies among codebooks within its decoding sequence, but the native delay additionally exposes same-frame cb0 to the temporal backbone itself.
Without this delay, cb0 and cb1--7 of the same physical frame would be generated from the same temporal state, leaving their same-frame dependency to the depth transformer rather than incorporating cb0 into the temporal representation.

DiffuPlex preserves this native delay pattern.
It masks only inputs that would not yet have been generated or observed at the backbone wake that predicts a future position.
This distinction becomes important in our analysis of future speech generation in Appendix~\ref{asubsec:future_speech}.

\subsection{Masked Training Layout}
\label{asubsec:masked_training_layout}

\begin{figure*}[t]
    \centering
    \includegraphics[width=\textwidth]{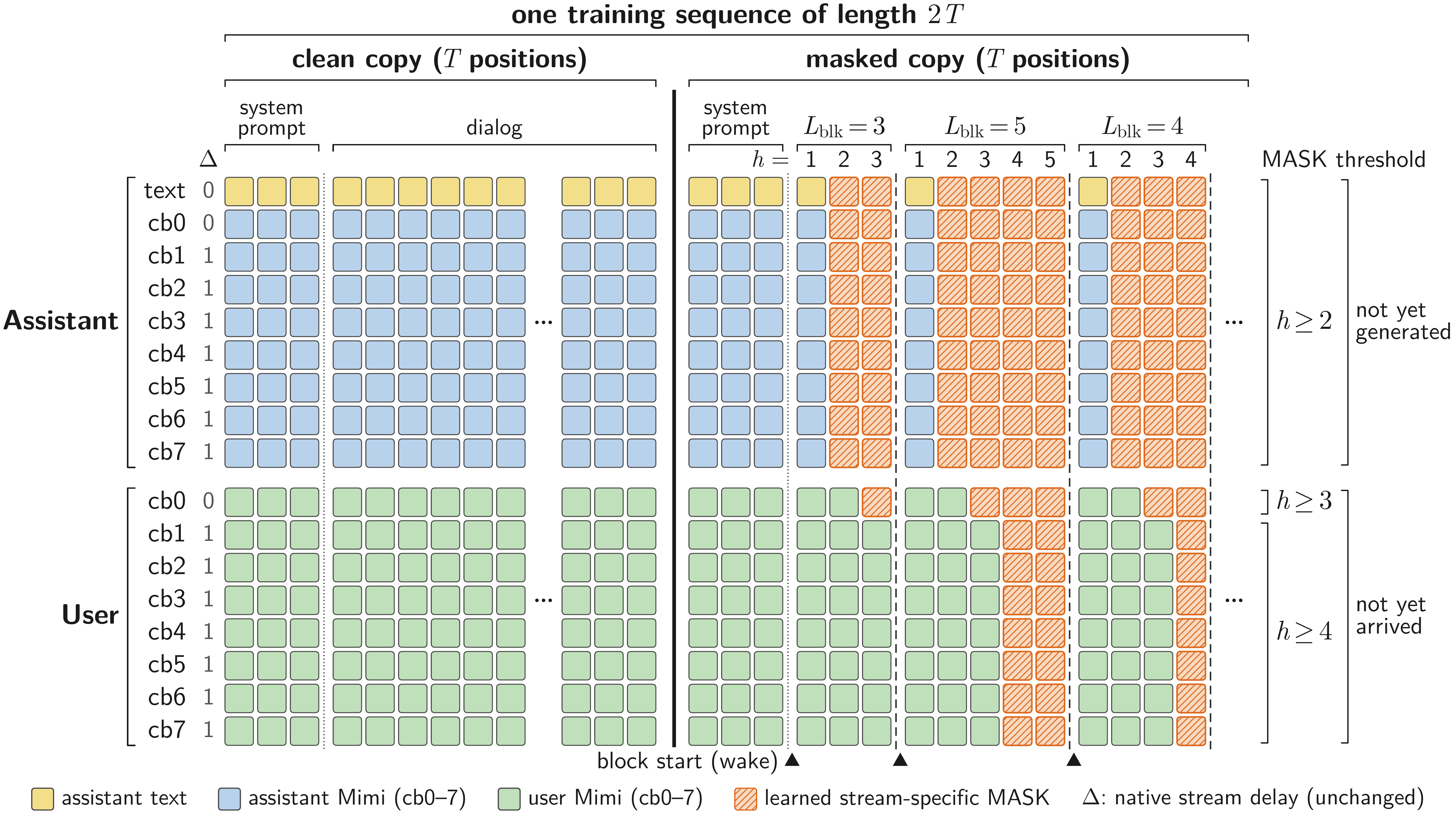}
    \caption{
    Training sequence layout for masked future prediction.
    A clean and masked copy of the same 17-stream sequence are concatenated to form a sequence of length $2T$.
    The masked dialog region is tiled by variable-length blocks, with lookahead $h$ reset at every block start.
    Inputs unavailable at that backbone wake are replaced by learned stream-specific MASK tokens: assistant streams from $h\geq2$, user cb0 from $h\geq3$, and user cb1--7 from $h\geq4$.
    Native stream delays are unchanged.
    The staircase denotes stream-wise input availability rather than an attention mask; temporal attention remains causal.
    }
    \label{afig:training_layout}
\end{figure*}

Figure~\ref{afig:training_layout} shows the exact training layout used for masked future prediction.
For each training example, we concatenate a clean copy of the original sequence with a masked copy of the same sequence, producing a sequence of length $2T$.
The clean copy preserves the realized interaction.
In the masked copy, only the dialog region is modified; the system-prompt region remains unchanged.

The masked dialog region is divided into consecutive blocks.
For each block, its length is sampled independently as
\begin{equation}
L_{\mathrm{blk}}
\sim
\mathcal{U}\{1,\ldots,W\},
\label{aeq:block_length}
\end{equation}
where $W=16$ for the main model.
The final block is truncated if it reaches the end of the dialog region.
Within every block, the lookahead index resets to
$h=1,\ldots,L_{\mathrm{blk}}$.
This construction exposes the model to different future distances from a backbone wake while conditioning on a varying amount of realized interaction available at each block boundary.

For each lookahead, we replace only inputs that would not yet be causally available when that future position is predicted.
The masking schedule is
\begin{equation}
\begin{aligned}
h=1 &: \quad \text{no streams are masked},\\
h\ge2 &: \quad \text{assistant text and assistant cb0--7 are masked},\\
h\ge3 &: \quad \text{user cb0 is additionally masked},\\
h\ge4 &: \quad \text{user cb1--7 are additionally masked}.
\end{aligned}
\label{aeq:mask_schedule}
\end{equation}
Each stream uses its own learned MASK embedding.
The asymmetry between assistant and user streams reflects their different availability at inference.
Future assistant outputs have not yet been generated and are therefore masked from $h=2$.
By contrast, user audio observed during the current 80ms interaction frame becomes available as input only at subsequent model positions.
Under the native delay pattern, user cb0 remains available at $h=2$ and is masked from $h=3$, while user cb1--7 remain available through $h=3$ and are masked from $h=4$, at which point all user streams are masked.
The native delays in Eq.~\ref{aeq:stream_delays} are left unchanged.

The staircase in Figure~\ref{afig:training_layout} therefore describes \emph{input availability}, not an attention mask.
Temporal attention remains causal.
For a position in the masked copy, attention includes the realized context preceding its block from the clean copy and masked positions up to the current position within the same block.
It does not attend to later positions in the current block or to positions from preceding blocks in the masked copy.
Instead, interaction preceding the current block boundary is provided by the corresponding clean context.
Each block therefore represents a fresh future prediction from the realized interaction at its boundary, rather than a continuation of the preceding masked block.
At $h=1$, no input is replaced by MASK, so the input is identical to the original next-frame autoregressive path.

\section{Implementation Details}
\label{asec:implementation}

\subsection{Data Processing}
\label{asubsec:data_processing}

We train DiffuPlex on Seamless Interaction~\citep{agrawal2025seamlessinteractiondyadicaudiovisual}, a large corpus of face-to-face dyadic conversations recorded on separate speaker channels.
The source dataset contains more than 4,000h of interaction from more than 4,000 participants and includes both naturalistic conversations and improvised scenarios.
We process the two speaker channels jointly so that their original conversational timing, including pauses, overlaps, and interruptions, is preserved. 
We first resample all audio to 24kHz.
Each channel is transcribed with Qwen3-ASR-1.7B~\citep{shi2026qwen3asrtechnicalreport}, using English as the fixed language, and the resulting transcript is aligned to the waveform at the word level using Qwen3-ForcedAligner-0.6B~\citep{shi2026qwen3asrtechnicalreport}.
We use Silero VAD~\citep{SileroVAD} to identify speech activity and encode each channel with the same Mimi codec used by the backbone, producing eight codebooks at 12.5Hz.
The resulting aligned representation therefore retains the original 80ms interaction grid used by \texttt{personaplex-rl-seamless}.

We apply session-level filtering before constructing training examples.
We first remove sessions containing residual non-Latin transcription.
We then filter sessions for microphone leakage using a transcript-similarity threshold of 0.30 and an overlap threshold of 0.20.
We additionally remove one-sided conversations in which either speaker has no transcribed words, sessions with low speech density where speech occupies less than 60\% of the active region, sessions with an active region shorter than 30s, sessions containing an intermediate silence longer than 10s, and sessions with strong speaker imbalance where either speaker accounts for at least 75\% of the total speaking duration.
After filtering, 42,613 sessions remain and 21,717 are excluded.

From the retained sessions, we reserve 200 session-disjoint conversations, approximately 14h, for validation.
The remaining training split contains 42,413 sessions totaling 2,728.5h of conversation from 3,516 participants.
It contains 28,242 naturalistic sessions totaling 1,628.7h and 14,171 improvised sessions totaling 1,099.8h.
We use both speaker assignments as training examples when valid voice references are available, resulting in 84,183 role-conditioned training examples.

We additionally construct a voice-reference pool for each speaker.
Candidate references are between 2s and 10s long, contain at least three ASR words, and are normalized to $-24$ LUFS.
We retain up to six references per channel and mark a reference as clean when it does not overlap speech from the other channel.
During training, we preferentially select a clean reference that lies outside the sampled interaction window, so that voice conditioning does not expose speech from the target region.

\subsection{Training}
\label{asubsec:training}

We optimize the distillation objective in Eq.~\ref{eq:distill} over positions belonging to the dialog region.
System-prompt positions, padding, and positions without defined targets under the backbone's native stream delays are excluded from the loss.
Following the backbone objective, we use $w_q=1$ for the semantic codebook of each speaker and $w_q=0.02$ for cb1--7.
The distribution $\bar p_{\mathrm{text}}$ groups the full assistant-text distribution $p_{\mathrm{text}}$ into the three interaction states PAD, EPAD, and lexical.
Because most text positions are non-lexical, matching $p_{\mathrm{text}}$ alone can underemphasize the coarse interaction-state distinctions that govern full-duplex behavior, including when the assistant remains silent, begins speaking, backchannels, or yields to user speech.
We therefore additionally distill $\bar p_{\mathrm{text}}$ to preserve these conversational dynamics, assigning its KL term equal weight to the KL on $p_{\mathrm{text}}$.
The losses are pooled over all valid positions rather than normalized separately by future lookahead.

We train the stream-specific MASK embeddings and future-gated LoRA adapters described in Section~\ref{subsec:masked_future}.
LoRA is applied to both the temporal backbone and the depth transformer with rank 128 and scaling factor 2.0, and the MASK vectors are initialized with norm-matched random directions.
Each sampled training window contains at most 2,047 code frames; after inserting the backbone's initial token and applying the native stream-delay construction, this yields 2,048 temporal model positions, corresponding to approximately 163.8s at 12.5Hz.
The clean and masked copies are concatenated as described in Appendix~\ref{asubsec:masked_training_layout}, and we do not prepend additional dialog context outside the sampled window.
We construct the training prompt in the same form used by PersonaPlex: a speaker voice reference is followed by six silence frames, the persona text prompt, and another six silence frames, while the user side is filled with the backbone's sine-wave conditioning during this priming region.
When several valid voice references are available, we preferentially use one outside the sampled dialog window.

We optimize for 526 steps using AdamW~\citep{loshchilov2019decoupledweightdecayregularization} with learning rate $10^{-4}$, weight decay $0.1$, and gradient clipping at $1.0$.
We use a OneCycle learning-rate schedule with a 15\% warmup followed by cosine decay, bfloat16 precision, gradient checkpointing, and random seed 0.
The per-GPU batch size is 4 with two gradient-accumulation steps on four NVIDIA RTX A6000 GPUs, giving an effective batch size of 32.
Training processes 16,832 sampled windows drawn from 8,480 distinct sessions, corresponding to approximately 702h of conversational audio.
We save checkpoints every 263 steps and use the final checkpoint at step 526 for all main experiments.

\subsection{Inference}
\label{asubsec:inference}

For all main experiments, we use a future horizon of $W=16$, a confidence threshold of $\tau_{\mathrm{conf}}=0.80$, and a revision threshold of $\tau_{\mathrm{rev}}=-0.5$.
As described in Section~\ref{subsec:rolling_draft_generation}, the ordinary next-frame prediction at $h=1$ is always retained.
DiffuPlex-\textsc{Listen} retains positions beyond $h=1$ only while the sampled assistant text state remains PAD and requires a minimum candidate prefix of $m=6$, whereas DiffuPlex-\textsc{Speak} permits PAD, EPAD, and lexical states with $m=2$.
If the candidate prefix does not reach the policy-specific minimum, we retain only $h=1$.
Thus, \textsc{Listen} advances conservatively through predicted PAD states, while \textsc{Speak} can additionally consume confidently predicted EPAD and lexical states.

Apart from the retention and revision rules, we preserve the original backbone sampling procedure.
Assistant audio codebooks are sampled with temperature $0.8$ and top-$k=250$, while assistant text uses temperature $0.7$ and top-$k=25$.
Following Section~\ref{subsec:rolling_draft_generation}, retention confidence is computed from the probability of the sampled assistant text token under the logits before temperature scaling and top-$k$ filtering.
We use random seed 0 for all main evaluations.
We also retain the official PersonaPlex priming procedure: the \texttt{SC0001.wav} voice prompt is encoded first, followed by six silence frames, the system text prompt, and another six silence frames.
Task-specific system prompts are provided with their corresponding evaluation protocols in Appendix~\ref{asec:evaluation_details}.
After priming, incoming user audio is encoded by Mimi at 12.5Hz following the autoregressive backbone.

At each backbone wake, DiffuPlex predicts a future block and selects a contiguous prefix for retention.
The assistant audio codebooks of the retained frames are decoded in parallel across temporal positions, while remaining autoregressive over codebooks within each frame, then placed in a playback queue and emitted at their original 80ms interaction times.
The corresponding future user distributions are stored only for consistency checking; actual incoming user frames, rather than these predictions, form the realized interaction used at the next backbone wake.
The temporal backbone wakes again when the retained prefix is exhausted or when newly observed user input triggers revision of the unplayed future.

The revision check uses the distributions stored at the backbone wake that produced the current draft.
For lookahead $h$, let $x_h$ denote the observed user semantic token, $q_h$ its stored predicted distribution, $y_h$ the sampled assistant text token, and $P_h$ the corresponding stored assistant text distribution.
At $h=1$, $x_1$ is the user semantic token for the frame that has just arrived at the wake boundary. Because of the backbone's autoregressive shift, $q_1$ predicts this token from earlier context, even though the observed $x_1$ is already available to deeper draft positions in the same wake. We therefore use $q_1(x_1)$ only to measure how surprising the newly arrived boundary frame was, rather than to verify a future prediction. Starting from $h=2$, the corresponding user token has not yet arrived when the wake is computed, so $q_h$ is a genuine prediction of future user input that can be checked once $x_h$ is later observed. For these future positions, we let $H_0$ denote that the observed user input remains consistent with the current draft and $H_1$ that the unplayed future should be revised.
By Bayes' rule,
\begin{equation}
\frac{P(H_1\mid x_h)}{P(H_0\mid x_h)}
=
\frac{P(H_1)}{P(H_0)}
\frac{p(x_h\mid H_1)}{p(x_h\mid H_0)}.
\label{aeq:revision_bayes_impl}
\end{equation}

Under $H_0$, the stored user prediction directly provides the likelihood of the received token,
\begin{equation}
p(x_h\mid H_0)=q_h(x_h).
\end{equation}
Under $H_1$, however, no corresponding user distribution is available: obtaining one would require generating an alternative future after revision with an additional temporal-backbone wake.
Doing so would defeat the purpose of deciding whether to revise using only information already available from the current draft.
We therefore use a position-independent reference likelihood,
\begin{equation}
p(x_h\mid H_1)\approx\xi.
\end{equation}
This does not require $\xi$ to be a calibrated likelihood under an explicit alternative future.
Because the same positive constant is used at every position, it only sets the reference scale against which $q_h(x_h)$ is compared; in log space, its effect is a constant offset that can be absorbed into the revision threshold.
Thus, the user-side evidence varies only with how surprising the received token is under the stored prediction.

We use the assistant distribution stored at the same wake as a proxy for the assistant-side prior odds:
\begin{equation}
\frac{P(H_1)}{P(H_0)}
\approx
\frac{P_h(\mathcal{A}_h(y_h))}
     {P_h(y_h)},
\end{equation}
where $\mathcal{A}_h(y_h)$ denotes the valid alternatives to the sampled assistant text token.
A larger ratio indicates that the assistant distribution assigned more mass to valid alternatives relative to the token selected in the draft.

Combining these approximations gives
\begin{equation}
\frac{P(H_1\mid x_h)}{P(H_0\mid x_h)}
\approx
\frac{P_h(\mathcal{A}_h(y_h))}
     {P_h(y_h)}
\frac{\xi}{q_h(x_h)}.
\end{equation}
Taking log odds,
\begin{equation}
\log
\frac{P(H_1\mid x_h)}{P(H_0\mid x_h)}
\approx
-\log q_h(x_h)
+
\log P_h(\mathcal{A}_h(y_h))
-
\log P_h(y_h)
+
\log\xi.
\end{equation}

Let $\tilde{\tau}_{\mathrm{rev}}$ denote the decision threshold on these approximate log odds.
Since $\xi$ is constant across positions, we absorb $\log\xi$ into the threshold by defining
\begin{equation}
\tau_{\mathrm{rev}}
=
\tilde{\tau}_{\mathrm{rev}}-\log\xi.
\end{equation}
The implemented revision score is therefore
\begin{equation}
S_h
=
-\log q_h(x_h)
+
\log P_h(\mathcal{A}_h(y_h))
-
\log P_h(y_h),
\label{aeq:revision_score_impl}
\end{equation}
and revision is triggered when
\begin{equation}
S_h \geq \tau_{\mathrm{rev}}.
\end{equation}

We construct $\mathcal{A}_h(y_h)$ according to the assistant text-state transitions of the backbone.
Let $s_{h-1}$ denote the preceding assistant text state.
At the first position of a draft, $s_{h-1}$ is the last realized assistant state; at later positions, it is the state of the preceding drafted token $y_{h-1}$.
The valid alternatives are defined as follows:
\begin{itemize}
    \item \textbf{PAD.}
    If $s_{h-1}$ is PAD, the next state can only remain PAD or transition to EPAD, since lexical generation must be preceded by EPAD.
    The alternative mass is therefore the probability assigned to the other valid state.

    \item \textbf{EPAD.}
    If $s_{h-1}$ is EPAD, the next token must be lexical.
    For a sampled lexical token $y_h$, the alternatives are all other lexical tokens, with total probability mass
    \begin{equation}
    P_h(\mathcal{A}_h(y_h))
    =
    1-P_h(\mathrm{PAD})-P_h(\mathrm{EPAD})-P_h(y_h).
    \end{equation}

    \item \textbf{Lexical.}
    If $s_{h-1}$ is lexical, the next token may be lexical, PAD, or EPAD.
    Every token other than the sampled $y_h$ is therefore a valid alternative, giving
    \begin{equation}
    P_h(\mathcal{A}_h(y_h))
    =
    1-P_h(y_h).
    \end{equation}
\end{itemize}

Computing Eq.~\ref{aeq:revision_score_impl} requires only the stored distributions and the newly observed user token and therefore does not itself invoke the temporal backbone.
If revision is triggered, all assistant frames that have already been emitted and all user frames that have already been observed remain fixed.
Only the unplayed portion of the draft is discarded.
The stored future predictions are cleared, the unplayed region is returned to the masked state, and the next backbone wake predicts a new future from the updated realized interaction.
Revision never modifies audio that has already been played.

For the signal-decomposition ablations in Appendix~\ref{asubsec:revision_decomposition}, we calibrate the user-only and assistant-only revision thresholds on a fixed subset of Full-Duplex-Bench v1.5~\citep{11463576} to approximately match the average number of interaction frames per backbone wake ($c$) of the full revision score.
These calibrated variants are compared only on Full-Duplex-Bench v1.0~\citep{11433838}, which is not used for threshold selection.
Full-Duplex-Bench v1.5 is not used to select any other inference hyperparameters.
Finally, as an implementation sanity check, setting $W=1$ disables future prediction and reduces DiffuPlex to ordinary frame-by-frame generation.
Under matched priming, streaming user encoding, sampling, and random seed, we verify token-level equivalence between this path and the official autoregressive inference implementation.

\subsection{Deferred-Execution Baselines}
\label{asubsec:deferred_baselines}

The deferred-execution baselines test whether backbone calls can be reduced without predicting future interaction.
Rather than generating a future block as DiffuPlex does, these methods retain the original autoregressive backbone and decide only whether its next update can be postponed.
All three baselines use the same frozen \texttt{personaplex-rl-seamless} backbone, with no MASK embeddings or LoRA adapters, and share the same deferred-execution mechanism described below.
They differ only in the controller signal used to determine when deferral is permitted.
Deferral is permitted only while the assistant is silent: we require both the previously emitted assistant text token to be PAD and the corresponding decoded assistant-audio frame to have RMS below a silence threshold $\tau_{\mathrm{sil}}$.
The audio condition is necessary because PAD tokens can also occur within spoken regions; using the text stream alone can therefore incorrectly classify an ongoing utterance as silence.
We set $\tau_{\mathrm{sil}}=2.74670\times10^{-4}$ from the official PersonaPlex silence codebooks by streaming-decoding 2,000 silence frames, discarding the first 25 frames to remove decoder warm-up effects, taking the 99th percentile of the remaining RMS values ($2.19736\times10^{-4}$), and adding a 25\% margin.
No evaluation audio is used to determine this threshold.

While execution is deferred, the assistant text stream is fixed to PAD and the official silence audio codebooks are used for the assistant streams, respecting the backbone's native stream delays.
Incoming user audio continues to be encoded by Mimi and is stored normally.
When the controller wakes the backbone, the accumulated known user inputs and imposed assistant-silence trajectory are processed causally in a single temporal-backbone call, whose final position produces the next autoregressive prediction and avoids an additional redundant backbone call.
We cap each deferred segment by a maximum run length $R\in\{2,4,6,8\}$ interaction frames.
The count includes the current backbone wake; for example, $R=6$ consists of the current wake followed by at most five deferred frames.

\noindent\textbf{\textsc{PAD}.}
The \textsc{PAD} controller uses only the backbone's own assistant-text prediction.
At an eligible silent backbone wake, if the sampled assistant text token is PAD with probability at least $0.80$, the controller starts a deferred segment, using the same sampled-token probability definition as the DiffuPlex confidence criterion.
Because no new backbone prediction is produced during the deferred frames, the segment continues until the maximum run length $R$ is reached.
Thus, for $R=6$, the qualifying wake is followed by five deferred frames, after which the backbone is forced to wake again and produce a new autoregressive prediction.
If the PAD condition is not satisfied at a wake, no new deferral is started.
This baseline therefore tests whether confident current-frame silence alone is sufficient to postpone future autoregressive updates without an external speech or turn detector.

\noindent\textbf{\textsc{Silero}.}
The \textsc{Silero} controller uses \textsc{Silero VAD} v6.2.1~\citep{SileroVAD} in streaming mode on the incoming user audio.
Audio is resampled to 16kHz mono and processed causally in 512-sample (32ms) chunks using the official \texttt{VADIterator}.
We use a speech threshold of $0.5$, the released hysteresis exit threshold of $0.35$, \texttt{min\_silence\_duration\_ms}$=200$, and \texttt{speech\_pad\_ms}$=0$.
When the user is speaking, the backbone remains deferred while \textsc{Silero} continues to process the incoming audio.
The backbone wakes at whichever occurs first: \textsc{Silero} reports that the user has been silent for at least 200ms, or the deferred segment reaches the maximum run length $R$.
Thus, if speech continues throughout an $R=6$ segment, the qualifying wake is followed by five deferred frames before the backbone is forced to wake again.
All \textsc{Silero} decisions are causal: we act only when the streaming VAD reports a transition, without shifting the decision back to an earlier estimated speech boundary.

\noindent\textbf{\textsc{Smart Turn}.}
The \textsc{Smart Turn} controller augments the same \textsc{Silero} frontend with \textsc{Smart Turn} v3.2\footnote{\url{https://huggingface.co/pipecat-ai/smart-turn-v3}}, using the released INT8 CPU model \texttt{smart-turn-v3.2-cpu.onnx}.
At each \textsc{Silero}-detected speech end, \textsc{Smart Turn} is evaluated on the most recent user-audio context, truncated to the latest 8s and zero-padded at the beginning when shorter.
If the completion probability is at least $0.5$, the backbone wakes; otherwise, it remains deferred.
Deferral then continues until a later speech end produces a completion probability of at least $0.5$, the maximum run length $R$ is reached, or 3s of continuous silence forces a wake.
For the primary \textsc{Smart Turn} baseline, we account for the controller's measured inference latency: a completion decision is applied at the first 80ms interaction frame after its result becomes available.
If the user resumes speaking before that result is available, the decision is discarded as stale.

\textsc{Silero} and \textsc{Smart Turn} decisions are precomputed before timed generation and replayed at their corresponding interaction times, so that controller execution does not affect GPU timing.
We therefore exclude controller inference from the reported streaming-model runtime and report its CPU cost separately.
For the primary \textsc{Smart Turn} configuration, the measured controller cost is 13.90 CPU-ms per interaction second, approximately 80\% of which comes from the continuously running \textsc{Silero} frontend.

\section{Evaluation Details}
\label{asec:evaluation_details}

We provide the complete evaluation protocols for the interaction, general-capability, and runtime results reported in the main paper.
Unless otherwise stated, all systems use the inference configuration in Appendix~\ref{asubsec:inference}.
Model names and system prompts follow the corresponding backbone or benchmark conventions described below.

\subsection{Full-Duplex Benchmarks}
\label{asubsec:full_duplex_benchmarks}

We evaluate full-duplex interaction using Full-Duplex-Bench v1.0~\citep{11433838} and v1.5~\citep{11463576}.
Unlike the single-turn spoken-language benchmarks considered later, these evaluations preserve continuous full-duplex generation throughout the input, since when the assistant chooses to speak, remain silent, backchannel, or yield is itself part of the behavior being measured.

\paragraph{Full-Duplex-Bench v1.0.}
Full-Duplex-Bench v1.0 contains 727 clips spanning four interaction behaviors.
Pause handling consists of 216 Candor~\citep{doi:10.1126/sciadv.adf3197} clips and 137 synthetic clips, smooth turn-taking contains 119 Candor clips, backchanneling contains 55 ICC~\citep{umair-etal-2024-large} clips, and user interruption contains 200 synthetic clips.
For each example, the user waveform is streamed through Mimi and the assistant generates freely until the input waveform ends; assistant activity is never suppressed during the interaction.

We follow the system-prompt convention of \texttt{personaplex-rl-seamless}.
Pause handling, smooth turn-taking, and backchanneling use
\texttt{You enjoy having a good conversation.},
whereas user interruption uses
\texttt{You are a wise and friendly teacher. Answer questions or provide advice in a clear and engaging way.}
We use the official \texttt{SC0001.wav} voice prompt and the priming procedure described in Appendix~\ref{asubsec:inference}.

We use the official benchmark evaluation scripts and metric definitions.
Turn-Over Rate (TOR) measures whether the assistant takes a conversational turn; the official evaluator counts an output as a turn when an ASR transcription is present and either its duration is at least 1s or it contains more than three words.
Lower TOR is preferred for pause handling and backchanneling, where a full assistant response is generally undesirable, whereas higher TOR is preferred for smooth turn-taking and user interruption.
For smooth turn-taking and interruption, latency measures the onset of the first assistant word relative to the end of the corresponding user input.
Backchannel frequency measures how often the model produces backchannels, and backchannel JSD measures the Jensen-Shannon divergence between their timing distribution and the reference ICC distribution.
For user interruption, response appropriateness (Judge) is additionally scored on the official 1--5 scale.
We use \texttt{nvidia/parakeet-tdt-0.6b-v2}~\citep{pmlr-v202-xu23g} for ASR-dependent metrics and \textit{gpt-5.6-luna}\footnote{\url{https://developers.openai.com/api/docs/models/gpt-5.6-luna}} with reasoning disabled for the interruption judge, while retaining the official prompt.

\paragraph{Full-Duplex-Bench v1.5.}
Full-Duplex-Bench v1.5 contains 498 examples covering background speech (100), talking to another person (100), user backchannels (98), and user interruptions (200).
Each example provides an input containing the target overlap event together with a corresponding clean input in which that event is absent.
We therefore run each system separately on both \texttt{input.wav} and \texttt{clean\_input.wav} and apply the official paired evaluation protocol to the resulting outputs.
Background speech, talking-to-other, and user-backchannel examples use the conversational system prompt above, while user interruption uses the teacher prompt.

The benchmark measures how assistant behavior changes in response to the overlap event.
Its official evaluation includes response-behavior distributions, stop and response latency, and changes in speech characteristics before and after the event, including UTMOSv2~\citep{10832315}, speaking rate, pitch statistics, and intensity statistics.
For the LLM-based behavior classification, we use \textit{gpt-5.6-luna} with reasoning disabled while retaining the official evaluation prompt.
We apply the benchmark's paired pre/post comparisons as provided by the official evaluator.
Complete Full-Duplex-Bench v1.5 results are reported in Appendix~\ref{asubsec:additional_full_duplex}.

\subsection{General Spoken-Language Benchmarks}
\label{asubsec:general_benchmarks}

Full-duplex interaction metrics test whether DiffuPlex preserves conversational behavior, but do not establish whether future prediction changes the backbone's general response capability.
We therefore additionally evaluate VoiceBench~\citep{chen-etal-2026-voicebench} and OpenAudioBench~\citep{li2025baichuanaudiounifiedframeworkendtoend} under a common single-turn protocol.
VoiceBench contains 7,783 examples from nine tasks: AlpacaEval (636), CommonEval (200), WildVoice (1,000), SD-QA with the U.S. accent split (553), OpenBookQA (455), MMSU (3,074), BBH (1,000), IFEval (345), and AdvBench (520).
We exclude the 46-example MT-Bench subset because it requires a multi-turn evaluation protocol that does not match the single-turn setting used here.
For OpenAudioBench, we evaluate the four English subsets, Llama Questions (300), Web Questions (1,000), TriviaQA (1,000), and AlpacaEval (199), yielding 2,499 examples.
We exclude the remaining Reasoning QA subset, which is released in Chinese, because the backbone and our speech-to-speech transcription pipeline are English-only.

Both benchmarks assume that the complete spoken instruction is received before response generation begins.
We adapt them to the continuously running full-duplex backbone following the Moshi~\citep{defossez2024moshispeechtextfoundationmodel} evaluation pipeline in VoiceBench.
While the instruction waveform is streamed normally through Mimi, the assistant output streams are constrained to silence, and free generation begins immediately after the final instruction frame.
This prevents incidental full-duplex behavior, such as backchannels produced during the question, from being interpreted as part of the benchmark answer while leaving the subsequent response unconstrained.
Because the backbone does not emit an EOS event for spoken generation, we append a 163.84s silent user tail, corresponding to 2,048 interaction frames, after each instruction.
Once assistant speech has begun, generation terminates after 8.0s of continuous assistant silence; if this condition is never reached, generation ends at the end of the appended tail.
The silence counter begins only after the instruction has finished.
Both benchmarks use the conversational system prompt \texttt{You enjoy having a good conversation.} and the sampling configuration in Appendix~\ref{asubsec:inference}.

We evaluate both the model's native text output and the speech ultimately presented to the user.
For speech-to-text (s$\rightarrow$t), we directly score the assistant text stream from the free-generation region after removing the backbone's special text tokens, without using ASR.
For speech-to-speech (s$\rightarrow$s), we remove the instruction-aligned prefix from the generated assistant audio, transcribe the remaining speech with \texttt{nvidia/parakeet-tdt-0.6b-v2}, and apply the original benchmark scorers to the resulting transcription.
For judge-based tasks, we use \textit{gpt-5.6-luna} with reasoning disabled while retaining the official task prompts and aggregation procedures.
Complete per-task results for both output modalities are reported in Appendix~\ref{asubsec:additional_general}.

\subsection{Runtime Measurement}
\label{asubsec:runtime_measurement}

We measure runtime on a fixed subset of 60 Full-Duplex-Bench v1.0 clips using a single NVIDIA RTX A6000 GPU.
The subset contains 15 examples from each of the four interaction categories.
Within each category, we select examples to cover the range of assistant-silence fractions observed under the autoregressive \texttt{personaplex-rl-seamless} reference; for pause handling, we additionally include examples from both the Candor and synthetic sources.
This subset is selected once and reused unchanged for every system, so the runtime evaluation does not depend on each system's own speaking behavior.

We distinguish three notions of runtime.
\emph{Streaming wall time} measures elapsed time through the deployment-path streaming loop and therefore includes Python and CPU control, synchronization, and GPU execution.
\emph{E2E GPU time} accumulates GPU execution for the complete streaming inference pipeline, including the speech codec path.
\emph{Core LM time} measures GPU execution of the language-model generation core alone.
All speedups are computed relative to the autoregressive \texttt{personaplex-rl-seamless} reference under the same measurement path, excluding file writing, ASR, judging, and benchmark scoring.
To separate how often the temporal backbone runs from the cost of each wake, we define the number of interaction frames advanced per backbone wake as
\begin{equation}
c
=
\frac{\text{total interaction frames}}
     {\text{total backbone wakes}}.
\end{equation}
The relative Core LM time per wake reported in Table~\ref{tab:runtime_results} is the average Core LM time per backbone wake normalized by the autoregressive reference.
This distinction is important because a DiffuPlex wake processes a future block and can therefore be more expensive than a single autoregressive wake, while still reducing total Core LM time by invoking the backbone less frequently.

GPU totals are measured with CUDA timing while avoiding unnecessary synchronization.
Wake-frame latency is measured separately on the streaming critical path, with synchronization when the temporal backbone is invoked, and the initial CUDA graph capture is excluded.
We report the p99 wake-frame latency together with the fraction
whose wall-clock latency exceeds 80ms, the duration of one interaction frame.
For the deferred-execution baselines in Appendix~\ref{asubsec:deferred_baselines}, controller decisions are computed causally in advance and replayed during the timed generation loop so that controller execution does not perturb GPU timing.
\textsc{Silero} and \textsc{Smart Turn} inference is therefore excluded from streaming wall time and measured separately.
The primary \textsc{Smart Turn} pipeline requires approximately 13.90ms of controller processing per interaction second in our setup, with approximately 80\% of this time spent in the continuously running \textsc{Silero} frontend.
The runtime comparison therefore separates backbone streaming cost from this additional CPU-side controller computation.
Runtime measurements for the main systems are reported in
Table~\ref{tab:runtime_results}, with deferred-execution results
in Appendix~\ref{asubsec:additional_runtime}.

\subsection{Statistical Testing}
\label{asubsec:statistical_testing}

For Full-Duplex-Bench v1.0, we compare DiffuPlex-\textsc{Listen} and DiffuPlex-\textsc{Speak} separately with the autoregressive \texttt{personaplex-rl-seamless} backbone using paired statistical tests over ten interaction metrics per policy: the two pause-handling TORs; backchannel TOR, frequency, and JSD; smooth-turn TOR and latency; and interruption TOR, response score, and latency, yielding 20 comparisons in total. We use the exact McNemar test for binary clip-level outcomes, including TOR, and a two-sided Wilcoxon signed-rank test for paired continuous or ordinal measurements. We jointly correct the resulting 20 tests using the Holm procedure and report both unadjusted and Holm-adjusted results in Appendix~\ref{asubsec:paired_significance}. A non-significant result after correction is interpreted only as an absence of detected evidence of a difference under these tests, not as evidence of formal equivalence.

\section{Human Evaluation}
\label{asec:human_evaluation}

\begin{figure*}[p]
    \centering
    \includegraphics[width=\textwidth]{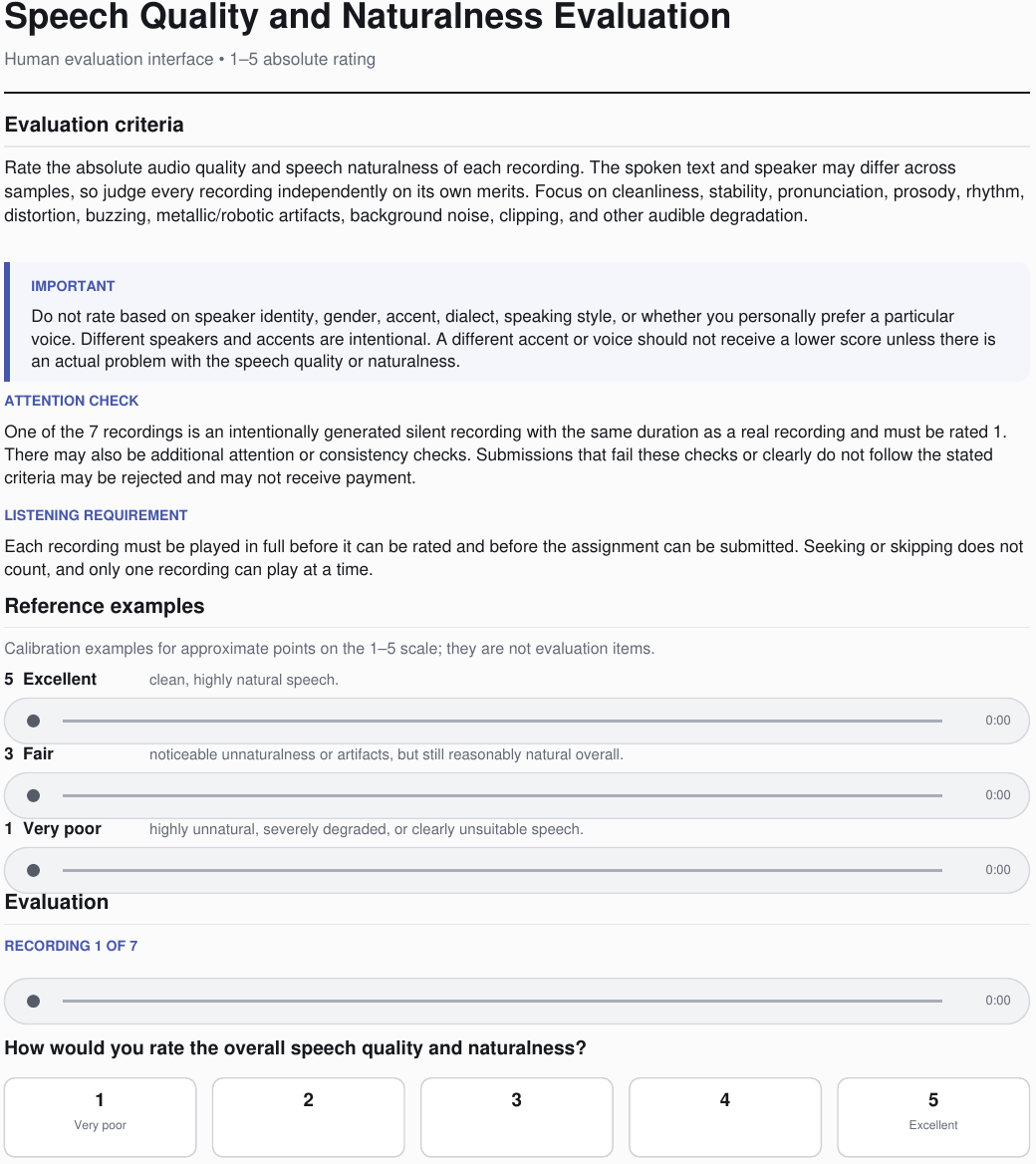}
    \caption{
    NMOS evaluation interface.
    Evaluators rate the speech quality and naturalness of the assistant response following the user interruption on a five-point scale.
    }
    \label{afig:nmos_interface}
\end{figure*}
\clearpage

\begin{figure*}[p]
    \centering
    \includegraphics[width=\textwidth]{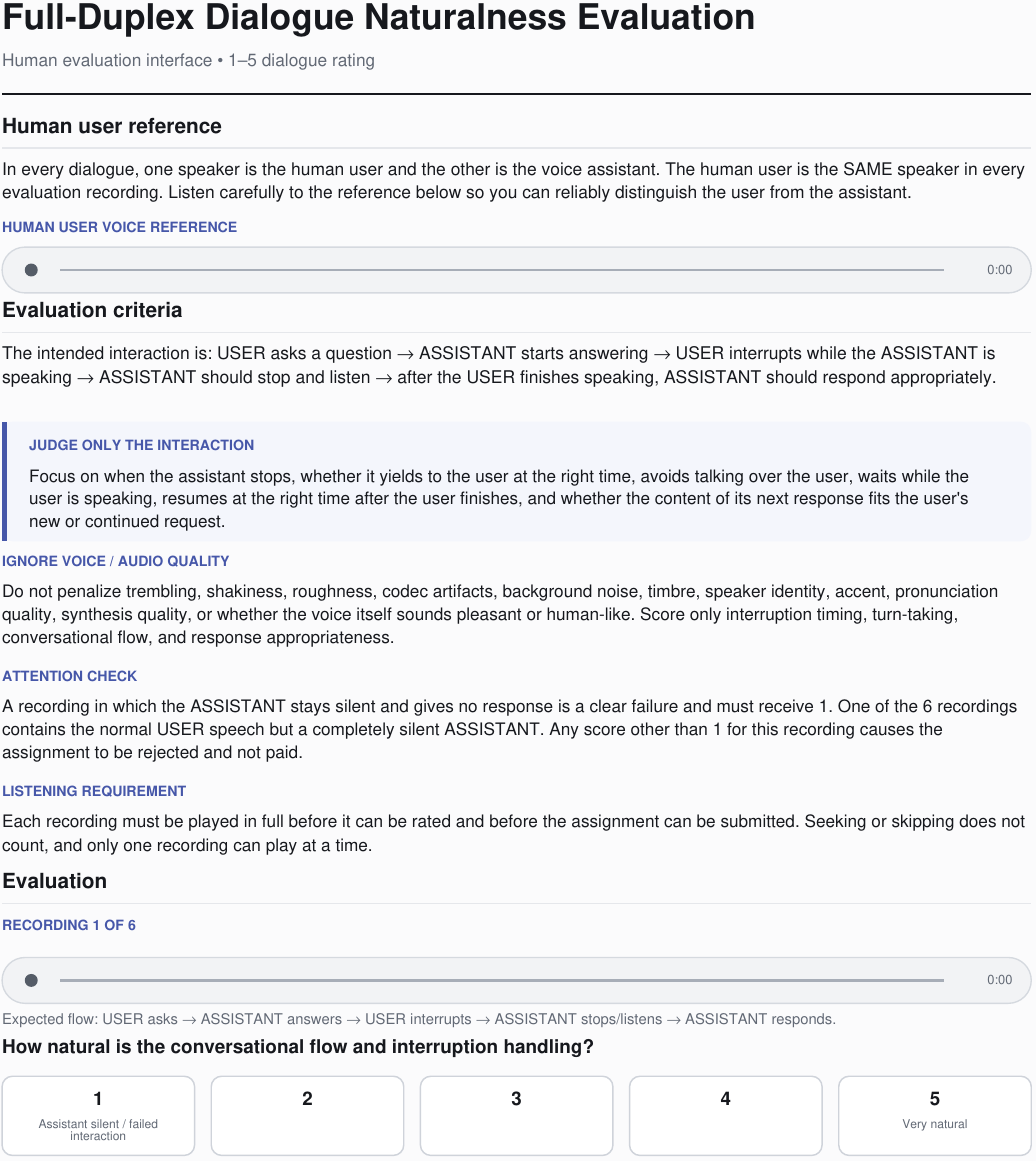}
    \caption{
    DMOS evaluation interface.
    Evaluators rate the complete interaction on a five-point scale while focusing on interruption handling and conversational flow.
    }
    \label{afig:dmos_interface}
\end{figure*}
\clearpage

We conduct two complementary human evaluations on Amazon Mechanical Turk (MTurk) using user-interruption examples from Full-Duplex-Bench v1.0.
Naturalness mean opinion score (NMOS) evaluates only the assistant speech following the interruption, whereas dialog mean opinion score (DMOS) evaluates the complete interaction, including how the assistant yields to the interruption and resumes afterward.
For each assignment, outputs from the compared systems correspond to the same underlying interaction, with presentation order randomized and model identities hidden.

\paragraph{Speech naturalness.}
For NMOS, evaluators hear only the assistant response following the interruption and answer, ``How would you rate the overall speech quality and naturalness?'', on a five-point scale from 1 (\textit{Very poor}) to 5 (\textit{Excellent}).
They are instructed to judge speech quality and naturalness, including pronunciation, prosody, rhythm, and audio artifacts, while ignoring speaker identity, accent, speaking style, and personal voice preference.
Before rating, workers are provided with fixed reference recordings illustrating approximate scores of 5, 3, and 1 for scale calibration.
Each assignment includes two attention checks: an intentionally silent recording and a known score-1 calibration item, both of which must receive a score of 1.
Assignments that fail either check are discarded and recollected from a new worker.
The interface requires every recording to be played in full before submission.
The final NMOS analysis contains 200 valid assignments from 120 workers.

\paragraph{Conversational interaction.}
For DMOS, evaluators hear the complete interaction and answer, ``How natural is the conversational flow and interruption handling?'', on a five-point scale from 1 (\textit{Very unnatural}) to 5 (\textit{Very natural}).
They are instructed to judge whether the assistant yields at an appropriate time, avoids talking over the user, waits while the user is speaking, resumes appropriately, and produces a response that fits the user's continued or revised request.
Because DMOS is intended to isolate interaction behavior, evaluators are explicitly instructed to ignore acoustic quality, including synthesis artifacts, timbre, speaker identity, accent, and pronunciation.
Each assignment contains one attention-check interaction with the normal user channel but a completely silent assistant channel, which must receive a score of 1.
Assignments that fail this check are discarded and recollected from a new worker.
The interface requires every recording to be played in full before submission.
The final DMOS analysis contains 180 valid assignments from 84 workers.

\begin{table*}[t]
\centering
\caption{
Paired Full-Duplex-Bench v1.0 comparisons against the autoregressive \texttt{personaplex-rl-seamless} backbone.
AR and DiffuPlex values are computed over the paired clips used by each test and may therefore differ from the aggregate values in the main table.
$\Delta$ denotes DiffuPlex minus AR.
Holm correction is applied jointly over all 20 comparisons.
$\dagger$ denotes a binary comparison for which the two systems have identical clip-level outcomes.
}
\label{atab:paired_significance}
\small
\begin{tabular}{lllrcccrr}
\toprule
Policy & Metric & Test & $n$ & AR & DiffuPlex & $\Delta$ & Raw $p$ & Holm $p$ \\
\midrule
\textsc{Listen}
& Pause C & McNemar & 216 & 0.3009 & 0.3426 & +0.0417 & 0.2221 & 1.0000 \\
& Pause S & McNemar & 137 & 0.2993 & 0.2774 & -0.0219 & 0.6776 & 1.0000 \\
& BC TOR & McNemar & 55 & 0.2182 & 0.1091 & -0.1091 & 0.2379 & 1.0000 \\
& BC Freq & Wilcoxon & 55 & 0.1634 & 0.1402 & -0.0231 & 0.0266 & 0.5320 \\
& BC JSD & Wilcoxon & 55 & 0.7405 & 0.7534 & +0.0130 & 0.0859 & 1.0000 \\
& Turn TOR & McNemar$^\dagger$ & 119 & 1.0000 & 1.0000 & 0.0000 & 1.0000 & 1.0000 \\
& Turn Lat. & Wilcoxon & 119 & 0.1242 & 0.1732 & +0.0490 & 0.3609 & 1.0000 \\
& Int. TOR & McNemar & 200 & 0.9950 & 0.9900 & -0.0050 & 1.0000 & 1.0000 \\
& Int. Lat. & Wilcoxon & 197 & 0.2412 & 0.3070 & +0.0658 & 0.0535 & 1.0000 \\
& Int. Judge & Wilcoxon & 197 & 4.2843 & 4.3299 & +0.0457 & 0.6279 & 1.0000 \\
\midrule
\textsc{Speak}
& Pause C & McNemar & 216 & 0.3009 & 0.3472 & +0.0463 & 0.1934 & 1.0000 \\
& Pause S & McNemar & 137 & 0.2993 & 0.2628 & -0.0365 & 0.4049 & 1.0000 \\
& BC TOR & McNemar & 55 & 0.2182 & 0.1455 & -0.0727 & 0.4545 & 1.0000 \\
& BC Freq & Wilcoxon & 55 & 0.1634 & 0.1543 & -0.0091 & 0.4666 & 1.0000 \\
& BC JSD & Wilcoxon & 55 & 0.7405 & 0.7378 & -0.0027 & 0.8210 & 1.0000 \\
& Turn TOR & McNemar$^\dagger$ & 119 & 1.0000 & 1.0000 & 0.0000 & 1.0000 & 1.0000 \\
& Turn Lat. & Wilcoxon & 119 & 0.1242 & 0.1559 & +0.0317 & 0.6327 & 1.0000 \\
& Int. TOR & McNemar$^\dagger$ & 200 & 0.9950 & 0.9950 & 0.0000 & 1.0000 & 1.0000 \\
& Int. Lat. & Wilcoxon & 199 & 0.2806 & 0.3465 & +0.0659 & 0.1104 & 1.0000 \\
& Int. Judge & Wilcoxon & 199 & 4.2864 & 4.3568 & +0.0704 & 0.3263 & 1.0000 \\
\bottomrule
\end{tabular}
\end{table*}

\section{Additional Results}
\label{asec:additional_results}

\subsection{Paired Significance Results}
\label{asubsec:paired_significance}

\begin{table*}[t]
\centering
\caption{
Full-Duplex-Bench v1.5 behavioral response distributions and interaction latencies.
Stop and response latencies are reported in seconds.
}
\label{atab:fdb_v15_behavior}
\small
\begin{tabular}{llrrr}
\toprule
Scenario & Class / Metric
& AR backbone
& DiffuPlex-\textsc{Listen}
& DiffuPlex-\textsc{Speak} \\
\midrule
User interruption
& Respond $\uparrow$     & 0.94  & 0.95  & 0.96 \\
& Resume $\downarrow$    & 0.04  & 0.04  & 0.04 \\
& Uncertain $\downarrow$ & 0.00  & 0.01  & 0.00 \\
& Unknown $\downarrow$   & 0.02  & 0.00  & 0.00 \\
& Stop (s) $\downarrow$  & 0.443 & 0.459 & 0.446 \\
& Resp. (s) $\downarrow$ & 0.368 & 0.417 & 0.455 \\
\midrule
User backchannel
& Respond $\downarrow$   & 0.11  & 0.06  & 0.06 \\
& Resume $\uparrow$      & 0.74  & 0.92  & 0.91 \\
& Uncertain $\downarrow$ & 0.01  & 0.01  & 0.00 \\
& Unknown $\downarrow$   & 0.13  & 0.01  & 0.03 \\
& Stop (s) $\uparrow$    & 0.586 & 0.556 & 0.598 \\
& Resp. (s) $\downarrow$ & 0.603 & 0.598 & 0.730 \\
\midrule
Talking to other
& Respond $\downarrow$   & 0.60  & 0.56  & 0.54 \\
& Resume $\uparrow$      & 0.26  & 0.31  & 0.34 \\
& Uncertain $\uparrow$   & 0.05  & 0.01  & 0.03 \\
& Unknown $\downarrow$   & 0.09  & 0.12  & 0.09 \\
& Stop (s) $\uparrow$    & 0.528 & 0.515 & 0.547 \\
& Resp. (s) $\downarrow$ & 0.387 & 0.322 & 0.364 \\
\midrule
Background speech
& Respond $\downarrow$   & 0.49  & 0.46  & 0.49 \\
& Resume $\uparrow$      & 0.39  & 0.45  & 0.35 \\
& Uncertain $\uparrow$   & 0.01  & 0.01  & 0.04 \\
& Unknown $\downarrow$   & 0.11  & 0.08  & 0.12 \\
& Stop (s) $\uparrow$    & 0.710 & 0.630 & 0.680 \\
& Resp. (s) $\downarrow$ & 0.362 & 0.291 & 0.452 \\
\bottomrule
\end{tabular}
\end{table*}
\begin{table*}[t]
\centering
\caption{
Speech-feature changes before and after user interruption for trials classified as \textsc{Respond}, following the Full-Duplex-Bench v1.5 evaluation protocol.
Each cell reports pre-overlap $\rightarrow$ post-overlap, followed by the post-minus-pre change and paired $t$-test $p$-value.
The number of valid paired trials varies slightly across systems and features because \textsc{Respond} filtering and the benchmark's MAD-based outlier removal ($k=3.5$) are applied separately to each system.
}
\label{atab:fdb_v15_acoustic}
\scriptsize
\resizebox{\textwidth}{!}{
\begin{tabular}{lcccccc}
\toprule
Model
& WPM
& Pitch Mean
& Pitch SD
& Intensity Mean
& Intensity SD
& UTMOSv2 \\
\midrule

AR backbone
& \shortstack{$141.05 \rightarrow 220.05$\\ $(+79.00,\;p<.001)$}
& \shortstack{$238.45 \rightarrow 236.77$\\ $(-1.68,\;p=.134)$}
& \shortstack{$46.24 \rightarrow 46.30$\\ $(+0.06,\;p=.950)$}
& \shortstack{$59.10 \rightarrow 60.48$\\ $(+1.38,\;p<.001)$}
& \shortstack{$17.93 \rightarrow 15.98$\\ $(-1.95,\;p<.001)$}
& \shortstack{$2.829 \rightarrow 2.858$\\ $(+0.029,\;p=.429)$}
\\

\midrule
DiffuPlex-\textsc{Listen}
& \shortstack{$140.79 \rightarrow 219.90$\\ $(+79.11,\;p<.001)$}
& \shortstack{$238.32 \rightarrow 234.53$\\ $(-3.79,\;p=.001)$}
& \shortstack{$44.40 \rightarrow 42.58$\\ $(-1.82,\;p=.065)$}
& \shortstack{$58.64 \rightarrow 60.25$\\ $(+1.61,\;p<.001)$}
& \shortstack{$18.58 \rightarrow 16.22$\\ $(-2.37,\;p<.001)$}
& \shortstack{$2.697 \rightarrow 2.812$\\ $(+0.115,\;p<.001)$}
\\

\midrule
DiffuPlex-\textsc{Speak}
& \shortstack{$141.76 \rightarrow 220.04$\\ $(+78.28,\;p<.001)$}
& \shortstack{$239.46 \rightarrow 236.36$\\ $(-3.09,\;p=.023)$}
& \shortstack{$47.57 \rightarrow 46.64$\\ $(-0.93,\;p=.367)$}
& \shortstack{$58.02 \rightarrow 60.22$\\ $(+2.21,\;p<.001)$}
& \shortstack{$18.21 \rightarrow 16.08$\\ $(-2.13,\;p<.001)$}
& \shortstack{$2.549 \rightarrow 2.635$\\ $(+0.086,\;p=.024)$}
\\

\bottomrule
\end{tabular}
}
\end{table*}

Table~\ref{atab:paired_significance} reports the 20 paired Full-Duplex-Bench v1.0 comparisons defined in Appendix~\ref{asubsec:statistical_testing}.
Each comparison uses the same clips for the autoregressive backbone and DiffuPlex, so the test asks whether the two systems behave differently on matched examples rather than comparing independent sets of outputs.
For binary outcomes such as TOR, we use the exact McNemar test, which examines how often the two systems disagree on the same clips.
For continuous or ordinal quantities such as latency, backchannel frequency, JSD, and interruption score, we use the two-sided Wilcoxon signed-rank test, which compares the paired differences across clips without assuming that those differences are normally distributed.
The raw $p$-value measures the evidence for a difference in each individual comparison, while the Holm-adjusted $p$-value additionally accounts for testing 20 hypotheses jointly.
We use the adjusted values when determining statistical significance.
None of the 20 comparisons remains significant after Holm correction, so these tests do not detect a statistically significant difference between DiffuPlex and the autoregressive backbone on any evaluated metric.
This should not be interpreted as evidence that the systems are formally equivalent.

\begin{table*}[t]
\centering
\caption{
Complete VoiceBench results.
Values follow the benchmark's official task-specific scoring and are reported on the common percentage scale used for aggregation.
Overall is the aggregate reported in the main paper.
}
\label{atab:voicebench_full}
\scriptsize
\resizebox{\textwidth}{!}{
\begin{tabular}{llrrrrrrrrrr}
\toprule
Model & Mode & AlpacaEval & CommonEval & WildVoice & SD-QA & OpenBookQA & MMSU & BBH & IFEval & AdvBench & Overall \\
\midrule
AR backbone
& $s\!\rightarrow\!t$ & 51.99 & 43.73 & 36.19 & 20.34 & 25.71 & 24.89 & 48.60 & 12.99 & 23.08 & 31.95 \\
DiffuPlex-\textsc{Listen}
& $s\!\rightarrow\!t$ & 50.91 & 45.00 & 36.82 & 22.42 & 25.05 & 25.05 & 49.30 & 12.00 & 20.77 & 31.93 \\
DiffuPlex-\textsc{Speak}
& $s\!\rightarrow\!t$ & 49.88 & 44.03 & 35.92 & 19.89 & 22.64 & 25.57 & 47.10 & 13.82 & 21.92 & 31.20 \\
\midrule
AR backbone
& $s\!\rightarrow\!s$ & 52.05 & 44.43 & 36.17 & 17.81 & 24.62 & 24.11 & 47.80 & 13.08 & 24.42 & 31.61 \\
DiffuPlex-\textsc{Listen}
& $s\!\rightarrow\!s$ & 50.37 & 44.23 & 36.58 & 20.16 & 25.71 & 24.43 & 48.40 & 12.25 & 21.54 & 31.52 \\
DiffuPlex-\textsc{Speak}
& $s\!\rightarrow\!s$ & 49.15 & 43.33 & 35.75 & 17.18 & 24.40 & 24.95 & 48.40 & 13.28 & 22.88 & 31.04 \\
\bottomrule
\end{tabular}
}
\end{table*}
\begin{table*}[t]
\centering
\caption{
OpenAudioBench results.
The four task columns retain their original score ranges.
Overall first normalizes each task to a percentage scale and then averages across tasks.
}
\label{atab:oab_full}
\small
\resizebox{\textwidth}{!}{
\begin{tabular}{llrrrrr}
\toprule
Model & Mode & Llama Questions $(0\!-\!100)$ & Web Questions $(0\!-\!10)$ & TriviaQA $(0\!-\!10)$ & AlpacaEval $(1\!-\!10)$ & Overall (\%) \\
\midrule
AR backbone
& $s\!\rightarrow\!t$ & 54.20 & 2.95 & 1.92 & 3.97 & 35.66 \\
DiffuPlex-\textsc{Listen}
& $s\!\rightarrow\!t$ & 57.60 & 3.07 & 1.92 & 3.93 & 36.71 \\
DiffuPlex-\textsc{Speak}
& $s\!\rightarrow\!t$ & 55.53 & 3.07 & 1.86 & 3.85 & 35.82 \\
\midrule
AR backbone
& $s\!\rightarrow\!s$ & 49.33 & 2.96 & 1.75 & 3.97 & 34.03 \\
DiffuPlex-\textsc{Listen}
& $s\!\rightarrow\!s$ & 46.73 & 3.02 & 1.80 & 3.92 & 33.53 \\
DiffuPlex-\textsc{Speak}
& $s\!\rightarrow\!s$ & 46.07 & 2.84 & 1.62 & 3.78 & 32.12 \\
\bottomrule
\end{tabular}
}
\end{table*}

\subsection{Full-Duplex-Bench v1.5}
\label{asubsec:additional_full_duplex}

Tables~\ref{atab:fdb_v15_behavior} and~\ref{atab:fdb_v15_acoustic} report the Full-Duplex-Bench v1.5 behavioral and speech-feature analyses under the protocol described in Appendix~\ref{asubsec:full_duplex_benchmarks}.
Across the four overlap conditions, DiffuPlex shows broadly similar response and resume patterns to the autoregressive backbone.
For user interruptions, the response rate is $0.94$ for the backbone and $0.95$ and $0.96$ for DiffuPlex-\textsc{Listen} and DiffuPlex-\textsc{Speak}, respectively.
For user backchannels, DiffuPlex resumes the interrupted response more often than the backbone ($0.92$ and $0.91$ versus $0.74$) while producing fewer new responses ($0.06$ versus $0.11$).
The talking-to-other and background-speech conditions show smaller, mixed shifts in response and resume behavior.

Table~\ref{atab:fdb_v15_acoustic} follows the benchmark's speech-feature analysis for user-interruption trials classified as \textsc{Respond}.
All three systems exhibit a large post-interruption increase in speaking rate, together with higher mean intensity and lower intensity variability.
Mean pitch decreases significantly for both DiffuPlex policies but not for the autoregressive backbone, while pitch variability does not change significantly for any system.
UTMOSv2 remains statistically unchanged for the autoregressive backbone and increases slightly for both DiffuPlex policies.
These tests evaluate pre/post changes within each system; they do not test whether the magnitude of those changes differs significantly across systems.

\begin{table*}[t]
\centering
\caption{
Complete deferred-execution runtime results.
$c_{60}$ is measured on the fixed runtime subset and $c_{727}$ over the complete Full-Duplex-Bench v1.0 evaluation.
All speedups are relative to the autoregressive backbone.
For \textsc{Silero} and \textsc{Smart Turn}, controller decisions are precomputed and replayed during the timed loop, so controller inference is excluded from these wall-clock measurements.
}
\label{atab:runtime_deferred}
\scriptsize
\resizebox{\textwidth}{!}{
\begin{tabular}{lrrrrrrrr}
\toprule
System
& $c_{60}$ & $c_{727}$
& Wall (s) & Wall speedup
& E2E GPU (s) & E2E speedup
& Core LM (s) & Core speedup \\
\midrule
AR backbone
& 1.000 & 1.000
& 745.7 & 1.000$\times$
& 745.3 & 1.000$\times$
& 656.8 & 1.000$\times$ \\
\midrule
\textsc{PAD} $R=2$
& 1.531 & 1.447
& 527.8 & 1.413$\times$
& 526.9 & 1.415$\times$
& 433.0 & 1.517$\times$ \\
\textsc{PAD} $R=4$
& 2.163 & 1.911
& 405.7 & 1.838$\times$
& 404.5 & 1.842$\times$
& 308.2 & 2.132$\times$ \\
\textsc{PAD} $R=6$
& 2.525 & 2.202
& 363.0 & 2.054$\times$
& 361.9 & 2.060$\times$
& 264.7 & 2.482$\times$ \\
\textsc{PAD} $R=8$
& 2.805 & 2.383
& 337.6 & 2.209$\times$
& 336.4 & 2.216$\times$
& 239.1 & 2.747$\times$ \\
\midrule
\textsc{Silero} $R=2$
& 1.416 & 1.346
& 561.1 & 1.329$\times$
& 560.2 & 1.330$\times$
& 468.0 & 1.403$\times$ \\
\textsc{Silero} $R=4$
& 1.789 & 1.629
& 467.2 & 1.596$\times$
& 466.1 & 1.599$\times$
& 371.5 & 1.768$\times$ \\
\textsc{Silero} $R=6$
& 1.955 & 1.753
& 432.2 & 1.725$\times$
& 431.1 & 1.729$\times$
& 339.8 & 1.933$\times$ \\
\textsc{Silero} $R=8$
& 2.055 & 1.823
& 418.8 & 1.781$\times$
& 417.7 & 1.784$\times$
& 324.0 & 2.027$\times$ \\
\midrule
\textsc{Smart Turn} $R=2$
& 1.494 & 1.414
& 537.3 & 1.388$\times$
& 536.4 & 1.390$\times$
& 443.8 & 1.480$\times$ \\
\textsc{Smart Turn} $R=4$
& 1.964 & 1.772
& 434.1 & 1.718$\times$
& 433.1 & 1.721$\times$
& 338.8 & 1.938$\times$ \\
\textsc{Smart Turn} $R=6$
& 2.211 & 1.942
& 396.7 & 1.880$\times$
& 395.6 & 1.884$\times$
& 301.3 & 2.180$\times$ \\
\textsc{Smart Turn} $R=8$
& 2.355 & 2.038
& 380.6 & 1.960$\times$
& 379.4 & 1.964$\times$
& 283.6 & 2.316$\times$ \\
\midrule
DiffuPlex-\textsc{Listen}
& 1.977 & 1.910
& 512.4 & 1.455$\times$
& 511.1 & 1.458$\times$
& 408.4 & 1.608$\times$ \\
DiffuPlex-\textsc{Speak}
& 2.264 & 2.230
& 469.1 & 1.590$\times$
& 467.8 & 1.593$\times$
& 365.5 & 1.797$\times$ \\
\bottomrule
\end{tabular}
}
\end{table*}

\subsection{VoiceBench and OpenAudioBench}
\label{asubsec:additional_general}

Tables~\ref{atab:voicebench_full} and~\ref{atab:oab_full} provide the per-task results underlying the aggregate VoiceBench and OpenAudioBench scores reported in the main paper.
As defined in Appendix~\ref{asubsec:general_benchmarks}, speech-to-text ($s\!\rightarrow\!t$) directly evaluates the native assistant text stream, whereas speech-to-speech ($s\!\rightarrow\!s$) evaluates a transcription of the generated assistant audio.
DiffuPlex-\textsc{Listen} remains close to the autoregressive backbone across both benchmarks, while DiffuPlex-\textsc{Speak} shows mixed changes across modalities, with its largest aggregate reduction on OpenAudioBench speech-to-speech.
The changes are spread across multiple tasks rather than being driven by a single benchmark category.

\subsection{Deferred-Execution Runtime Results}
\label{asubsec:additional_runtime}

The detailed runtime decomposition for the autoregressive backbone, model baselines, and DiffuPlex is reported in Table~\ref{tab:runtime_results} in the main paper.
Table~\ref{atab:runtime_deferred} reports the corresponding runtime measurements for the deferred-execution baselines.
Larger deferral limits can produce higher raw speedups by forcing longer periods of assistant silence between autoregressive backbone updates; the resulting interaction trade-offs are analyzed in the main paper.
For \textsc{Silero} and \textsc{Smart Turn}, controller decisions are replayed during timed generation, so controller inference is excluded from the reported runtime.
Measured separately, the \textsc{Smart Turn} controller pipeline requires approximately 13.90ms of CPU processing per interaction second, with approximately 80\% of this cost attributable to the continuously running Silero frontend.

\section{Additional Ablations}
\label{asec:additional_ablations}

The main paper focuses on the two inference components that directly determine whether a predicted future can be used during live interaction: confidence-based retention and user-aware revision.
Here, we broaden the analysis to the training objective, future-gated adaptation, model and training scale, train-test horizon mismatch, inference thresholds, future attention, and the two signals used by the revision rule.
Unless otherwise stated, all ablations are evaluated on the 727 clips of Full-Duplex-Bench v1.0 with the same prompts, sampling configuration, and seed as the main experiments.

For the sweep tables below, $c$ denotes the number of interaction frames advanced per backbone wake.
We additionally report the pooled LM speedup recorded during the ablation sweep.
This quantity is useful for comparing configurations within the sweep, but is distinct from the controlled wall-clock, E2E GPU, and Core LM measurements reported in Table~\ref{tab:runtime_results}.
Metric definitions follow Appendix~\ref{asubsec:full_duplex_benchmarks}.

\subsection{Training Objective and Architecture}
\label{asubsec:backbone_ablation}

We first vary components that determine how the masked future predictor is learned.
The default model freezes \texttt{personaplex-rl-seamless}, applies LoRA only for lookaheads $h\geq2$, and distills the clean autoregressive distributions into the masked path.
We compare this model with several changes to the training objective and backbone.
\emph{No gate} applies the learned adapters at every lookahead, including $h=1$, removing the structural guarantee that the ordinary next-frame path uses the original frozen backbone.
\emph{CE only} replaces the teacher distributions for the masked path with hard data targets and removes the three-way text-state distillation term.
We additionally remove either the token-level text KL or the three-way PAD/EPAD/lexical KL while retaining the remaining distillation terms, and train the same DiffuPlex adaptation on the original PersonaPlex backbone rather than the post-trained \texttt{personaplex-rl-seamless} checkpoint.
Table~\ref{atab:architecture_ablation} reports the resulting efficiency and interaction metrics for both inference policies.

\begin{table*}[t]
\centering
\caption{
Training-objective and architecture ablations on Full-Duplex-Bench v1.0.
The shaded rows report the autoregressive PersonaPlex and \texttt{personaplex-rl-seamless} references evaluated under the same protocol.
The future gate leaves the $h=1$ autoregressive path unchanged; removing it (No gate) allows the learned adapters to modify this path.
}
\label{atab:architecture_ablation}
\scriptsize
\resizebox{\textwidth}{!}{
\begin{tabular}{lcc cc ccc cc ccc}
\toprule
& \multicolumn{2}{c}{Efficiency}
& \multicolumn{2}{c}{Pause}
& \multicolumn{3}{c}{Backchannel}
& \multicolumn{2}{c}{Turn}
& \multicolumn{3}{c}{Interruption} \\
\cmidrule(lr){2-3}\cmidrule(lr){4-5}\cmidrule(lr){6-8}
\cmidrule(lr){9-10}\cmidrule(lr){11-13}
Variant
& $c$ & LM$\times$
& S$\downarrow$ & C$\downarrow$
& TOR$\downarrow$ & Freq. & JSD$\downarrow$
& TOR$\uparrow$ & Lat.$\downarrow$
& TOR$\uparrow$ & Judge$\uparrow$ & Lat.$\downarrow$ \\
\midrule

\rowcolor{gray!12}
\texttt{personaplex-rl-seamless} (AR)
& 1.00 & 1.00
& .299 & .301 & .218 & .163 & .740
& 1.000 & .124 & .995 & 4.286 & .281 \\

\midrule
Default / \textsc{Listen}
& 1.91 & 1.56 & .277 & .343 & .109 & .140 & .753
& 1.000 & .173 & .990 & 4.328 & .329 \\
Default / \textsc{Speak}
& 2.23 & 1.77 & .263 & .347 & .145 & .154 & .738
& 1.000 & .156 & .995 & 4.357 & .347 \\

\midrule
No gate / \textsc{Listen}
& 1.86 & 1.67 & .292 & .301 & .091 & .176 & .729
& .983 & .202 & 1.000 & 4.340 & .296 \\
No gate / \textsc{Speak}
& 2.11 & 1.89 & .263 & .282 & .109 & .162 & .738
& .975 & .253 & 1.000 & 4.265 & .265 \\

\midrule
CE only / \textsc{Listen}
& 1.76 & 1.44 & .277 & .352 & .091 & .136 & .758
& 1.000 & .182 & .985 & 4.365 & .512 \\
CE only / \textsc{Speak}
& 2.07 & 1.65 & .270 & .324 & .091 & .180 & .729
& 1.000 & .196 & .980 & 4.332 & .327 \\

\midrule
w/o token-level text KL / \textsc{Listen}
& 1.91 & 1.56 & .292 & .347 & .109 & .161 & .738
& 1.000 & .139 & .985 & 4.259 & .388 \\
w/o token-level text KL / \textsc{Speak}
& 2.24 & 1.78 & .270 & .343 & .054 & .219 & .720
& 1.000 & .121 & .990 & 4.202 & .293 \\

\midrule
w/o three-way text-state KL / \textsc{Listen}
& 1.89 & 1.54 & .277 & .389 & .182 & .165 & .737
& 1.000 & .161 & .980 & 4.250 & .280 \\
w/o three-way text-state KL / \textsc{Speak}
& 2.26 & 1.79 & .255 & .324 & .054 & .142 & .747
& 1.000 & .178 & .990 & 4.318 & .369 \\

\specialrule{0.8pt}{0.5ex}{0.15ex}
\specialrule{0.4pt}{0pt}{0.5ex}

\rowcolor{gray!12}
PersonaPlex (AR)
& 1.00 & --
& .307 & .343 & .200 & .058 & .824
& .941 & .285 & .930 & 4.328 & .218 \\

\midrule
PersonaPlex backbone / \textsc{Listen}
& 1.76 & 1.44 & .577 & .630 & .400 & .104 & .743
& .992 & .170 & .905 & 4.243 & 1.277 \\
PersonaPlex backbone / \textsc{Speak}
& 2.02 & 1.61 & .606 & .620 & .327 & .112 & .744
& .966 & .154 & .920 & 4.071 & 1.347 \\

\bottomrule
\end{tabular}}
\end{table*}

The future gate primarily protects the original next-frame path.
Without it, the model still reduces backbone wakes, but turn-taking TOR changes from $1.000$ to $.983$ for \textsc{Listen} and $.975$ for \textsc{Speak}, with corresponding turn latencies of $.202$s and $.253$s.
The gate therefore provides a structural separation between the unchanged autoregressive prediction at $h=1$ and the learned masked-future behavior at deeper lookaheads.
Replacing teacher distillation with hard-target cross-entropy also shortens the usable future horizon: $c$ decreases from $1.91$ to $1.76$ for \textsc{Listen} and from $2.23$ to $2.07$ for \textsc{Speak}.
The interaction changes are not uniform across metrics, but the reduced advancement and larger \textsc{Listen} interruption latency ($.512$s versus $.329$s) are consistent with a benefit from matching the backbone distribution rather than relying only on hard future targets.
Removing either text KL term produces mixed changes across the interaction metrics.
Both ablations reduce the interruption Judge score for both policies, with the largest drop from $4.357$ to $4.202$ when removing the token-level text KL for \textsc{Speak}.
The final model retains both text KL terms.

The PersonaPlex-backbone variant separates the ability to advance multiple frames from the ability to preserve the behavior of a particular backbone.
DiffuPlex still reaches $c=1.76$ for \textsc{Listen} and $2.02$ for \textsc{Speak} on PersonaPlex, showing that wake reduction transfers to this backbone.
However, pause and interruption behavior shifts substantially; for example, interruption TOR becomes $.905$ and $.920$, with latencies of $1.277$s and $1.347$s.
Thus, wake reduction transfers to PersonaPlex, but preservation of the underlying interaction behavior does not transfer automatically.

\subsection{Model and Training Scale}
\label{asubsec:training_scale}

\begin{table*}[t]
\centering
\caption{
Model and training-scale ablations.
Matched-$W$ rows use the same horizon during training and inference.
Increasing adapter capacity or training duration does not produce a monotonic change in the usable future horizon.
}
\label{atab:scale_ablation}
\scriptsize
\resizebox{\textwidth}{!}{
\begin{tabular}{lcc cc ccc cc ccc}
\toprule
& \multicolumn{2}{c}{Efficiency}
& \multicolumn{2}{c}{Pause}
& \multicolumn{3}{c}{Backchannel}
& \multicolumn{2}{c}{Turn}
& \multicolumn{3}{c}{Interruption} \\
\cmidrule(lr){2-3}\cmidrule(lr){4-5}\cmidrule(lr){6-8}
\cmidrule(lr){9-10}\cmidrule(lr){11-13}
Variant
& $c$ & LM$\times$
& S$\downarrow$ & C$\downarrow$
& TOR$\downarrow$ & Freq. & JSD$\downarrow$
& TOR$\uparrow$ & Lat.$\downarrow$
& TOR$\uparrow$ & Judge$\uparrow$ & Lat.$\downarrow$ \\
\midrule
Default / \textsc{Listen}
& 1.91 & 1.56 & .277 & .343 & .109 & .140 & .753 & 1.000 & .173 & .990 & 4.328 & .329 \\
Default / \textsc{Speak}
& 2.23 & 1.77 & .263 & .347 & .145 & .154 & .738 & 1.000 & .156 & .995 & 4.357 & .347 \\
\midrule
Train $W=8$ / \textsc{Listen}
& 1.86 & 1.58 & .248 & .366 & .109 & .146 & .751 & 1.000 & .242 & .985 & 4.371 & .344 \\
Train $W=8$ / \textsc{Speak}
& 2.21 & 1.82 & .255 & .356 & .291 & .149 & .740 & .992 & .202 & .990 & 4.308 & .315 \\
Train $W=32$ / \textsc{Listen}
& 1.90 & 1.43 & .277 & .356 & .200 & .146 & .749 & 1.000 & .159 & .980 & 4.327 & .342 \\
Train $W=32$ / \textsc{Speak}
& 2.23 & 1.63 & .292 & .356 & .182 & .144 & .744 & 1.000 & .112 & .985 & 4.320 & .361 \\
\midrule
LoRA rank 32 / \textsc{Listen}
& 1.92 & 1.58 & .226 & .241 & .145 & .127 & .760 & .958 & .206 & .990 & 4.328 & .427 \\
LoRA rank 32 / \textsc{Speak}
& 2.25 & 1.78 & .219 & .231 & .109 & .126 & .762 & .966 & .231 & .990 & 4.338 & .370 \\
LoRA rank 512 / \textsc{Listen}
& 1.88 & 1.44 & .248 & .259 & .036 & .117 & .765 & .992 & .132 & .995 & 4.467 & .318 \\
LoRA rank 512 / \textsc{Speak}
& 2.26 & 1.67 & .248 & .306 & .164 & .124 & .765 & 1.000 & .160 & .985 & 4.244 & .333 \\
\midrule
263 steps ($\sim$351h) / \textsc{Listen}
& 1.90 & 1.55 & .277 & .375 & .145 & .150 & .744 & 1.000 & .153 & .980 & 4.357 & .313 \\
263 steps ($\sim$351h) / \textsc{Speak}
& 2.25 & 1.79 & .285 & .347 & .109 & .159 & .744 & .992 & .271 & .990 & 4.298 & .439 \\
1,052 steps ($\sim$1,404h) / \textsc{Listen}
& 1.88 & 1.54 & .292 & .347 & .073 & .160 & .736 & 1.000 & .103 & .990 & 4.278 & .326 \\
1,052 steps ($\sim$1,404h) / \textsc{Speak}
& 2.25 & 1.78 & .270 & .315 & .091 & .157 & .736 & 1.000 & .119 & .985 & 4.289 & .323 \\
\bottomrule
\end{tabular}}
\end{table*}

We next vary three training-scale axes: the maximum future block used during training, LoRA rank, and the number of optimization steps.
The default uses $W=16$, rank 128, and 526 optimization steps.
For the training-horizon variants, we use the same $W$ at inference as during training, yielding matched $8\!\rightarrow\!8$ and $32\!\rightarrow\!32$ settings; train-inference horizon mismatch is examined separately in Appendix~\ref{asubsec:inference_horizon}.
To examine the effect of training exposure, the 263-step and 1,052-step variants use schedules whose learning-rate decay is defined to terminate at their respective final steps, rather than taking intermediate checkpoints from a longer run.
Table~\ref{atab:scale_ablation} reports the resulting efficiency and interaction metrics.
Across the rank and training-duration sweeps, the number of frames advanced per wake remains relatively stable, while individual interaction metrics vary in both directions.
Increasing LoRA rank from 128 to 512 or doubling training from 526 to 1,052 steps does not yield a consistent improvement, whereas rank 32 retains a similar $c$ but shifts turn-taking behavior.

The matched-horizon comparison shows a different trade-off.
Training and inference with $W=8$ yields similar advancement but larger shifts on several interaction metrics, including \textsc{Speak} backchannel TOR and turn latency.
Increasing both training and inference horizons to $W=32$ likewise provides little additional advancement, while the larger proposed block increases per-wake LM cost and reduces the pooled LM speedup.
The default $W=16$ setting lies between these two trade-offs.

\begin{table*}[t]
\centering
\caption{
Train-inference horizon mismatch on Full-Duplex-Bench v1.0.
The retained advancement $c$ changes little across most combinations even when the maximum proposed horizon changes substantially.
}
\label{atab:horizon_ablation}
\scriptsize
\resizebox{\textwidth}{!}{
\begin{tabular}{lcc cc ccc cc ccc}
\toprule
& \multicolumn{2}{c}{Efficiency}
& \multicolumn{2}{c}{Pause}
& \multicolumn{3}{c}{Backchannel}
& \multicolumn{2}{c}{Turn}
& \multicolumn{3}{c}{Interruption} \\
\cmidrule(lr){2-3}\cmidrule(lr){4-5}\cmidrule(lr){6-8}
\cmidrule(lr){9-10}\cmidrule(lr){11-13}
Train $\rightarrow$ infer
& $c$ & LM$\times$
& S$\downarrow$ & C$\downarrow$
& TOR$\downarrow$ & Freq. & JSD$\downarrow$
& TOR$\uparrow$ & Lat.$\downarrow$
& TOR$\uparrow$ & Judge$\uparrow$ & Lat.$\downarrow$ \\
\midrule
$8\rightarrow16$ / \textsc{Listen}
& 1.89 & 1.55 & .307 & .347 & .127 & .162 & .736 & 1.000 & .200 & 1.000 & 4.300 & .372 \\
$8\rightarrow16$ / \textsc{Speak}
& 2.25 & 1.79 & .292 & .333 & .091 & .146 & .745 & 1.000 & .182 & .995 & 4.226 & .429 \\
$8\rightarrow32$ / \textsc{Listen}
& 1.91 & 1.44 & .299 & .389 & .073 & .146 & .745 & 1.000 & .157 & .995 & 4.332 & .400 \\
$8\rightarrow32$ / \textsc{Speak}
& 2.27 & 1.66 & .255 & .384 & .218 & .153 & .741 & .992 & .160 & .985 & 4.284 & .349 \\
\midrule
$16\rightarrow8$ / \textsc{Listen}
& 1.86 & 1.58 & .365 & .394 & .145 & .155 & .742 & 1.000 & .162 & .990 & 4.293 & .335 \\
$16\rightarrow8$ / \textsc{Speak}
& 2.20 & 1.81 & .350 & .380 & .091 & .158 & .741 & 1.000 & .155 & .990 & 4.328 & .318 \\
$16\rightarrow32$ / \textsc{Listen}
& 1.88 & 1.42 & .285 & .356 & .127 & .142 & .743 & .992 & .170 & 1.000 & 4.435 & .523 \\
$16\rightarrow32$ / \textsc{Speak}
& 2.24 & 1.64 & .299 & .347 & .145 & .144 & .745 & .992 & .146 & .980 & 4.398 & .316 \\
\midrule
$32\rightarrow8$ / \textsc{Listen}
& 1.87 & 1.59 & .307 & .347 & .145 & .144 & .749 & 1.000 & .152 & .990 & 4.247 & .351 \\
$32\rightarrow8$ / \textsc{Speak}
& 2.19 & 1.81 & .314 & .329 & .127 & .148 & .746 & 1.000 & .157 & .985 & 4.340 & .342 \\
$32\rightarrow16$ / \textsc{Listen}
& 1.87 & 1.54 & .314 & .361 & .182 & .142 & .745 & 1.000 & .149 & .995 & 4.337 & .344 \\
$32\rightarrow16$ / \textsc{Speak}
& 2.23 & 1.77 & .285 & .370 & .164 & .140 & .750 & 1.000 & .112 & .980 & 4.250 & .326 \\
\bottomrule
\end{tabular}}
\end{table*}

\subsection{Inference Horizon}
\label{asubsec:inference_horizon}

The training horizon need not equal the maximum horizon used at inference, so we evaluate all cross-horizon combinations among $W\in\{8,16,32\}$.
Table~\ref{atab:horizon_ablation} reports the resulting efficiency and interaction metrics.
Across the mismatch sweep, \textsc{Listen} remains near $c=1.86$--$1.91$ and \textsc{Speak} near $c=2.19$--$2.27$ despite changing the inference cap from 8 to 32.
The maximum proposed horizon is therefore not the same as the amount of future that is actually used: increasing $W$ exposes more candidate future positions, while the confidence and revision rules continue to determine how much of that future is ultimately consumed.

The per-wake LM cost of proposing the block nevertheless grows with $W$.
For the default $W=16$ training, increasing the inference horizon to $W=32$ leaves $c$ nearly unchanged but reduces the pooled LM speedup from $1.56\times$ to $1.42\times$ for \textsc{Listen} and from $1.77\times$ to $1.64\times$ for \textsc{Speak}.
Conversely, $W=8$ is cheaper but does not consistently improve interaction timing.
The default $W=16$ provides an intermediate horizon without proposing substantially more candidate positions than the retention policy typically uses.

\begin{table*}[t]
\centering
\caption{
Confidence-threshold ablation.
Lower thresholds retain longer predicted futures, whereas stricter thresholds wake the backbone more frequently.
The no-confidence intervention disables only the confidence threshold for \textsc{Listen}
(retaining the PAD and $m=6$ constraints), whereas \textsc{Speak} retains the full proposed future when confidence filtering is disabled, since its original $m=2$ minimum is non-binding.
}
\label{atab:confidence_ablation}
\scriptsize
\resizebox{\textwidth}{!}{
\begin{tabular}{lcc cc ccc cc ccc}
\toprule
& \multicolumn{2}{c}{Efficiency}
& \multicolumn{2}{c}{Pause}
& \multicolumn{3}{c}{Backchannel}
& \multicolumn{2}{c}{Turn}
& \multicolumn{3}{c}{Interruption} \\
\cmidrule(lr){2-3}\cmidrule(lr){4-5}\cmidrule(lr){6-8}
\cmidrule(lr){9-10}\cmidrule(lr){11-13}
Setting
& $c$ & LM$\times$
& S$\downarrow$ & C$\downarrow$
& TOR$\downarrow$ & Freq. & JSD$\downarrow$
& TOR$\uparrow$ & Lat.$\downarrow$
& TOR$\uparrow$ & Judge$\uparrow$ & Lat.$\downarrow$ \\
\midrule
\textsc{Listen}, no confidence filtering
& 2.11 & 1.64 
& .197 & .343
& .164 & .191 & .724
& .966 & .170
& .980 & 4.398 & .314 \\
\textsc{Listen}, $\tau_{\mathrm{conf}}=0.70$
& 1.99 & 1.62 & .270 & .319 & .182 & .140 & .750 & 1.000 & .191 & .985 & 4.345 & .342 \\
\textsc{Listen}, $\tau_{\mathrm{conf}}=0.80$
& 1.91 & 1.56 & .277 & .343 & .109 & .140 & .753 & 1.000 & .173 & .990 & 4.328 & .329 \\
\textsc{Listen}, $\tau_{\mathrm{conf}}=0.90$
& 1.80 & 1.48 & .263 & .366 & .145 & .148 & .743 & 1.000 & .089 & .990 & 4.384 & .485 \\
\midrule
\textsc{Speak}, no confidence filtering
& 2.80 & 2.07 & .321 & .347 & .236 & .191 & .729 & 1.000 & .209 & 1.000 & 3.550 & .281 \\
\textsc{Speak}, $\tau_{\mathrm{conf}}=0.70$
& 2.39 & 1.87 & .241 & .315 & .145 & .162 & .741 & .992 & .218 & .995 & 4.266 & .385 \\
\textsc{Speak}, $\tau_{\mathrm{conf}}=0.80$
& 2.23 & 1.77 & .263 & .347 & .145 & .154 & .738 & 1.000 & .156 & .995 & 4.357 & .347 \\
\textsc{Speak}, $\tau_{\mathrm{conf}}=0.90$
& 2.07 & 1.67 & .285 & .366 & .109 & .138 & .758 & 1.000 & .166 & .985 & 4.330 & .378 \\
\bottomrule
\end{tabular}}
\end{table*}

\subsection{Confidence and Retention}
\label{asubsec:confidence_retention}

The confidence threshold $\tau_{\mathrm{conf}}$ controls how aggressively a predicted future is retained before subsequent user evidence is observed.
The main experiments use $\tau_{\mathrm{conf}}=0.80$; we vary it from $0.70$ to $0.90$ and additionally evaluate policy-specific no-confidence variants.
For \textsc{Listen}, the confidence threshold is disabled while the PAD restriction and $m=6$ minimum are retained. For \textsc{Speak}, the confidence threshold is disabled while the original $m=2$ minimum is kept; because all assistant states are permitted, this minimum is non-binding and the full proposed future is retained.
Table~\ref{atab:confidence_ablation} reports the resulting efficiency and interaction metrics.
As expected from the retention rule, lowering $\tau_{\mathrm{conf}}$ increases advancement per wake, while raising it decreases advancement.
This provides a direct inference-time control over how aggressively DiffuPlex uses its predicted future.

The no-confidence interventions increase advancement for both policies, with different effects under the two retention policies.
For \textsc{Listen}, where future retention remains restricted to PAD states, $c$ increases from $1.91$ to $2.11$ with comparatively modest changes in the interaction metrics.
For \textsc{Speak}, the no-confidence intervention increases $c$ from $2.23$ to $2.80$, but the interruption Judge score decreases from $4.357$ to $3.550$ and backchannel TOR increases from $.145$ to $.236$.
These results show that more aggressive future retention can increase advancement while altering interaction behavior.

\begin{table*}[t]
\centering
\caption{
Revision-rule ablation.
A larger $\tau_{\mathrm{rev}}$ requires stronger evidence before revising, while the no-revision variant always continues through the retained draft unless it is exhausted.
The uncalibrated User-only rows remove the assistant-side contribution while retaining the default revision threshold, and should not be interpreted as the calibrated User-only ablation in Appendix~\ref{asubsec:revision_decomposition}.
}
\label{atab:revision_ablation}
\scriptsize
\resizebox{\textwidth}{!}{
\begin{tabular}{lcc cc ccc cc ccc}
\toprule
& \multicolumn{2}{c}{Efficiency}
& \multicolumn{2}{c}{Pause}
& \multicolumn{3}{c}{Backchannel}
& \multicolumn{2}{c}{Turn}
& \multicolumn{3}{c}{Interruption} \\
\cmidrule(lr){2-3}\cmidrule(lr){4-5}\cmidrule(lr){6-8}
\cmidrule(lr){9-10}\cmidrule(lr){11-13}
Setting
& $c$ & LM$\times$
& S$\downarrow$ & C$\downarrow$
& TOR$\downarrow$ & Freq. & JSD$\downarrow$
& TOR$\uparrow$ & Lat.$\downarrow$
& TOR$\uparrow$ & Judge$\uparrow$ & Lat.$\downarrow$ \\
\midrule
\textsc{Listen}, $\tau_{\mathrm{rev}}=-1.0$
& 1.83 & 1.50 & .321 & .384 & .164 & .161 & .735 & 1.000 & .136 & 1.000 & 4.350 & .437 \\
\textsc{Listen}, $\tau_{\mathrm{rev}}=-0.5$
& 1.91 & 1.56 & .277 & .343 & .109 & .140 & .753 & 1.000 & .173 & .990 & 4.328 & .329 \\
\textsc{Listen}, $\tau_{\mathrm{rev}}=0$
& 1.95 & 1.60 & .321 & .352 & .036 & .157 & .739 & .992 & .125 & .995 & 4.236 & .446 \\
\textsc{Listen}, no revision
& 2.49 & 2.07 & .146 & .176 & .109 & .078 & .796 & .966 & .440 & .995 & 4.372 & .744 \\
\textsc{Listen}, User-only (uncalibrated)
& 1.00 & .79 & .431 & .417 & .145 & .145 & .743 & 1.000 & .089 & 1.000 & 4.335 & .307 \\
\midrule
\textsc{Speak}, $\tau_{\mathrm{rev}}=-1.0$
& 2.15 & 1.70 & .314 & .370 & .145 & .152 & .740 & 1.000 & .152 & 1.000 & 4.350 & .419 \\
\textsc{Speak}, $\tau_{\mathrm{rev}}=-0.5$
& 2.23 & 1.77 & .263 & .347 & .145 & .154 & .738 & 1.000 & .156 & .995 & 4.357 & .347 \\
\textsc{Speak}, $\tau_{\mathrm{rev}}=0$
& 2.32 & 1.84 & .241 & .329 & .127 & .144 & .741 & .983 & .185 & .980 & 4.306 & .315 \\
\textsc{Speak}, no revision
& 3.08 & 2.50 & .131 & .157 & .091 & .080 & .797 & .950 & .480 & .995 & 4.251 & .515 \\
\textsc{Speak}, User-only (uncalibrated)
& 1.00 & .78 & .431 & .403 & .109 & .148 & .739 & 1.000 & .091 & 1.000 & 4.345 & .265 \\
\bottomrule
\end{tabular}}
\end{table*}
\begin{table*}[t]
\centering
\caption{
Causal and bidirectional future attention on Full-Duplex-Bench v1.0.
The two attention variants advance a similar number of frames per wake.
Bidirectional attention changes individual interaction metrics in both directions rather than providing a consistent advantage.
}
\label{atab:future_attention}
\scriptsize
\resizebox{\textwidth}{!}{
\begin{tabular}{l c cc ccc cc ccc}
\toprule
& \multicolumn{1}{c}{Efficiency}
& \multicolumn{2}{c}{Pause}
& \multicolumn{3}{c}{Backchannel}
& \multicolumn{2}{c}{Turn}
& \multicolumn{3}{c}{Interruption} \\
\cmidrule(lr){2-2}\cmidrule(lr){3-4}\cmidrule(lr){5-7}
\cmidrule(lr){8-9}\cmidrule(lr){10-12}
Setting
& $c$
& S$\downarrow$ & C$\downarrow$
& TOR$\downarrow$ & Freq. & JSD$\downarrow$
& TOR$\uparrow$ & Lat.$\downarrow$
& TOR$\uparrow$ & Judge$\uparrow$ & Lat.$\downarrow$ \\
\midrule
Causal / \textsc{Listen}
& 1.910 & .277 & .343 & .109 & .140 & .753 & 1.000 & .173 & .990 & 4.328 & .329 \\
Bidirectional / \textsc{Listen}
& 1.884 & .299 & .343 & .109 & .148 & .752 & 1.000 & .156 & .990 & 4.374 & .432 \\
\midrule
Causal / \textsc{Speak}
& 2.230 & .263 & .347 & .145 & .154 & .738 & 1.000 & .156 & .995 & 4.357 & .347 \\
Bidirectional / \textsc{Speak}
& 2.223 & .307 & .324 & .127 & .143 & .748 & .983 & .169 & .985 & 4.284 & .327 \\
\bottomrule
\end{tabular}}
\end{table*}

\subsection{Revision Rule}
\label{asubsec:revision}

Confidence determines which future frames are initially retained, whereas revision determines whether an already retained draft remains usable after new user speech arrives.
We vary the revision threshold around the default $\tau_{\mathrm{rev}}=-0.5$ and additionally disable revision entirely; Table~\ref{atab:revision_ablation} reports the resulting efficiency and interaction metrics.
Lowering $\tau_{\mathrm{rev}}$ makes revision easier to trigger, whereas raising it requires stronger evidence and permits greater advancement.
This trade-off is clearest without revision: for \textsc{Speak}, $c$ increases from $2.23$ to $3.08$, but turn-taking TOR decreases from $1.000$ to $.950$ and turn latency increases from $.156$s to $.480$s; for \textsc{Listen}, turn latency similarly increases from $.173$s to $.440$s, while backchannel frequency decreases substantially for both policies.
Thus, retaining a predicted future alone is insufficient; invalidating its unplayed remainder is important for maintaining responsiveness while advancing multiple frames per wake.

For completeness, Table~\ref{atab:revision_ablation} also includes an uncalibrated User-only variant that removes the assistant-side contribution from the revision score, reducing the decision rule to $-\log q_h(x_h)\geq\tau_{\mathrm{rev}}$, where $q_h(x_h)$ is the probability assigned by the stored user prediction to the user semantic token that actually arrives.
Keeping the default $\tau_{\mathrm{rev}}=-0.5$ after this removal changes the score scale: because $-\log q_h(x_h)\geq0$, every eligible check triggers revision, reducing $c$ to $1.00$ for both policies and making inference slower than the autoregressive baseline.
We therefore use these rows only to illustrate threshold-scale mismatch; the controlled User-only comparison in Appendix~\ref{asubsec:revision_decomposition} recalibrates its threshold separately.

\subsection{Causal and Bidirectional Future Attention}
\label{asubsec:future_attention}

The default masked predictor retains causal temporal attention within each future block: a position may use the realized context before the block and masked positions up to its current lookahead, but not later masked positions.
We compare this design with a bidirectional variant in which masked positions within the same block may attend to one another in both directions, while leaving the realized context and stream-masking schedule unchanged.
Because a bidirectional representation depends on the full set of positions present in the block, we train this variant with a fixed future-block length of 16 and evaluate it with $W=16$; all other inference settings, including $\tau_{\mathrm{conf}}=0.80$, $\tau_{\mathrm{rev}}=-0.5$, the \textsc{Listen} and \textsc{Speak} retention policies, prompts, sampling configuration, and seed, remain unchanged.
Table~\ref{atab:future_attention} reports the resulting efficiency and interaction metrics.

The two attention patterns yield almost identical advancement: $c$ changes from $1.910$ to $1.884$ for \textsc{Listen} and from $2.230$ to $2.223$ for \textsc{Speak}.
Their interaction effects are mixed: for \textsc{Listen}, bidirectional attention slightly reduces turn latency and increases interruption Judge, but interruption latency increases from $.329$s to $.432$s; for \textsc{Speak}, some pause and backchannel metrics improve, while turn-taking TOR decreases from $1.000$ to $.983$ and interruption TOR from $.995$ to $.985$.
Bidirectional future attention therefore does not provide a consistent improvement at comparable advancement.
Causal attention remains the default because it matches the causal structure of the autoregressive backbone, naturally supports variable block lengths during training, and prevents an earlier future representation from depending on how many later masked positions happen to be included in the same block.

\subsection{Revision Signal Decomposition}
\label{asubsec:revision_decomposition}

The full revision score combines two quantities with different roles.
The User term,
\begin{equation}
S_h^{\mathrm{user}}
=
-\log q_h(x_h),
\end{equation}
measures how surprising the newly observed user semantic token is under the continuation predicted at the previous wake, while the Assistant term,
\begin{equation}
S_h^{\mathrm{asst}}
=
\log P_h(\mathcal{A}_h(y_h))
-
\log P_h(y_h),
\end{equation}
measures how weakly the assistant preferred the drafted text decision relative to its valid alternatives.
The default revision rule adds these two terms.
Because removing either term changes the numerical scale of the score, using the default $\tau_{\mathrm{rev}}=-0.5$ would not provide a meaningful decomposition.
For these experiments only, we calibrate the User-only and Assistant-only thresholds on a fixed subset of Full-Duplex-Bench v1.5 to approximately match the average frames per wake $c$ of the Full rule, and then evaluate the selected settings on Full-Duplex-Bench v1.0.
The final thresholds are $5.5$ and $4.5$ for User-only \textsc{Listen} and \textsc{Speak}, respectively, and $-3.5$ and $-3.0$ for Assistant-only \textsc{Listen} and \textsc{Speak}.
Full-Duplex-Bench v1.0 is not used for threshold selection, and Full-Duplex-Bench v1.5 is not used to select any other inference hyperparameters.
Table~\ref{atab:revision_decomposition} reports the final comparison.

\begin{table*}[t]
\centering
\caption{
Revision-signal decomposition on the complete Full-Duplex-Bench v1.0 test set.
Full uses both incoming user surprisal and assistant uncertainty.
User-only retains only $-\log q_h(x_h)$, while Assistant-only retains only the assistant-side alternative-odds term.
The realized test-set advancement $c$ is reported because the held-out calibration does not remain exactly matched on the final evaluation set.
}
\label{atab:revision_decomposition}
\scriptsize
\resizebox{\textwidth}{!}{
\begin{tabular}{lcc cc ccc cc ccc}
\toprule
& \multicolumn{2}{c}{Revision}
& \multicolumn{2}{c}{Pause}
& \multicolumn{3}{c}{Backchannel}
& \multicolumn{2}{c}{Turn}
& \multicolumn{3}{c}{Interruption} \\
\cmidrule(lr){2-3}\cmidrule(lr){4-5}\cmidrule(lr){6-8}
\cmidrule(lr){9-10}\cmidrule(lr){11-13}
Setting
& $\tau_{\rm rev}$ & $c$
& S$\downarrow$ & C$\downarrow$
& TOR$\downarrow$ & Freq. & JSD$\downarrow$
& TOR$\uparrow$ & Lat.$\downarrow$
& TOR$\uparrow$ & Judge$\uparrow$ & Lat.$\downarrow$ \\
\midrule
Full / \textsc{Listen}
& -0.5 & 1.910 & .277 & .343 & .109 & .140 & .753 & 1.000 & .173 & .990 & 4.328 & .329 \\
User-only / \textsc{Listen}
& 5.5 & 1.920 & .182 & .264 & .109 & .117 & .769 & .992 & .251 & .990 & 4.354 & .315 \\
Assistant-only / \textsc{Listen}
& -3.5 & 2.122 & .270 & .296 & .127 & .162 & .729 & 1.000 & .125 & .980 & 4.270 & .457 \\
\midrule
Full / \textsc{Speak}
& -0.5 & 2.230 & .263 & .347 & .145 & .154 & .738 & 1.000 & .156 & .995 & 4.357 & .347 \\
User-only / \textsc{Speak}
& 4.5 & 2.096 & .197 & .255 & .018 & .106 & .781 & .983 & .301 & .985 & 4.355 & .312 \\
Assistant-only / \textsc{Speak}
& -3.0 & 2.586 & .212 & .255 & .127 & .129 & .752 & 1.000 & .192 & .985 & 4.365 & .484 \\
\bottomrule
\end{tabular}}
\end{table*}

As shown in Table~\ref{atab:revision_ablation}, using the uncalibrated User-only rule with the default threshold collapses advancement to $c=1.00$ for both policies.
Calibration removes this large scale mismatch, although the realized advancement is not identical on the final test set.
User-only reaches $c=1.920$ for \textsc{Listen} and $2.096$ for \textsc{Speak}, while Assistant-only reaches $2.122$ and $2.586$, compared with $1.910$ and $2.230$ for the Full rule.
We therefore interpret Table~\ref{atab:revision_decomposition} jointly with each variant's realized $c$, rather than as a strictly matched-efficiency comparison.

The single-signal rules also produce distinct interaction profiles.
For \textsc{Listen}, User-only changes turn latency from $.173$s to $.251$s and backchannel frequency from $.140$ to $.117$ despite similar advancement.
For \textsc{Speak}, User-only is more conservative overall and changes backchannel TOR from $.145$ to $.018$ while backchannel frequency simultaneously decreases from $.154$ to $.106$, so the lower TOR should not be interpreted in isolation as uniformly better interaction.
Assistant-only permits greater advancement in both policies, accompanied by longer interruption latency: $.329$s to $.457$s for \textsc{Listen} and $.347$s to $.484$s for \textsc{Speak}, with interruption TOR changing from $.990$ to $.980$ and from $.995$ to $.985$, respectively.
These results suggest that the two terms capture different revision cues: user surprisal reacts to newly observed evidence that conflicts with the stored future, whereas assistant uncertainty reflects how strongly the model committed to that future when it was drafted.
The Full rule used in the main experiments combines both sources of evidence rather than relying on either signal in isolation.

\begin{figure}[t]
\centering

\begin{subfigure}[t]{0.49\linewidth}
    \centering
    \includegraphics[width=0.90\linewidth]{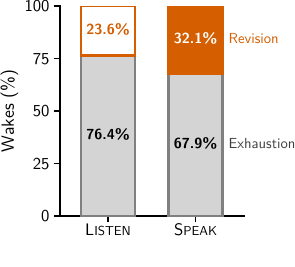}
    \caption{Wake termination}
\end{subfigure}
\hfill
\begin{subfigure}[t]{0.49\linewidth}
    \centering
    \includegraphics[width=0.90\linewidth]{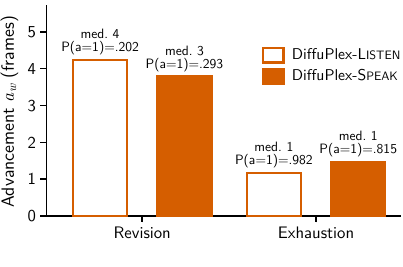}
    \caption{Advancement by termination reason}
\end{subfigure}

\vspace{1mm}

\begin{subfigure}[t]{0.49\linewidth}
    \centering
    \includegraphics[width=0.90\linewidth]{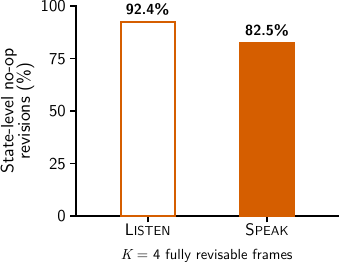}
    \caption{State-level no-op revisions}
\end{subfigure}
\hfill
\begin{subfigure}[t]{0.49\linewidth}
    \centering
    \includegraphics[width=0.90\linewidth]{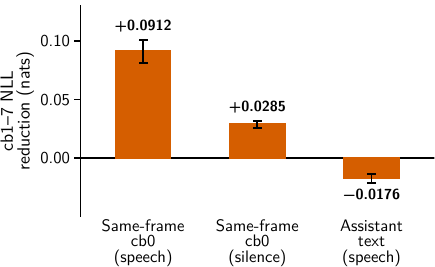}
    \caption{Future-speech prediction error}
\end{subfigure}

\caption{
Rolling-draft diagnostics.
Revision-triggered wakes typically occur after several frames have already been consumed, whereas exhaustion is dominated by one-frame advances.
State-level no-op revisions compare the next $K=4$ fully revisable frames.
Positive NLL reduction indicates improved future acoustic prediction.
}
\label{afig:rolling_diagnostics}
\end{figure}

\section{Rolling-Draft Analysis}
\label{asec:rolling_draft}

The average number of interaction frames advanced per backbone wake, $c$, summarizes the resulting reduction in sequential backbone calls but does not reveal why individual wake intervals are short or long.
We therefore analyze the relationship between the future retained at each wake, the frames actually consumed before the next wake, and the unplayed future that may be discarded by revision.
Figure~\ref{afig:rolling_diagnostics} summarizes these rolling-draft dynamics.

\subsection{Wake Dynamics and Revision}
\label{asubsec:wake_dynamics}

Let $r_w$ denote the retained prefix length selected at backbone wake $w$ after applying the confidence and policy constraints in Section~\ref{subsec:rolling_draft_generation}, and let $a_w$ denote the number of interaction frames actually realized before the temporal backbone next wakes.
Thus,
\[
1 \leq a_w \leq r_w \leq W,
\]
where $W$ is the proposed future horizon.
If the retained prefix is exhausted, $a_w=r_w$; if newly observed user input triggers revision first, $a_w<r_w$.
For the wake-level analyses below, we exclude wakes truncated only because the evaluation clip ends, while the main frames-per-wake metric $c$ continues to use the complete interaction.

Figure~\ref{afig:rolling_diagnostics}(a,b) separates wakes by whether they end through draft exhaustion or revision.
Revision accounts for 23.6\% of \textsc{Listen} wake terminations and 32.1\% of \textsc{Speak} terminations.
As summarized in the main results, revision-triggered wakes generally occur only after several retained frames have already been consumed, whereas exhaustion is dominated by short retained prefixes.
This shows that short wake intervals do not primarily arise from repeatedly retaining long futures and immediately discarding them; they more often arise because the retention rule conservatively selects only a short prefix in the first place.

Revision affects only interaction that has not yet been played.
Assistant frames already presented to the user and user frames already received remain part of the realized interaction.
Because of the backbone's native stream delays, if the observed user semantic token at lookahead $h$ triggers revision, the assistant text state associated with that boundary has already been selected.
We therefore begin before-after comparisons at $h+1$, the first assistant text state that is fully revisable.
We define a state-level no-op revision as one for which the PAD/EPAD/lexical sequence over the next $K=4$ fully revisable assistant frames (320ms) is unchanged after regeneration.
Figure~\ref{afig:rolling_diagnostics}(c) reports the resulting no-op rates: 92.4\% for \textsc{Listen} and 82.5\% for \textsc{Speak}.
Most revision triggers therefore do not alter the assistant's immediate interaction state, but a nontrivial subset does.
Together with the no-revision ablation in Appendix~\ref{asubsec:revision}, this suggests that revision serves as a safeguard against inconsistent drafts, even though many triggers leave the immediate assistant state unchanged.

\subsection{Future-Speech Mechanism}
\label{asubsec:future_speech}

DiffuPlex-\textsc{Speak} may consume lexical speech predicted several frames into the future, so the quality of these frames depends on how well the masked path predicts the acoustic codebooks.
The key difference from ordinary autoregressive generation comes from the native delay between the semantic audio codebook cb0 and the acoustic codebooks cb1--7.
For a physical audio frame, the autoregressive model has already observed that frame's cb0 when it forms the temporal representation used to predict cb1--7.
At a masked future position, however, the same-frame cb0 has not yet been generated and is replaced by MASK, so cb1--7 must be predicted from a temporal representation that lacks this semantic audio information.
We test whether this missing input contributes to the degradation by selectively giving it back.
As shown in Figure~\ref{afig:rolling_diagnostics}(d), restoring the same-frame cb0 during assistant speech and using it to sequentially generate the remaining acoustic codebooks cb1--7 reduces their NLL by $0.0912$, whereas the corresponding improvement during silence is only $0.0285$.
As a control, restoring assistant text instead of cb0 does not help and changes NLL by $\Delta=-0.0176$.
The effect is therefore not simply caused by masking any assistant-side input: it is specifically associated with losing the same-frame semantic audio signal that normally helps the backbone predict the acoustic codebooks.

\begin{figure}[t]
\centering
\includegraphics[width=\linewidth]{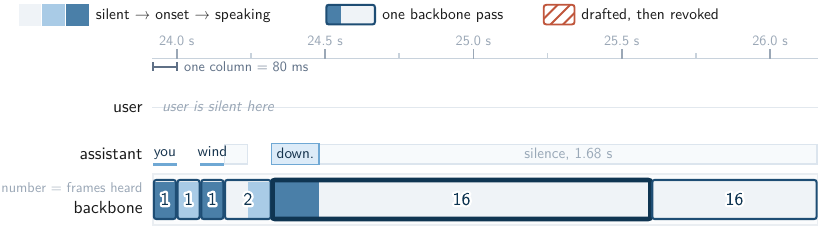}
\caption{
Long draft consumption during a stable silent interval.
After several short backbone wakes, DiffuPlex-\textsc{Speak} consumes two consecutive 16-frame drafts without revision.
The number inside each backbone pass indicates the number of frames realized before the next wake.
}
\label{afig:case_long}
\end{figure}

\begin{figure}[t]
\centering
\includegraphics[width=\linewidth]{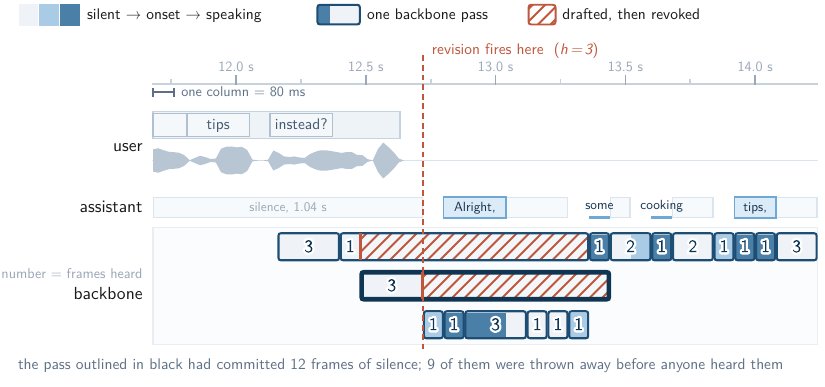}
\caption{
Revision after partial draft consumption.
For DiffuPlex-\textsc{Speak}, the highlighted backbone pass drafts 12 frames of assistant silence.
Revision fires after three frames have been consumed, discarding the remaining nine unplayed frames before a new assistant continuation is generated.
Hatched regions denote drafted frames that are revoked before playback.
}
\label{afig:case_revision}
\end{figure}

We further test this explanation by replaying DiffuPlex outputs through the original autoregressive temporal path.
The purpose of this experiment is to separate errors in \emph{what} DiffuPlex decided to say from errors in \emph{how} that content was acoustically realized.
Starting from a DiffuPlex-\textsc{Speak} output, we run the frozen \texttt{personaplex-rl-seamless} backbone frame by frame with $W=1$.
In the first replay condition, we keep both the text and semantic cb0 sequence generated by DiffuPlex fixed and regenerate only cb1--7; thus, the linguistic content and semantic audio decisions are unchanged, but the acoustic codebooks are produced with their ordinary autoregressive temporal conditioning.
In the second condition, we keep only the DiffuPlex text fixed and regenerate all audio codebooks cb0--7.
As reported in the main text, restoring the autoregressive path in the first condition already recovers most of the measured UTMOSv2 gap, while regenerating cb0 as well provides only a small additional gain.
Together with the NLL intervention above, this identifies missing same-frame semantic conditioning as one contributor, while the replay shows that restoring broader autoregressive audio-token conditioning recovers much of the measured quality gap.
These replay experiments are diagnostic analyses, not additional stages of DiffuPlex inference.

\begin{figure}[t]
\centering
\includegraphics[width=\linewidth]{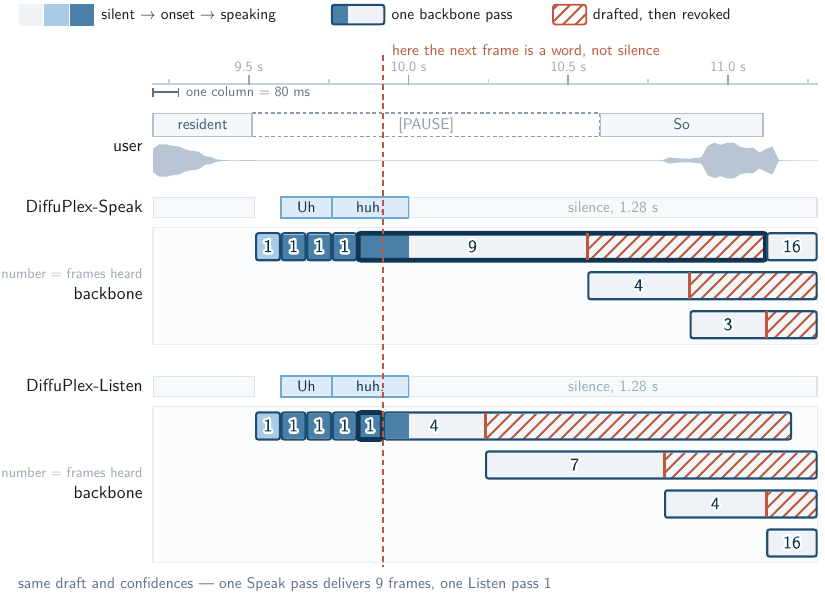}
\caption{
Different retention policies on the same interaction.
Because the prediction at $h=2$ is lexical, DiffuPlex-\textsc{Listen}
retains only the autoregressive frame ($r=a=1$), whereas
DiffuPlex-\textsc{Speak} retains $r=16$ and consumes $a=9$ before revision.
}
\label{afig:case_policy}
\end{figure}

\begin{figure}[t]
\centering
\includegraphics[width=\linewidth]{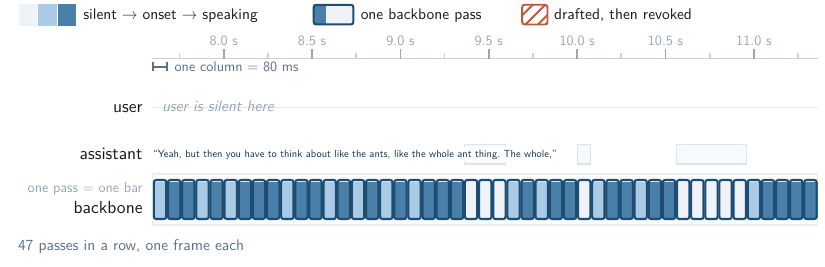}
\caption{
Repeated one-frame retention during a stable user context.
Although the user remains silent, DiffuPlex-\textsc{Speak} performs 47 consecutive backbone passes that each advance only one frame.
This example shows that limited acceleration can arise because the retention rule repeatedly admits only the immediate next frame, rather than because long drafts are repeatedly consumed and then revoked.
}
\label{afig:case_failure}
\end{figure}

\subsection{Qualitative Examples and Failure Cases}
\label{asubsec:qualitative}

Figures~\ref{afig:case_long}--\ref{afig:case_failure} illustrate four rollout behaviors.
Figure~\ref{afig:case_long} shows a stable silent interval in which several short wakes are followed by two consecutive backbone passes that each advance 16 frames without revision.
Figure~\ref{afig:case_revision} shows the opposite case: a backbone pass initially drafts 12 frames of assistant silence, but revision fires after three frames have been consumed, so the remaining nine unplayed frames are discarded before the assistant begins a new response.
These examples illustrate the two possible outcomes of a retained future: it may be consumed completely when the interaction remains stable, or only its unplayed suffix may be revoked when newly observed user input invalidates the draft.

Figure~\ref{afig:case_policy} compares the two retention policies on the same interaction.
At the marked boundary, the next predicted assistant state is lexical rather than silence.
DiffuPlex-\textsc{Speak} is therefore able to consume nine frames from a single retained draft before revision, whereas DiffuPlex-\textsc{Listen} advances only one frame from the corresponding wake.
Figure~\ref{afig:case_failure} shows a conservative-retention regime: despite a stable silent user context, the model executes 47 consecutive one-frame backbone passes.
Together, these examples show that limited acceleration can arise for two different reasons: a longer retained future may later be revised, or the retention rule may repeatedly admit only a single frame in the first place.

\section{Artifacts, Licenses, and Reproducibility}
\label{asec:licenses}

Table~\ref{atab:artifacts} lists the external datasets, models, and tools used in this work together with their licenses and public sources.
We use these artifacts for research purposes consistent with their intended use.

\begin{table*}[t]
\centering
\caption{
All external artifacts used in this work with their licenses and sources.
}
\label{atab:artifacts}
\small
\resizebox{\textwidth}{!}{
\begin{tabular}{lll}
\toprule
Artifact & License & URL \\
\midrule

\multicolumn{3}{l}{\textit{Backbone and Model Lineage}} \\
\texttt{personaplex-rl-seamless}
& CC BY-NC 4.0 + NVIDIA Open Model License
& \url{https://huggingface.co/kyutai/personaplex-rl-seamless} \\
PersonaPlex
& NVIDIA Open Model License
& \url{https://huggingface.co/nvidia/personaplex-7b-v1} \\
Moshi / Mimi
& CC BY 4.0
& \url{https://huggingface.co/kyutai/moshiko-pytorch-bf16} \\

\midrule
\multicolumn{3}{l}{\textit{Training Data}} \\
Seamless Interaction
& CC BY-NC 4.0
& \url{https://huggingface.co/datasets/facebook/seamless-interaction} \\

\midrule
\multicolumn{3}{l}{\textit{Evaluation Benchmarks}} \\
Full-Duplex-Bench v1.0
& CC BY-NC 4.0
& \url{https://github.com/DanielLin94144/Full-Duplex-Bench} \\
Full-Duplex-Bench v1.5
& CC BY-NC 4.0
& \url{https://github.com/DanielLin94144/Full-Duplex-Bench} \\
VoiceBench
& Apache 2.0
& \url{https://huggingface.co/datasets/hlt-lab/voicebench} \\
OpenAudioBench
& Not specified on dataset card
& \url{https://huggingface.co/datasets/baichuan-inc/OpenAudioBench} \\

\midrule
\multicolumn{3}{l}{\textit{Data Processing}} \\
Qwen3-ASR-1.7B
& Apache 2.0
& \url{https://huggingface.co/Qwen/Qwen3-ASR-1.7B} \\
Qwen3-ForcedAligner-0.6B
& Apache 2.0
& \url{https://huggingface.co/Qwen/Qwen3-ForcedAligner-0.6B} \\

\midrule
\multicolumn{3}{l}{\textit{Evaluation}} \\
\texttt{nvidia/parakeet-tdt-0.6b-v2}
& CC BY 4.0
& \url{https://huggingface.co/nvidia/parakeet-tdt-0.6b-v2} \\
UTMOSv2
& MIT
& \url{https://github.com/sarulab-speech/UTMOSv2} \\
\textit{gpt-5.6-luna}
& Proprietary (OpenAI API)
& \url{https://developers.openai.com/api/docs/models/gpt-5.6-luna} \\

\midrule
\multicolumn{3}{l}{\textit{Deferred-Execution Baselines}} \\
Silero VAD
& MIT
& \url{https://github.com/snakers4/silero-vad} \\
Smart Turn v3.2
& BSD 2-Clause
& \url{https://huggingface.co/pipecat-ai/smart-turn-v3} \\

\bottomrule
\end{tabular}}
\end{table*}

All main scored generations use seed 0 and retain the inference configuration used to produce each output.
Ablation runs additionally record the adapter path, backbone, inference configuration, and source-code revision, while analysis outputs retain their originating rollout and policy metadata.
Training configurations, checkpoints, generated audio, ASR transcripts, benchmark outputs, and rolling-draft analysis artifacts are stored separately so that each reported result can be traced to the configuration that produced it.

DiffuPlex is trained on four NVIDIA RTX A6000 GPUs.
Controlled runtime measurements use a single RTX A6000 as described in Appendix~\ref{asubsec:runtime_measurement}; no other jobs were running on the node during these measurements.
We additionally verify the autoregressive boundary case $W=1$: under matched priming, sampling, and random state, the rolling implementation produces the same token sequence as the official frame-by-frame \texttt{LMGen.step} path.
Selected generated audio and the qualitative examples are available on the anonymous project page at
\url{https://diffuplex.github.io/}.

\end{document}